\documentclass[10pt, conference, compsocconf]{IEEEtran}

\usepackage{booktabs} 
\usepackage{enumerate}
\usepackage{graphicx}
\usepackage{subfigure}
\usepackage{epstopdf}
\usepackage{tikz}
\usepackage{url}
\usepackage{pdfpages}
\usepackage{afterpage}
\usepackage{color, colortbl}
\usepackage{graphicx}

\usepackage{amsmath}
\usepackage{footmisc}
\usepackage{etoolbox}
\usepackage{listings}

\begin{document}


\title{The Impact of GPU DVFS on the Energy and Performance of Deep Learning: an Empirical Study}

\author{
	\IEEEauthorblockN{Zhenheng Tang, Yuxin Wang, Qiang Wang, Xiaowen Chu}
	\IEEEauthorblockA{Department of Computer Science\\
		Hong Kong Baptist University, Hong Kong\\
		\{zhtang, yxwang, qiangwang, chxw\}@comp.hkbu.edu.hk}
}


\maketitle
\begin{abstract}

Over the past years, great progress has been made in improving the computing power of general-purpose graphics processing units (GPGPUs), which facilitates the prosperity of deep neural networks (DNNs) in multiple fields like computer vision and natural language processing. A typical DNN training process repeatedly updates tens of millions of parameters, which not only requires huge computing resources but also consumes significant energy. In order to train DNNs in a more energy-efficient way, we empirically investigate the impact of GPU Dynamic Voltage and Frequency Scaling (DVFS) on the energy consumption and performance of deep learning. Our experiments cover a wide range of GPU architectures, DVFS settings, and DNN configurations. We observe that, compared to the default core frequency settings of three tested GPUs, the optimal core frequency can help conserve 8.7\%$\sim$23.1\% energy consumption for different DNN training cases. Regarding the inference, the benefits vary from 19.6\%$\sim$26.4\%. Our findings suggest that GPU DVFS has great potentials to help develop energy efficient DNN training/inference schemes. 

\end{abstract}

%
%
%


\begin{IEEEkeywords}
	Graphics Processing Units; Dynamic Voltage and Frequency Scaling; Deep Convolutional Neural Network;
	
\end{IEEEkeywords}

\maketitle

\section{Introduction}

Recent years witnessed the fast development of deep neural networks (DNN)\cite{lecun2015deep} that can achieve the state-of-art performance in many challenging AI problems, such as image recognition \cite{alexnet, googlenet, vggnet, resnet,imagenet}, object detection \cite{fast_rcnn, ssd,yolo} and natural language processing. 
However, this kind of successful applications heavily rely on the DNN training procedure, which requires a huge number of computational resources. 
Graphics processing units (GPUs) are currently the most widely used hardware to accelerate the training speed of DNNs. Different from the conventional CPUs, a high-end GPU board includes thousands of cores and a memory module with hundreds of Gigabytes of memory bandwidth. 

While most of the previous work addressed the model accuracy and training performance \cite{goyal2017,huai2018,you2018}, the energy consumption of those high-throughput GPU machines is usually overlooked. Large scale distributed systems \cite{lsddn,das2016distributed,goyal2017,deep_comp,li2014scaling,NIPS2014,chen2018adacomp} are being deployed to speed up the training of complex DNNs, but they also consume a significant amount of electricity. 
It becomes a critical issue to investigate the trade-off between training performance and energy consumption.

Dynamic voltage and frequency scaling (DVFS) is a widely used technique to balance the performance and power consumption of CPUs. In general, scaling up the CPU voltage/frequency can improve the performance but requires more power supply \cite{jiao2010power,mei2016survey,bridges2016understanding,eppminer}. 
Different from CPUs, GPUs have two sets of frequency domains, the core frequency ($f^{core}$) that controls the speed of ALU cores and other on-chip components, and the memory frequency ($f^{mem}$) that controls the SDRAM module. Since different GPU applications have different utilization of GPU cores and SDRAM \cite{mei2014benchmarking, mei2015tpds}, raising the frequency of the components with low utilization may bring no performance improvement but consume higher power. Furthermore, because energy consumption depends on the system power and running time, it is a non-trivial problem to understand how GPU DVFS affects the energy consumption of DNN training.

In this study, we empirically evaluate the performance and energy consumption of DNNs training under different GPU DVFS settings and investigate the impact and energy conservation opportunities of GPU DVFS. Our experiments cover a wide range of GPU architecture generations,
GPU DVFS settings, neural network configurations and convolution algorithms. Our major findings are listed as follows: 
\begin{enumerate}
\item Scaling up the GPU core frequency can improve the performance of DNN training and inference in varying degrees. Especially for the Turing GTX 2080Ti, the performance of different DNN training can achieve 17.4\%$\sim$38.2\% improvements by applying a 50\% higher core frequency than the default setting, while the performance of inference can be improved by 22.5\%$\sim$33.0\%. 
\item We observe that the default frequency settings are usually not optimal for energy efficiency. For the Pascal P100 and Volta V100 GPUs, the energy scaling curves with increasing core frequency generally show a valley trend and there exists a sweet spot. Compared to the default setting, the optimal core frequencies discovered by our experiments achieve 23.1\%, 14.5\% and 8.7\% average energy conservation for DNN training on three GPUs, respectively. For DNN inference, the average benefits are 26.4\%, 22,3\%, and 19.6\%. 
\item Three convolution algorithms, GEMM, FFT and Winograd, have varying degrees of energy conservation when applying GPU DVFS techniques. Compared to the default setting, the optimal core frequency brings an average 14.5\% energy savings for GEMM, 12.6\% for FFT and 15.8\% for Winograd.  
\end{enumerate}

The rest of this paper is organized as follows. Section \ref{sec:BM} introduces the background knowledge and related work of DNNs and GPU DVFS. Section \ref{sec:methodology} describes our experimental design and setup. Section \ref{sec:ER} demonstrates our experimental results and discusses the impact of GPU DVFS on the performance and energy consumption of different DNNs. Finally, Section \ref{sec:cc} concludes our work and discusses some future research directions.

\section{Background and Related Work} \label{sec:BM}
\subsection{Convolutional Neural Networks}
DNNs
have been rapidly developed as one of the most popular machine learning algorithms. Specifically, convolutional neural networks (CNNs) have achieved state-of-the-art performance in many AI applications. 
A typical CNN includes many convolutional layers \cite{alexnet, vggnet, resnet}. Some studies \cite{dl_survey} indicated that the computation of convolution layers usually dominate the training time. Besides, GPUs have been acknowledged as one of the most powerful devices to accelerate DNN training, whereas the downside 
is the huge energy consumption. It is important to develop not only fast but also energy efficient DNN training for GPUs. 

There are three popular implementations of the convolution operation. The first approach is transforming the convolution 
to matrix multiplication \cite{gemm}, which can then benefit from the highly optimized GPU library. The second approach is based on  Fourier transform \cite{fft}, which transforms the convolution operation in the spatial domain to point-wise multiplications in the Fourier domain. 
The last one is the Winograd algorithm  \cite{winograd}, which applies transforms to the input image and kernel to reduce the number of multiplications. NVIDIA's cuDNN library \cite{cudnn} implements all three algorithms. In addition to the exploration of the performance of different convolution algorithms in \cite{xia2016}, it is also important to investigate their energy efficiency. 

\subsection{GPU DVFS}
Recently NVIDIA has reinforced their GPUs with the extraordinary computational capability to meet the requirements of DNN training. For example, AutoML techniques often fully utilize hundreds or even thousands of GPUs to search for an efficient DNN structure with several weeks. E.g., Barret et. al \cite{nas_rl} adopted 800 GPUs to search for an efficient RNN for language modeling on PTB dataset.

DVFS is one of the most typical energy conservation techniques for traditional CPUs. Some previous GPU DVFS works indicated that GPUs have more complex energy scaling behaviors, and focused on how to balance the performance and energy efficiency of GPUs  \cite{jiao2010power,abe2014power,mei2016survey, bridges2016understanding, ge2013effects,qiang2018perf,wang2017power,lee2011,harmonia2015}. Mei et al. \cite{mei2017scheduling} and Chau et al. \cite{vincent2017scheduling} further adopted those DVFS-based energy conservation techniques to implement energy-efficient task scheduling for high-performance clusters. Recent papers \cite{ccbd2016, data2018, ee_dnn,icdcs2017} focused on the performance of scalability of DNN training on different software and hardware environments. Li et al. \cite{ee_dnn} and Cai et al. \cite{cai2017neuralpower} started to explore the energy characteristics of DNN processing on GPUs. We believe that it is essential to develop a deeper exploration of the impact of GPU DVFS on deep learning. 


\section{Methodology} \label{sec:methodology}
To conduct a solid exploration of the impact of GPU DVFS on deep learning, we design comparative experiments to cover different facets, including GPU architecture, DVFS setting, the structure of DNNs, and convolution algorithms. 
\subsection{Hardware Setup}
We perform our experiments on a single machine, which is equipped with an Intel i7 920 CPU and 8 GB main memory. We study three different GPUs, of which configurations are listed in Table \ref{tab:GPU_config}. The default frequency settings are bolded. 
Our experiments cover all the frequency options listed in Table \ref{tab:GPU_config}. 
Due to the limited support offered by the GPU vendor, we can only control the frequencies while the NVIDIA driver will automatically adjust the voltage accordingly.
We tune the GPU frequency setting with NVIDIA Inspector \cite{nvidiaInspector} and nvidia-smi \cite{nvidia-smi}. 
\begin{table}[ht]
	\centering
	\caption{Target GPU specifications}
	\label{tab:GPU_config}
	\begin{tabular}{|p{0.72in}|p{0.7in}|p{0.7in}|p{0.7in}|} \hline
		\textbf{Device} & \textbf{Tesla P100} & \textbf{Tesla V100} & \textbf{GTX 2080Ti} 	\\ \hline
		Architecture 		& Pascal	& Volta	& Turing	\\ \hline
		SMs/SM Cores 	   	& 16/128	& 28/128	& 72/64		\\  \hline
		Global mem.      & 16 GB	& 16 GB	& 12 GB \\ \hline
		Core freq. (MHz) & [544, 683, 810, 936, 1,063, 1,202, \textbf{1,328}] & [510, 652, 802, 945, 1,087, 1,237, \textbf{1,380}] & [950, 1,150, \textbf{1,350}, 1,550, 1,750, 1,950]\\ \hline
		Memory freq. (MHz) &  715 & 877 & [5,800, 6,300, \textbf{6,800}, 7,300]\\ \hline
	\end{tabular}
\end{table}
\begin{figure*}[t]
	\centering     
	\subfigure[The performance and energy of P100]
	{
		\includegraphics[width=0.31\linewidth]{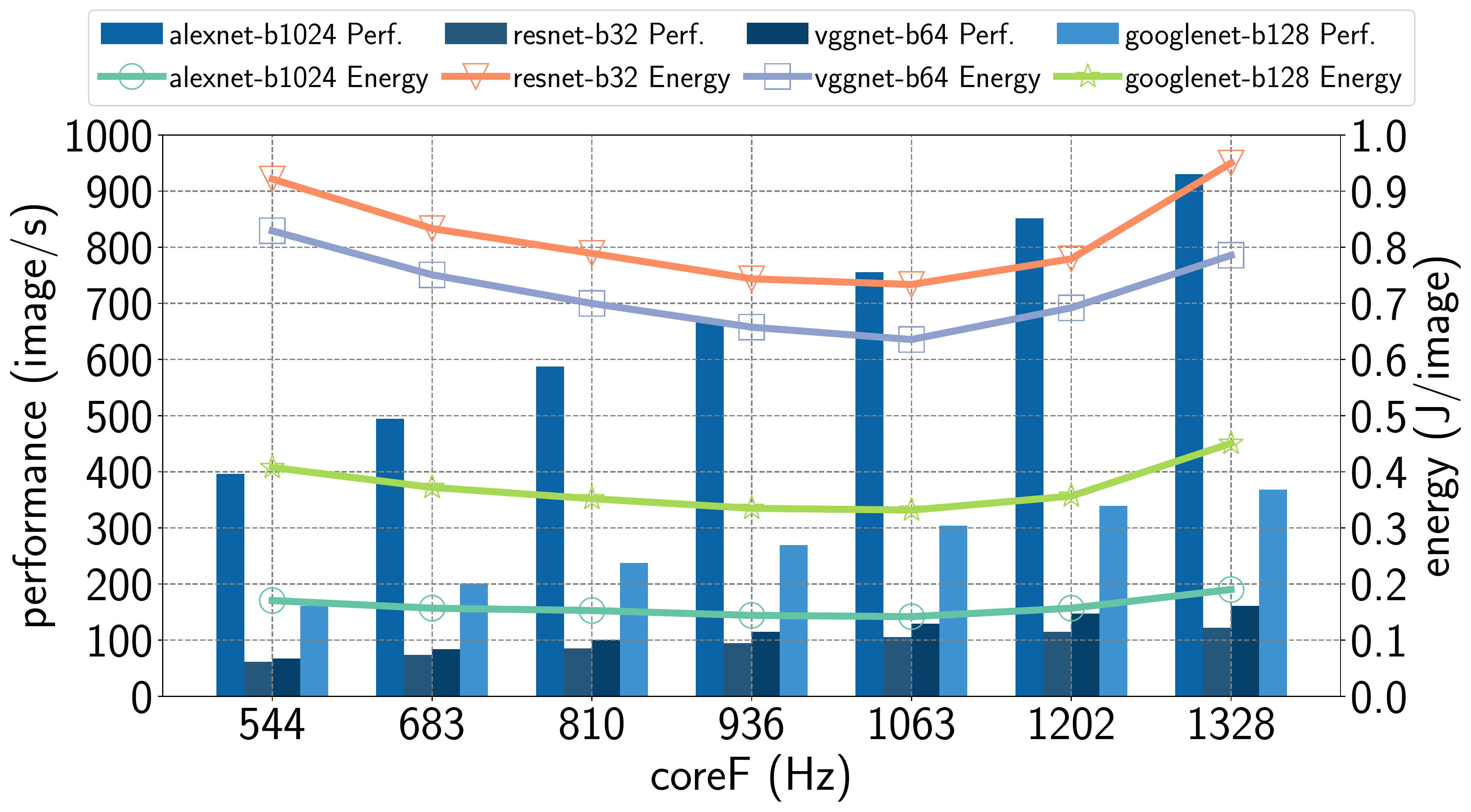}
		\label{fig:p100_alexnet_perf_energy}
	}
	\subfigure[The performance and energy of V100]
	{
		\includegraphics[width=0.31\linewidth]{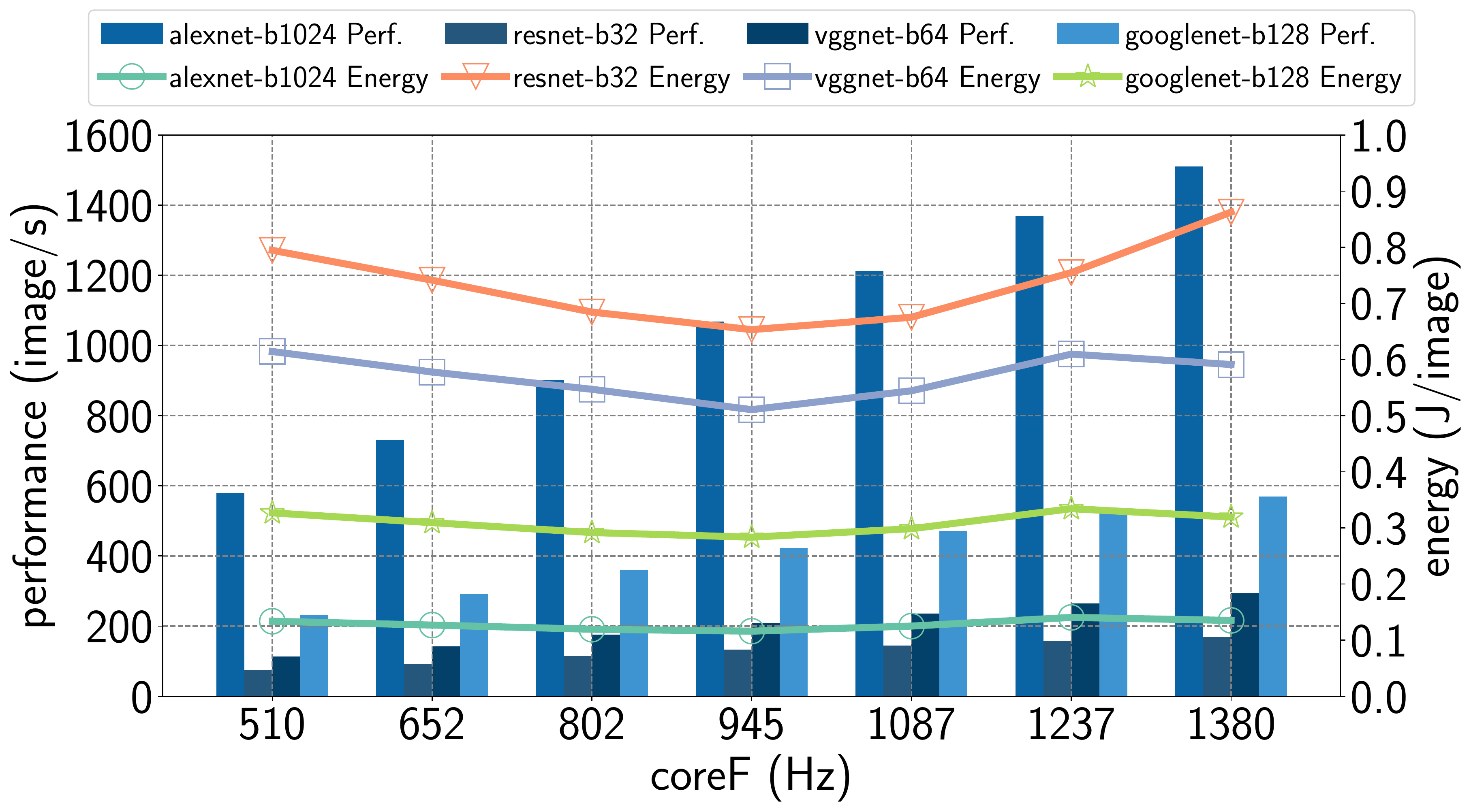}
		\label{fig:v100_alexnet_perf_energy}
	}
	\subfigure[The performance and energy of GTX 2080Ti]
	{
		\includegraphics[width=0.31\linewidth]{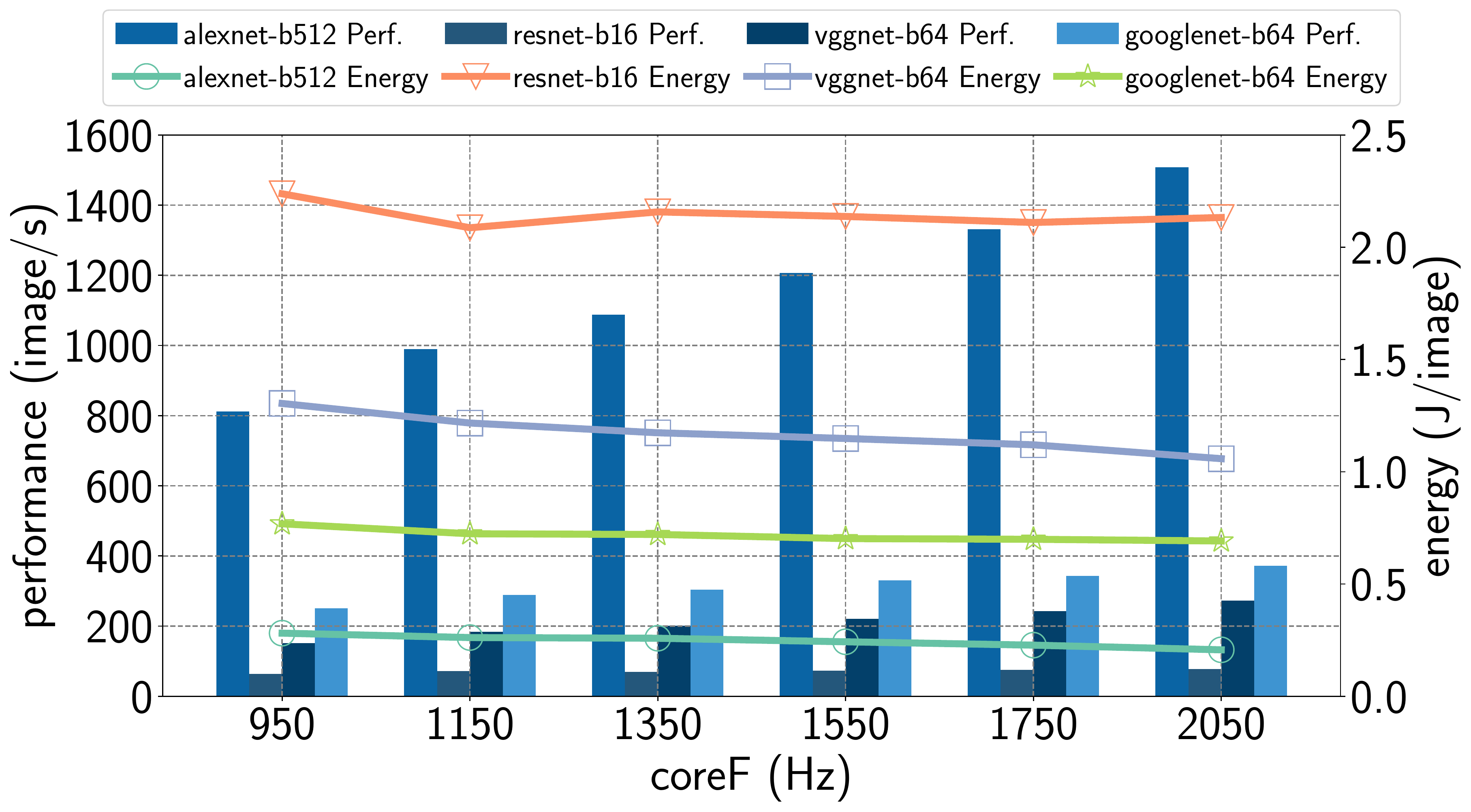}
		\label{fig:gtx2080ti_alexnet_perf_energy}
	}
	\caption{The impact of different core frequency settings on performance and energy consumptions of DNNs training.}
	\label{fig:dvfs_perf_energy}
\end{figure*}
\subsection{Network Setup}
We explore the impact on both the training and inference procedures of DNN. Caffe \cite{jia2014caffe} and TensorRT\footnote{https://developer.nvidia.com/tensorrt} are chosen as our training and inference implementations respectively. The CUDA version is 10.0 and the cuDNN version is 7.4.2 for both toolkits. We test four popular DNNs (i.e., AlexNet\cite{alexnet}, VggNet-16\cite{vggnet}, GoogleNet\cite{googlenet} and ResNet-50\cite{resnet}), 
and their setups are listed in Table \ref{tab:network_config}. Different batch sizes are tested for different DNNs according to the GPU memory availability. To explore the impact of DVFS on 
different convolution algorithms, we revise the Caffe source code to allow fixing the desired convolution algorithm. We test three algorithms, GEMM, FFT and Winograd. They are marked as ipc\_gemm, fft\_tile, winograd in the figures of Section \ref{sec:ER} respectively. 
\begin{table}[ht]
	\centering
	\caption{The experimental setup of neural networks}
	\label{tab:network_config}
	\begin{tabular}{|p{0.6in}|p{0.6in}|p{0.65in}|p{0.84in}|} \hline
		\textbf{Network} & \textbf{\# of layers} & \textbf{Parameters} & \textbf{Batch size} 	\\ \hline
		AlexNet 		& 8		& \~{}60 millions	& 128/256/512/1024	\\ \hline
		VggNet-16      & 16	& \~{}138 millions	& 16/32/64 \\ \hline
		GoogleNet 	   	& 22	& \~{}53 millions	& 16/32/64/128		\\  \hline
		ResNet-50 		& 50	& \~{}24 millions	& 8/16/32	\\ \hline
	\end{tabular}
\end{table}
\subsection{Performance and Power Measurements}
For DNN training, we define the performance, denoted by $Per$, as the processing images per second. We repeat the experiments for 120 times and record the average time of one training iteration and $Per$ can be obtained by dividing it by the batch size. The performance of inference is similar to training, except that it only records the time of forwarding. 
We measure the power consumption, denoted by $Pow$, by the NVIDIA management library (NVML)\cite{NVML} API. We implement a thread to sample the instantaneous power data during the training/inference procedure and the sampling interval is 2 ms. 
Since the thread may record those power data sampled before or after GPU execution, we intercept those power data within DNN processing from the sampling results and take the average value.
After obtaining all the performance and power data, we describe the energy consumption, denoted by $E$, with $\frac{Pow}{Per}$, which represents the average energy required by training/inferring a picture.

Notice that both core and memory frequency scaling are adopted to GTX 2080Ti. When exploring the effects of core frequency scaling, we calculate the geometric mean value among all the samples of each particular core frequency. The similar treatment is also used to explore the effects of memory frequency scaling and different convolution algorithms.



\section{Experimental Results}\label{sec:ER}
Due to the space limit, we only highlight some significant findings in the experimental results analysis. The complete experimental data can be found in the appendix. 

\begin{figure}[htbp]
	\centering     
	\subfigure[The performance and power of P100]
	{
		\includegraphics[width=0.48\linewidth]{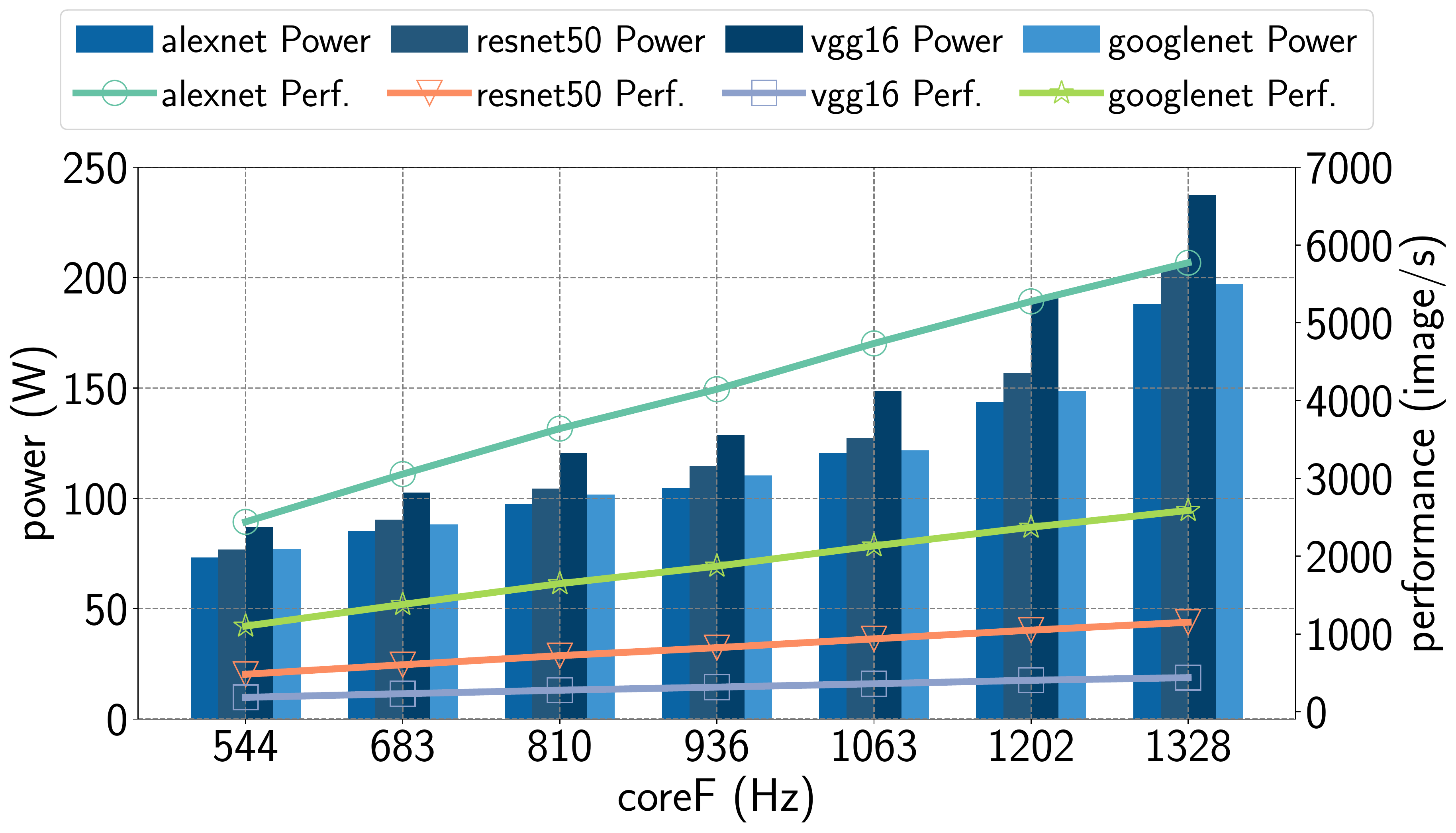}
		\label{fig:p100_infer_perf_pow}
	}
	\subfigure[The energy of P100]
	{
		\includegraphics[width=0.45\linewidth]{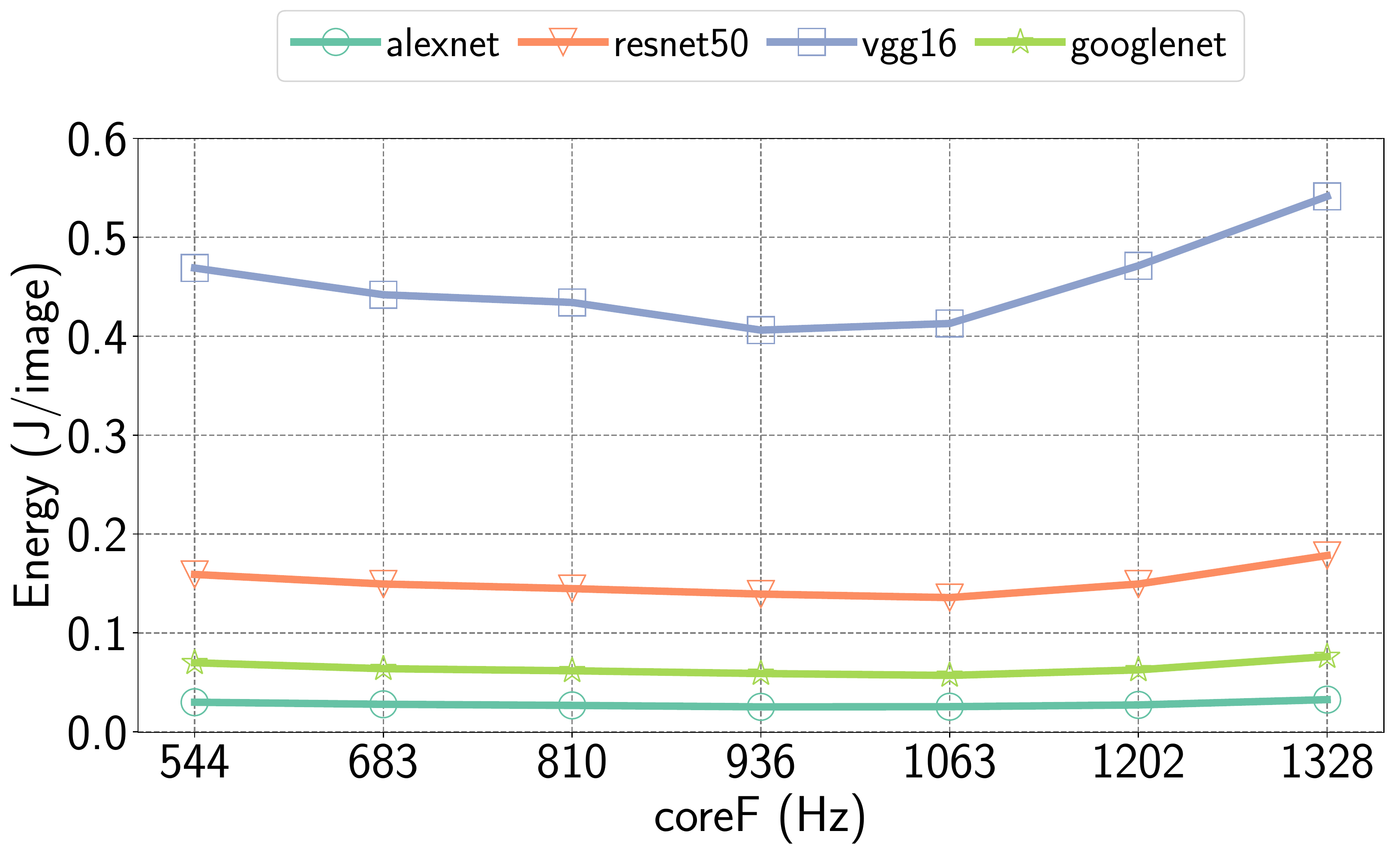}
		\label{fig:p100_infer_energy}
	}
	\subfigure[The performance and power of V100]
	{
		\includegraphics[width=0.48\linewidth]{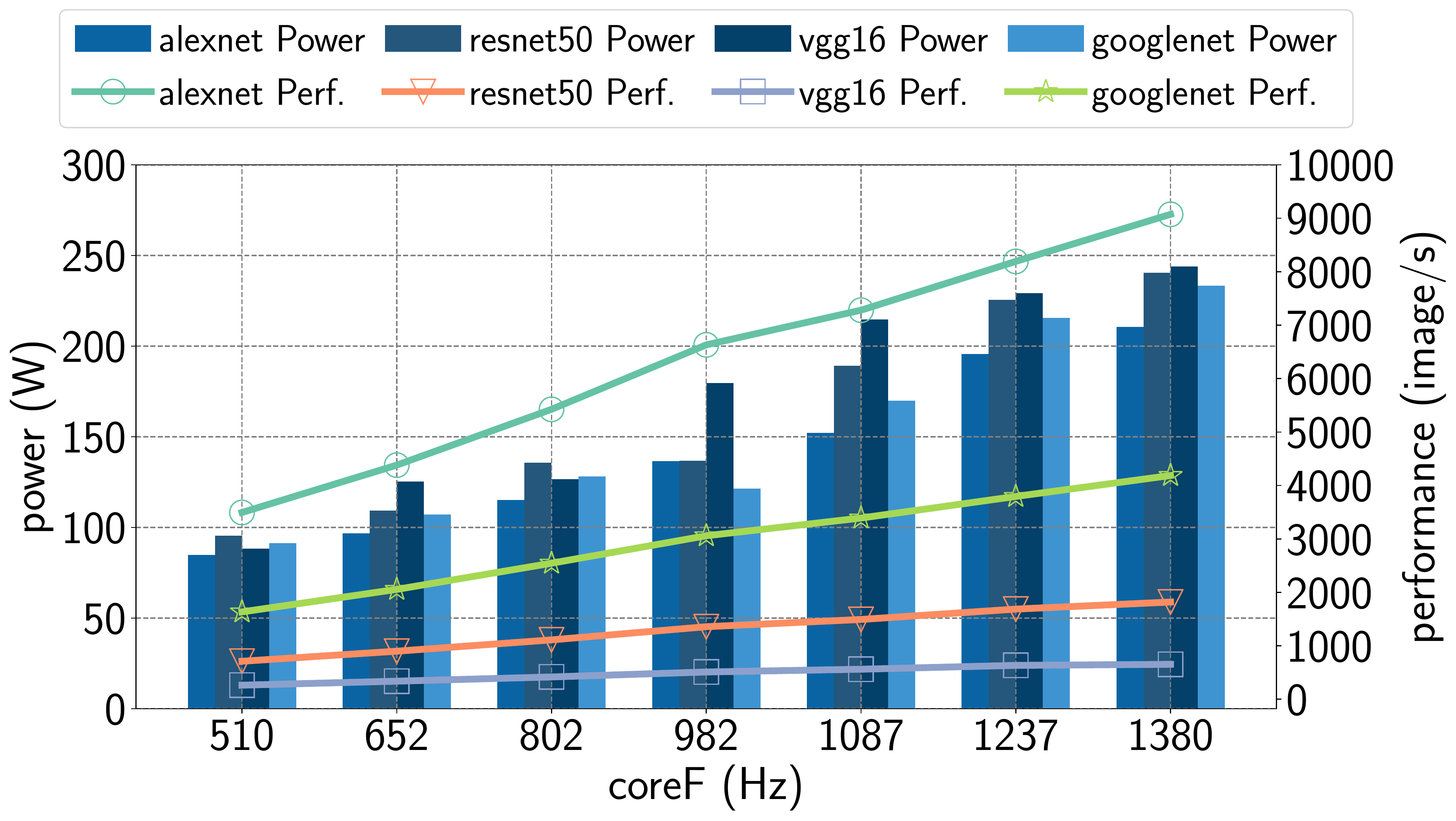}
		\label{fig:v100_infer_perf_pow}
	}
	\subfigure[The energy of V100]
	{
		\includegraphics[width=0.45\linewidth]{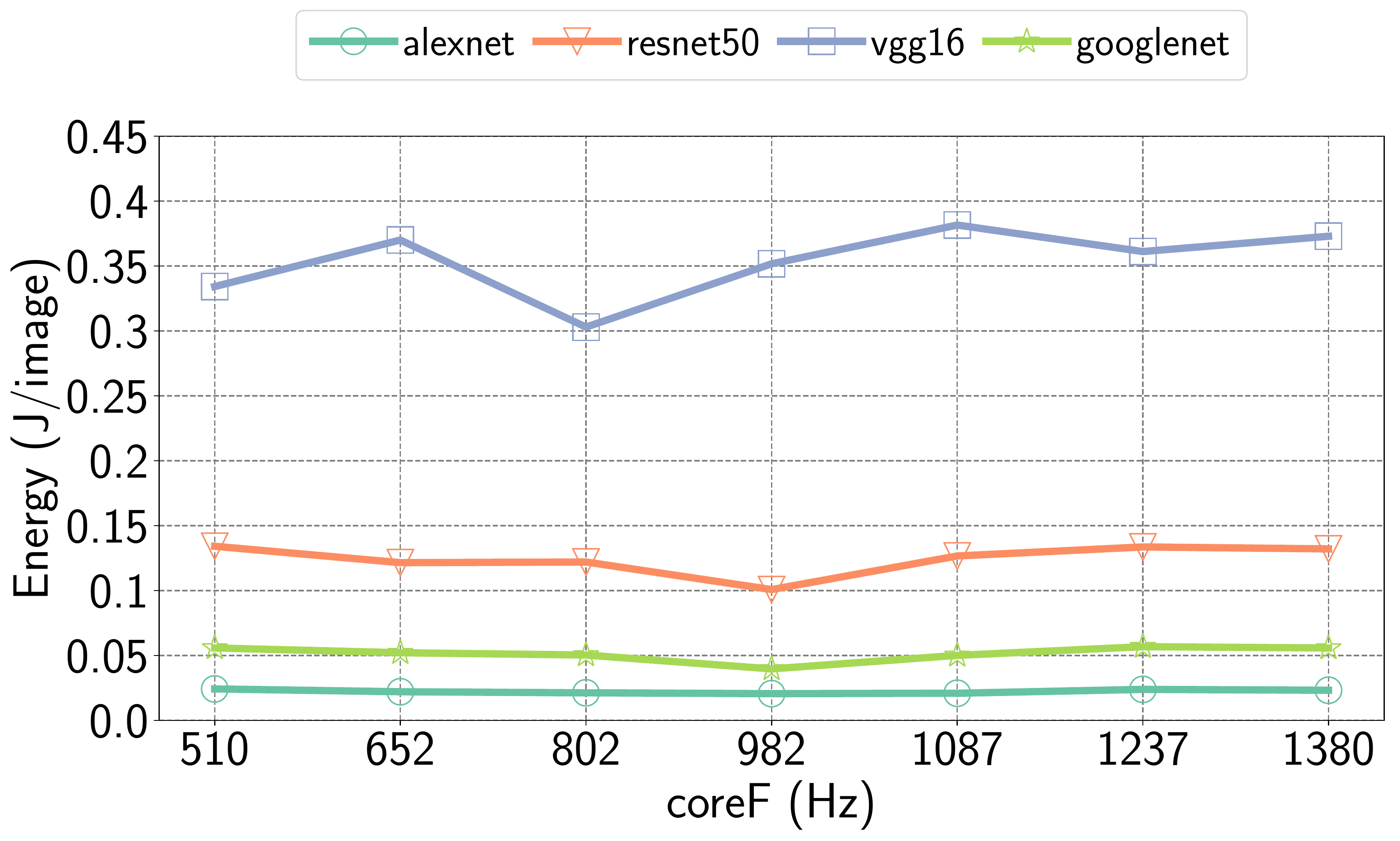}
		\label{fig:v100_infer_energy}
	}
	\qquad
	\subfigure[The performance and power of GTX 2080Ti]
	{
		\includegraphics[width=0.48\linewidth]{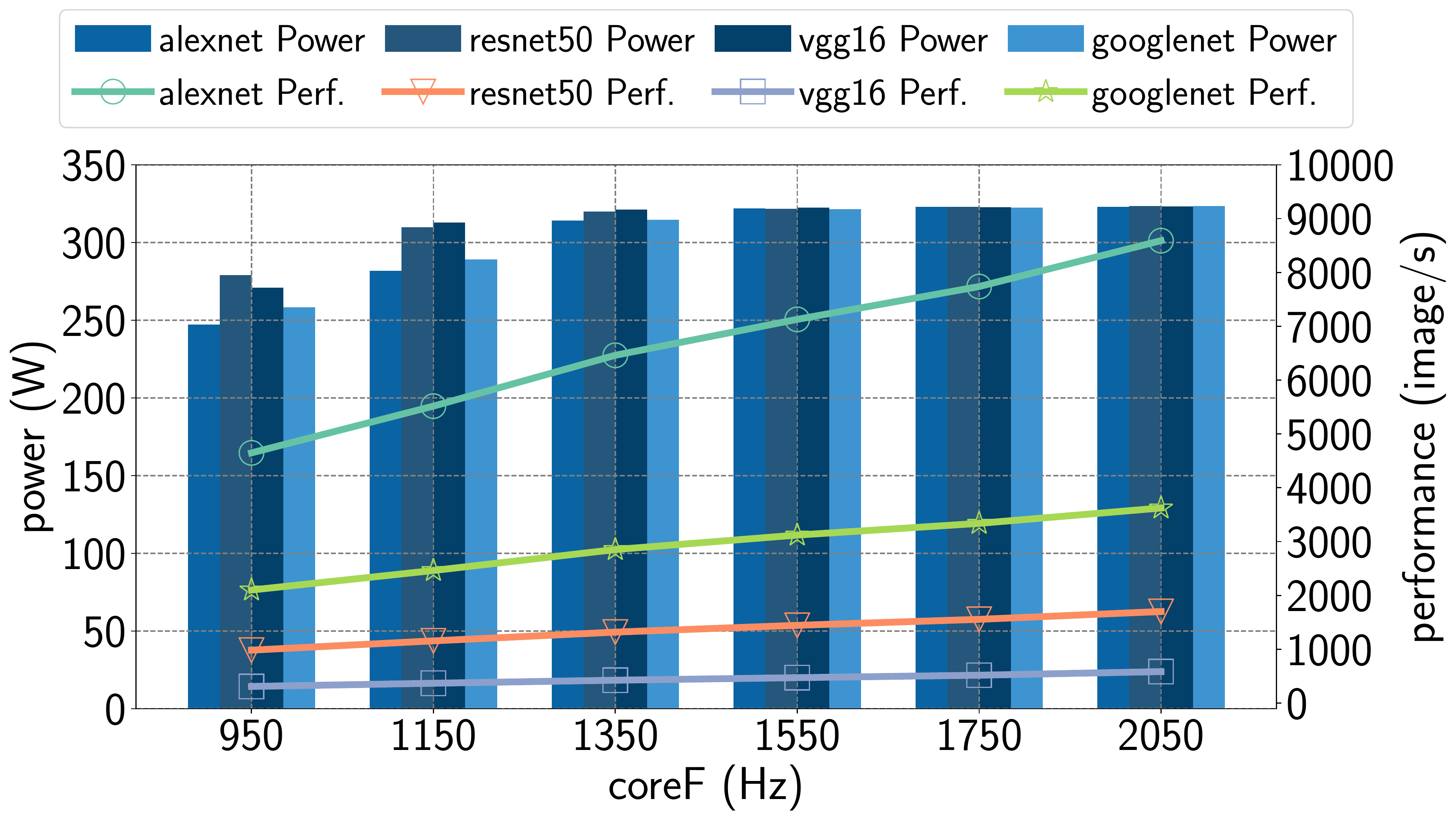}
		\label{fig:gtx2080ti_infer_perf_pow}
	}
	\subfigure[The energy of GTX 2080Ti]
	{
		\includegraphics[width=0.45\linewidth]{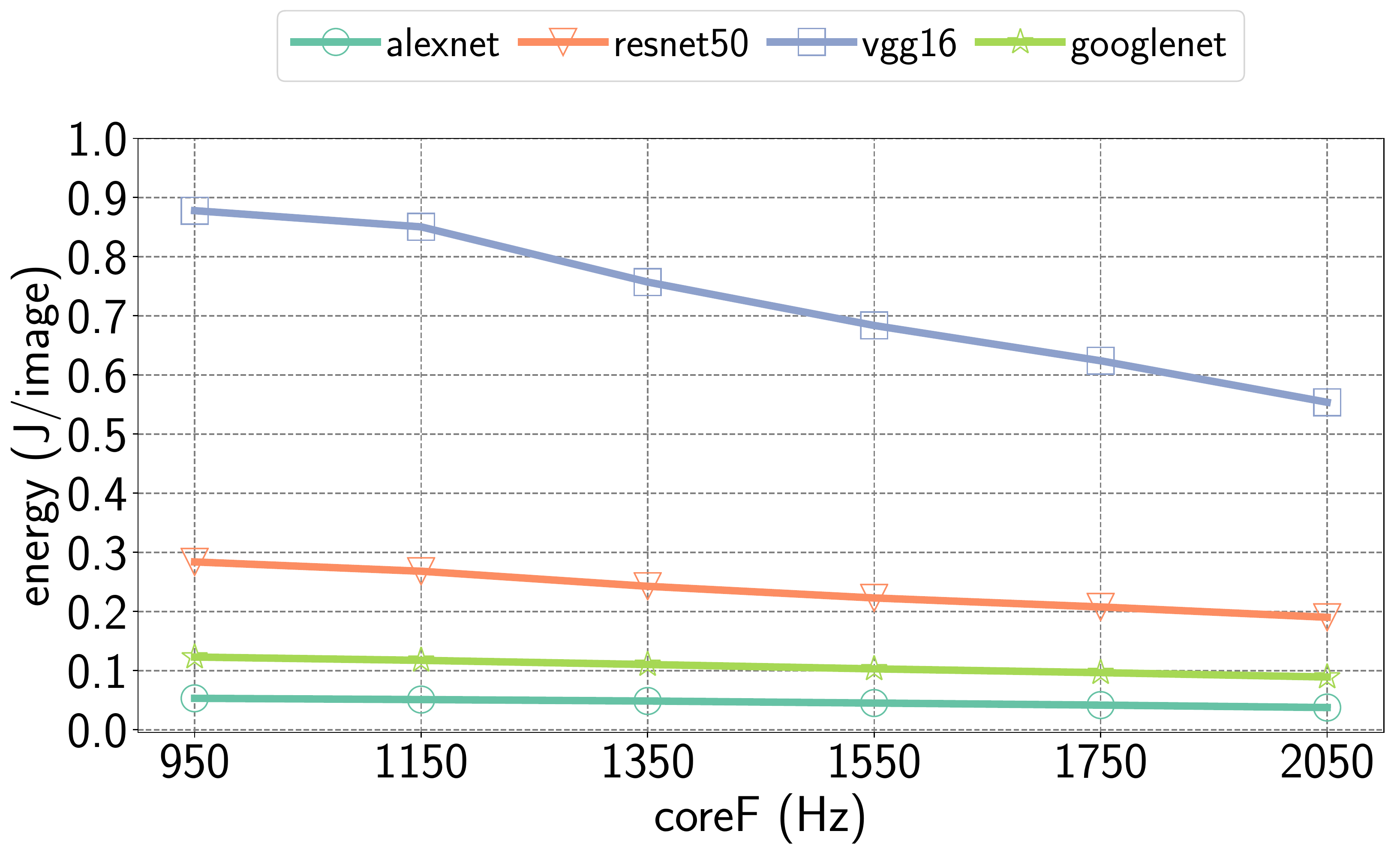}
		\label{fig:gtx2080ti_infer_energy}
	}
	\caption{The impact of different core frequency settings on performance, power and energy consumption of DNN inference.}
	\label{fig:inference_perf_pow_energy}
\end{figure}
\subsection{Impact of GPU DVFS on Performance and Energy Efficiency}
Figure \ref{fig:dvfs_perf_energy} shows the impact of different core frequency settings on the performance and energy consumption of DNN training. Some interesting phenomena are observed. First, scaling up the core frequency generally helps improve the training speed, especially for AlexNet and GoogleNet. 
The default core frequency of P100 and V100 are also the highest, which reasonably achieve the best performance. 
For GTX 2080Ti, scaling up the default 1350 MHz to 2050 MHz has 17.4\%$\sim$38.2\% performance improvements for different DNNs. Second, P100 and V100 generally perform a valley trend in the energy scaling curve with increasing core frequency. They achieve a sweet spot of the best energy efficiency in the middle core frequency level, while GTX 2080Ti seems to benefit more from a higher core frequency. The possible reason is that two Tesla GPUs have a dramatically increasing power consumption (refer to appendix) when the core frequency surpasses 1,000 MHz, while GTX 2080Ti does not have this issue. 

Different DNNs also demonstrate different performance and energy characteristics. The four DNNs have different numbers of convolution layers. 
It is reasonable that AlexNet always achieves the best performance and the best energy efficiency, while ResNet-50 has the lowest throughput and needs the largest energy consumption for each image processing. Notice that ResNet-50 shows the best convergence and classification accuracy among four DNNs. The progress of DNNs needs the support of GPU computing energy. 
\begin{table}[ht]
	\caption{Energy Conservation on DNN training/inference by the optimal core frequency: different CNNs}
	\begin{center}
		\begin{tabular}{|p{0.55in}|p{0.3in}|p{0.3in}|p{0.3in}|p{0.3in}|p{0.3in}|p{0.3in}|}
			\hline
			network & \multicolumn{3}{c|}{DNN training} & \multicolumn{3}{c|}{DNN inference} \\ \cline{2-7}
			& P100 & V100 & 2080Ti & P100 & V100 & 2080Ti \\ \hline
			AlexNet & 25.7\% & 7.7\% & 20.2\% & 28.7\% & 17.9\% & 21.3\% \\ \hline
			VggNet-16 & 19.1\% & 17.9\% & 9.2\% & 25.7\% & 18.9\% & 26.9\%\\ \hline
			GoogleNet & 24.3\% & 7.0\% & 2.3\% & 28.2\% & 27.8\% & 10.9\% \\ \hline
			ResNet-50 & 23.1\% & 25.3\% & 3.2\% & 23.1\% & 24.7\% & 19.5\% \\ \hline
		\end{tabular}
	\end{center}
	\label{tab:energy_saving_dnn}
\end{table}

Figure \ref{fig:inference_perf_pow_energy} illustrates the impact of different core frequency settings on DNN inference. It is observed that the benefits of GPU DVFS is similar to DNN training. Higher core frequency leads to better image processing throughput. For GTX 2080Ti, scaling up the default 1350 MHz to 2050 MHz has 22.5\%$\sim$33.0\% performance improvements for different DNNs. The energy curves of P100 and V100 achieve the best energy efficiency in the middle frequency zone, while GTX 2080Ti benefits more from a high frequency. 

It is also interesting to explore the energy saving by applying the optimal frequency setting compared to the default one. Table \ref{tab:energy_saving_dnn} concludes the results. Compared to the default setting, the optimal core frequency found in our experiments helps achieve remarkable energy conservation for DNNs training (23.1\% for P100, 14.5\% for V100 and 8.7\% for GTX 2080Ti on average). For DNNs inference, the average benefits are 26.4\%, 22,3\%, 19.6\% for three GPUs. 

\subsection{Impact of GPU DVFS on Convolution Algorithms}
\begin{figure}[htbp]
	\centering     
	\subfigure[The performance and power of P100]
	{
		\includegraphics[width=0.48\linewidth]{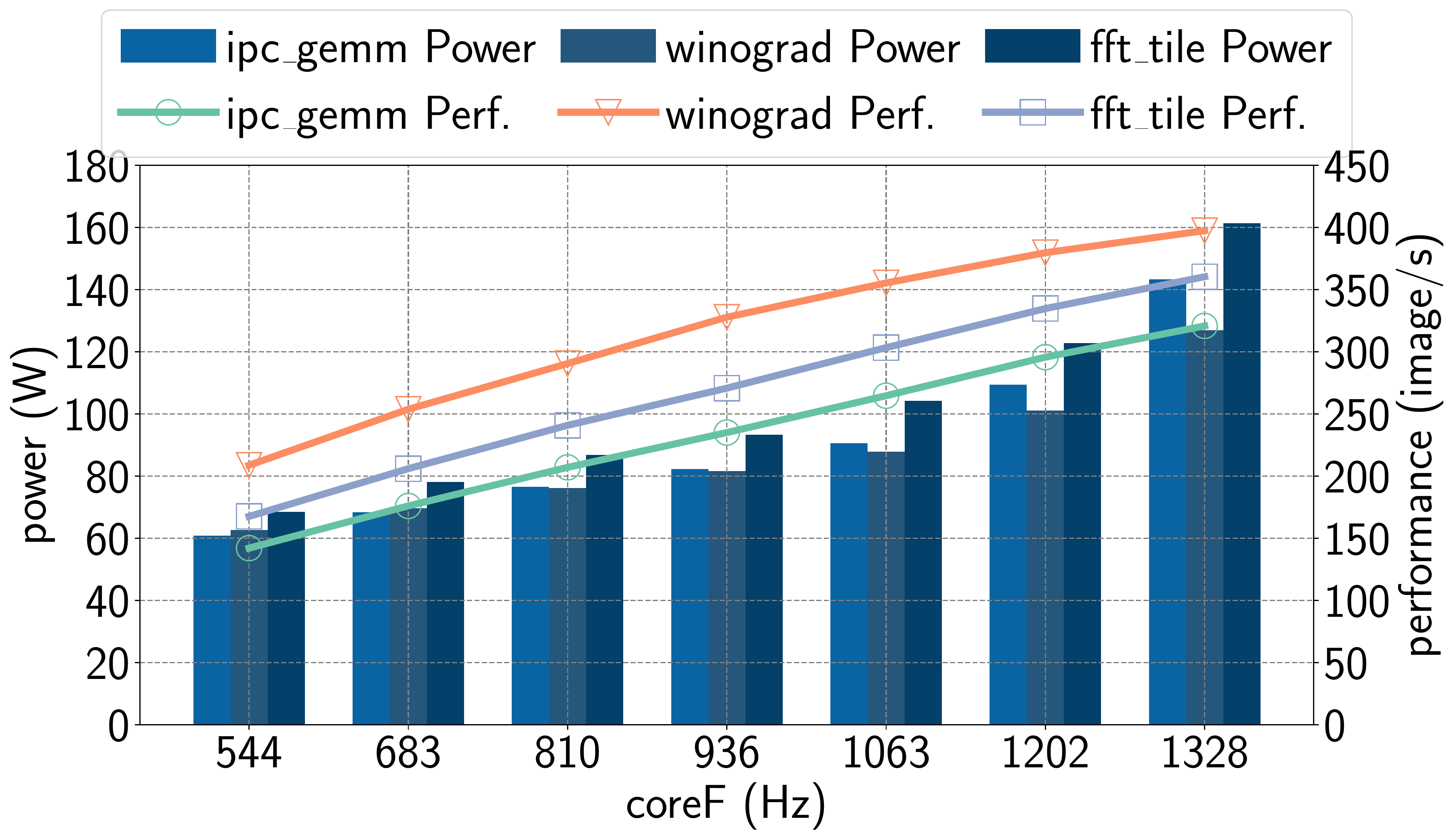}
		\label{fig:p100_conv_perf_pow}
	}
	\subfigure[The energy of P100]
	{
		\includegraphics[width=0.45\linewidth]{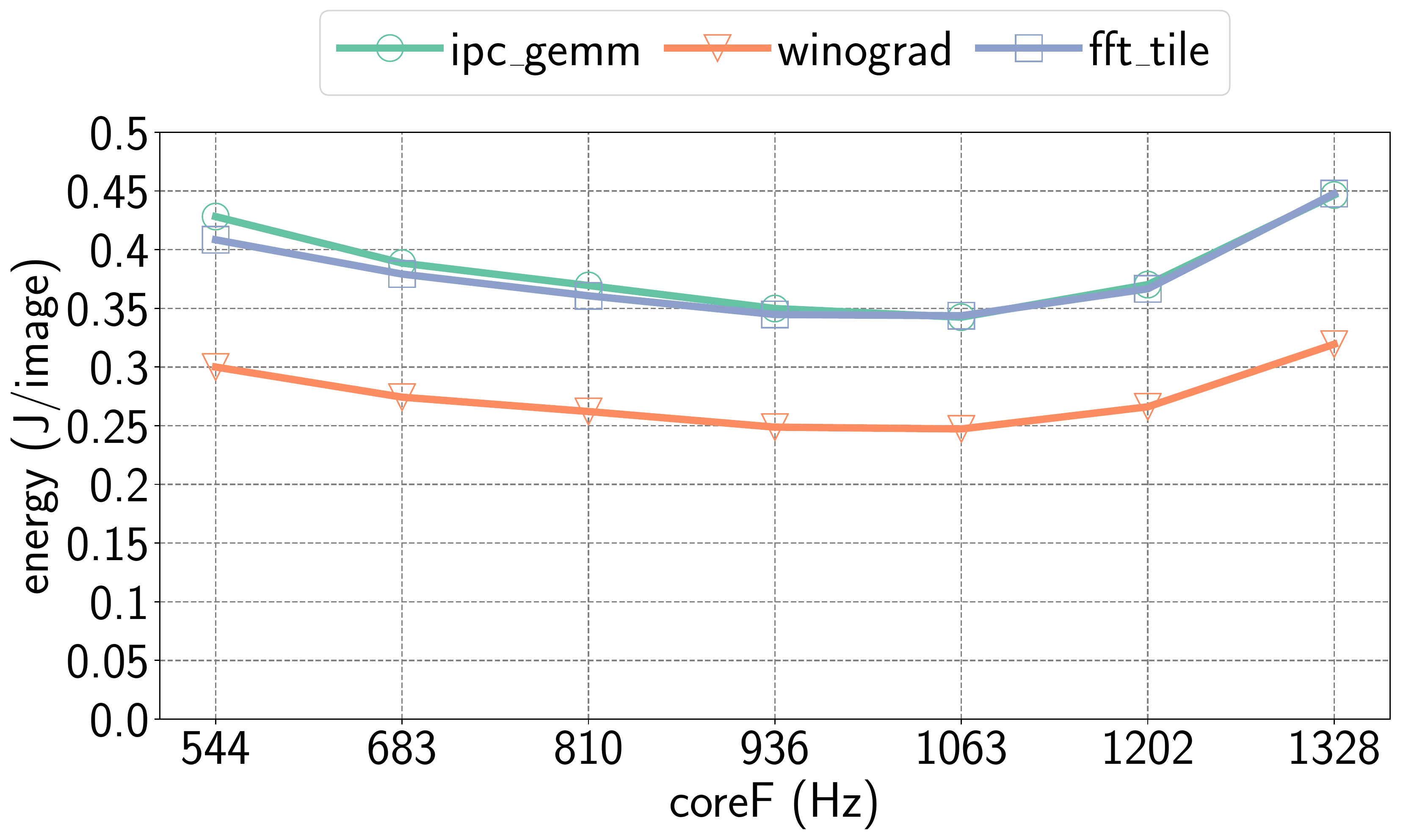}
		\label{fig:p100_conv_energy}
	}
	\subfigure[The performance and power of V100]
	{
		\includegraphics[width=0.48\linewidth]{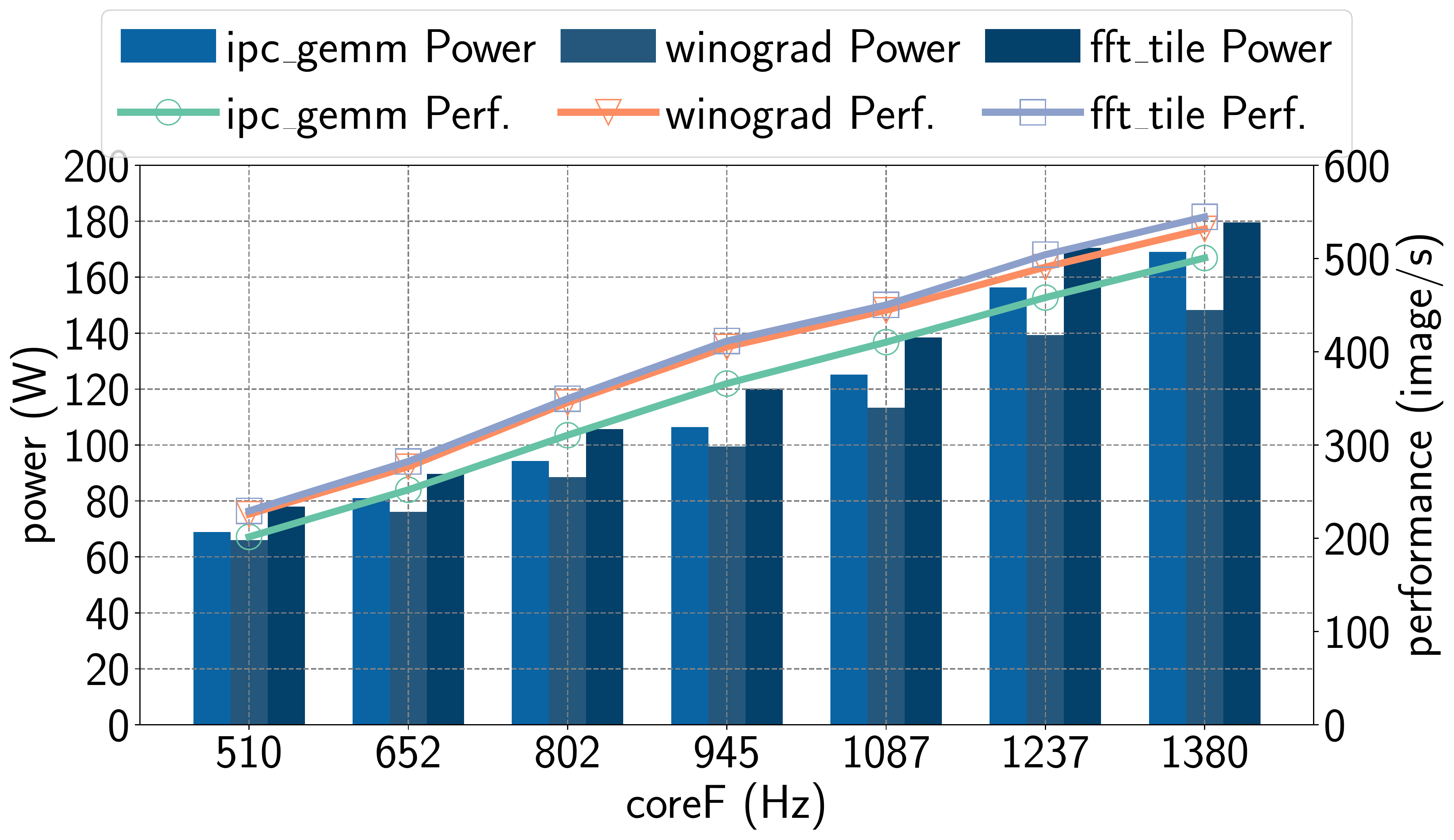}
		\label{fig:v100_conv_perf_pow}
	}
	\subfigure[The energy of V100]
	{
		\includegraphics[width=0.45\linewidth]{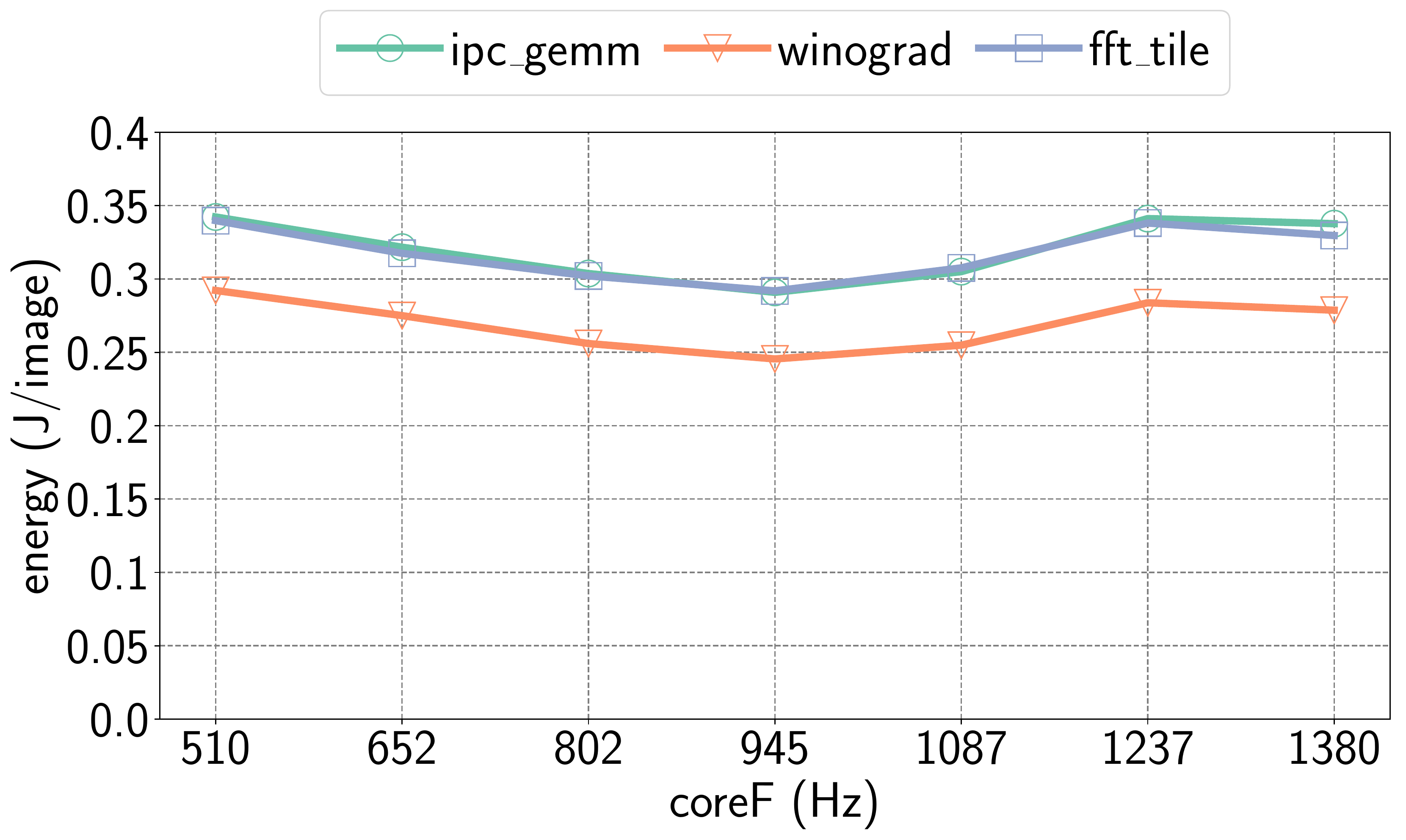}
		\label{fig:v100_conv_energy}
	}
	\subfigure[The performance and power of GTX 2080Ti]
	{
		\includegraphics[width=0.48\linewidth]{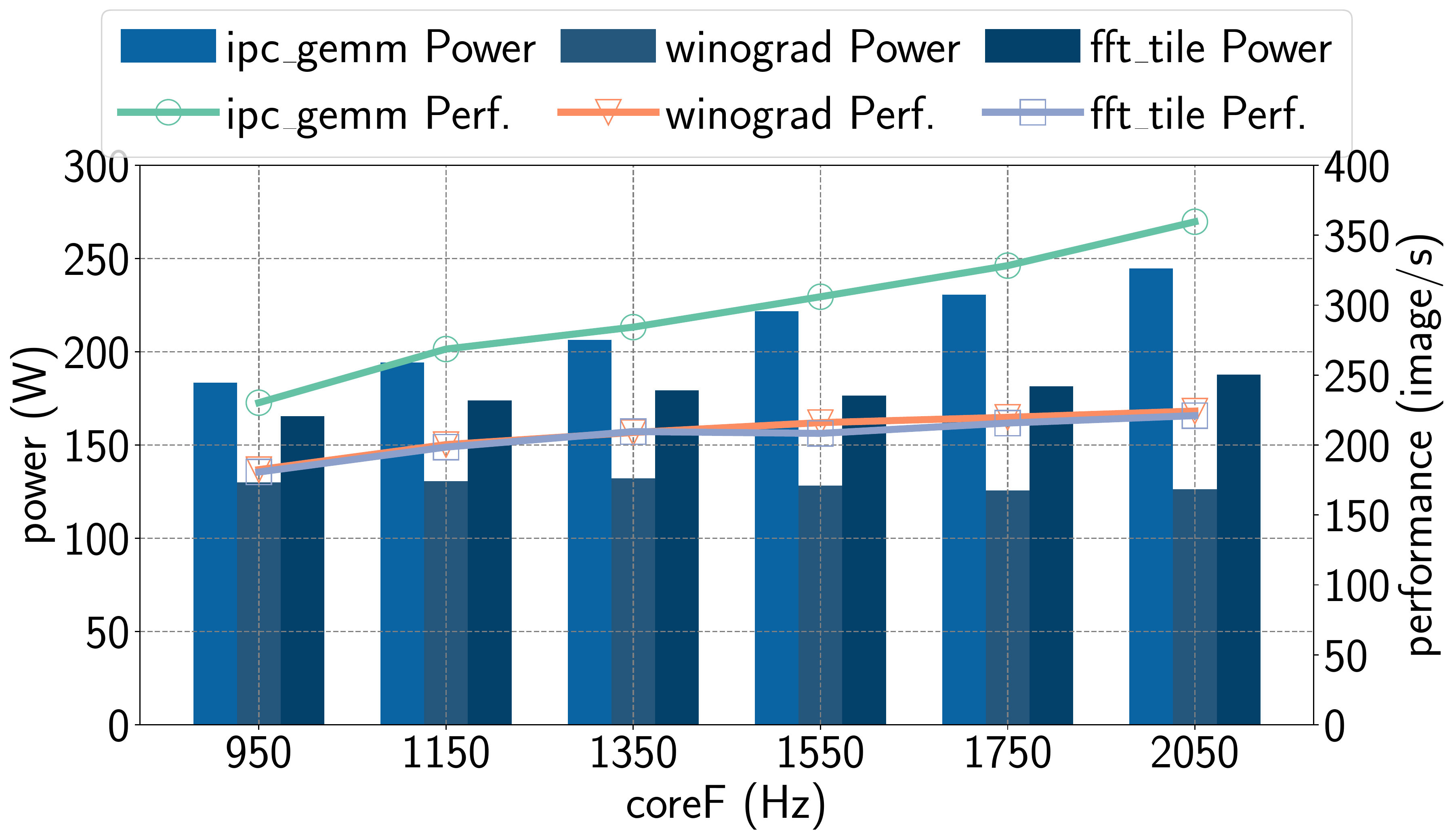}
		\label{fig:gtx2080ti_coreF_conv_perf_pow}
	}
	\subfigure[The energy of GTX 2080Ti]
	{
		\includegraphics[width=0.45\linewidth]{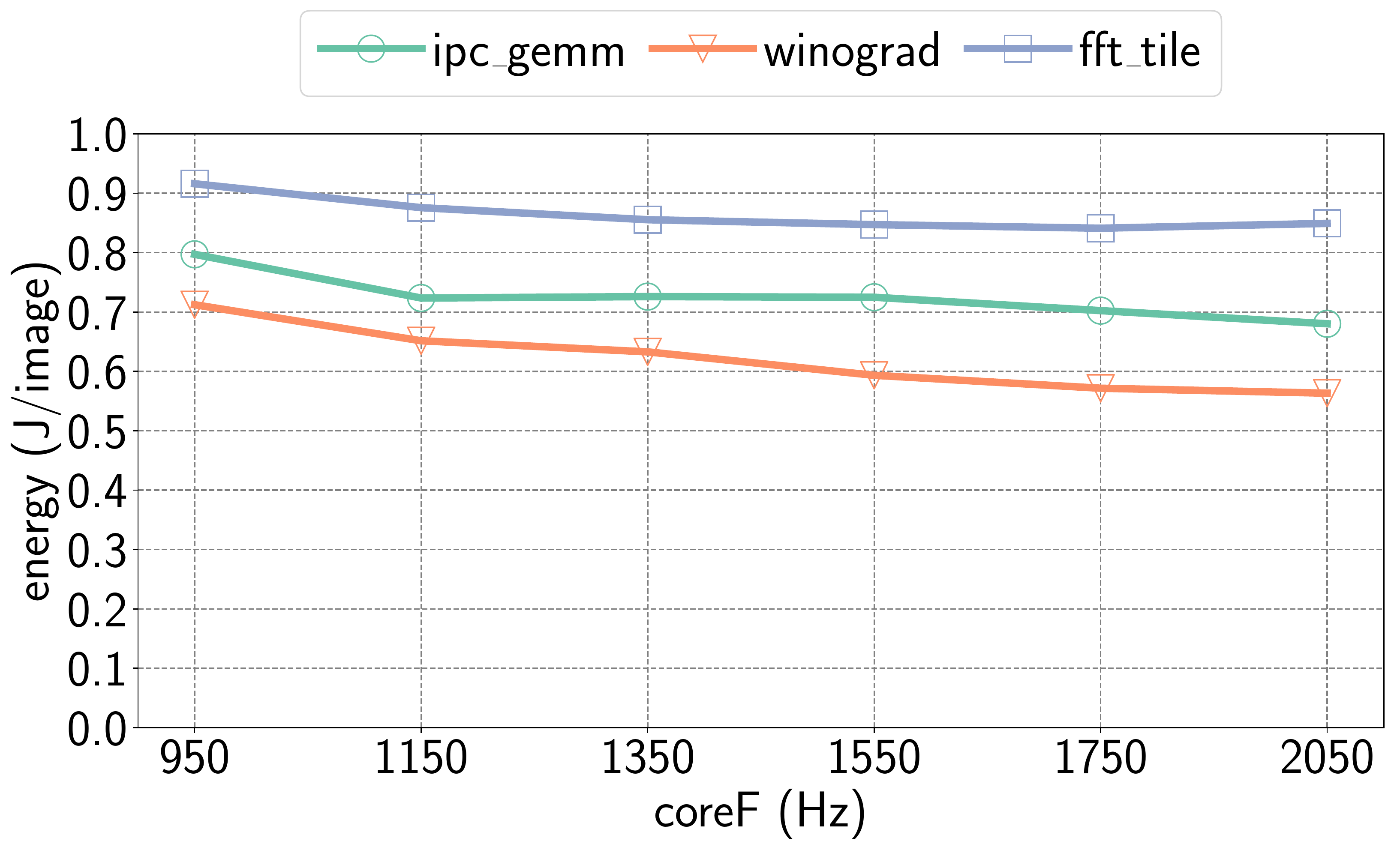}
		\label{fig:gtx2080ti_coreF_conv_energy}
	}
	\caption{The impact of different core frequency settings on performance, power and energy consumption of different convolution algorithms.}
	\label{fig:conv_algo_compare}
\end{figure}

Figure \ref{fig:conv_algo_compare} illustrates the impact of GPU DVFS on the performance and energy consumption of DNNs training when applying different convolution algorithms. For P100 and V100, the performance of three algorithms performs a similar linearly increasing trend with scaling up the core frequency. The energy consumption curves of them also show a valley trend and have a sweet spot on the middle-level core frequency. It can be interpreted by the fact that the power consumption of P100 and V100 have a larger jump when the core frequency surpasses 1,000 MHz. Different from P100 and V100, the performance of ipc\_gemm on GTX 2080Ti shows a higher acceleration rate than fft\_tile and Winograd. 
\begin{table}[ht]
	\caption{Energy Conservation on DNN training by the optimal core frequency: different convolution algorithms}
	\begin{center}
		\begin{tabular}{|p{0.55in}|p{0.3in}|p{0.3in}|p{0.3in}|}
			\hline
			Algorithm & P100 & V100 & 2080Ti \\ \hline
			GEMM & 23.3\% & 14.0\% & 6.3\%  \\ \hline
			FFT & 23.1\% & 11.5\% & 3.1\% \\ \hline
			Winograd & 25.1\% & 11.3\% & 11.0\%  \\ \hline
		\end{tabular}
	\end{center}
	\label{tab:energy_saving_conv}
\end{table}

Besides, we notice that the power consumption of fft\_tile and winograd have negligible changes when adjusting the core frequency. We also have explored the effects of memory frequency scaling and found it not significant. Thus, it is possible that the current implementations of those two convolution algorithms on Turing GPUs still cannot fully utilize the computational resources. Table \ref{tab:energy_saving_conv} concludes the energy conservation results of applying the optimal core frequency on three convolution algorithms, compared to the default setting. The average energy conservation is 14.5\% for GEMM, 12.6\% for FFT and 15.8\% for Winograd respectively.

\subsection{Discussion}
We have investigated the benefits brought by GPU DVFS for the performance improvement and energy conservation of DNN training/inference. First, the performance improvement brought by scaling up the core frequency depends on the GPU core utilization of the software. On the one hand, DNN training includes the data loading step that is not operated on GPUs. Whether the data loading latency can be well hidden significantly affects the GPU core utilization. On the other hand, the GPU kernels of tackling DNN training/inference mainly determine the GPU core utilization. Notice that for GTX 2080Ti, scaling up the default core frequency by 1.5$\times$ has 1.17$\sim$1.38$\times$ performance improvement for DNN training and 1.22$\sim$1.33$\times$ for DNN inference. Although Caffe, cuDNN and TensorRT have highly optimized implementations for DNN training/inference, the performance gap still exists.

Second, whether GPU DVFS helps conserve the energy consumption of DNN training/inference depends on the changing curves of both performance and power with the increase of the core/memory frequency. For example, as shown in Figure \ref{fig:v100_infer_perf_pow}, the performance of alexnet-b1024 on V100 is improved with an approximately equal ratio to the increase of the core frequency (nearly 83\% from $f^{core}$=510 MHz to $f^{core}$=945 MHz). Meanwhile, as shown in Figure \ref{fig:power_perf_appendix_p100_ipc_gemm_alexnet} in the appendix, the average power of V100 is only increased by 66\%. Thus, the energy curve has the lowest value at $f^{core}$=945 MHz. However, when $f^{core}\ge$1,000 MHz, 
the power consumption has a big jump since a higher core voltage is needed to support that frequency range, and then the energy consumption goes up again.
\section{Conclusion} \label{sec:cc}
In this paper, we investigate the impact of GPU DVFS on energy consumption and performance of DNN training and inference. Our experiments cover a wide range of GPU architectures, DVFS settings and CNNs. The results show that the optimal core frequency can not only help improve the DNN performance by up to 33\% but also conserve up to 23.1\% energy consumption of DNN training and 26.4\% of DNN inference. The observations suggest that GPU DVFS has great potentials to help develop energy efficient DNN processing schemes without significant performance degradation. 

There are two directions of our future explorations on energy efficient DNN training/inference. First, notice that the same voltage and frequency is applied throughout the feed-forwarding and back-propagation procedures. But different layers may have different energy conservation benefits from different DVFS settings. It is interesting to explore a layer-wise DVFS scheme for DNN training/inference to further reduce energy consumption. Second, considering a scheduling system for multiple DNN training tasks, GPU DVFS can perform as an effective technique to improve the system-wide throughput and decrease energy consumption. 

\section*{Acknowledgment}
The authors would like to thank the reviewers for their thorough and insightful comments and suggestions. The research was supported by Hong Kong RGC GRF grant HKBU 12200418.

\bibliographystyle{IEEEtran}
\bibliography{dvfs_dl.bbl}
\clearpage
\appendix
\section{Appendix}
In the appendix, we demonstrate the complete experimental results of four DNNs on three GPU cards, which are grouped by three different convolution algorithms: GEMM, Winograd, and FFT. We present how the performance and power change with respect to the GPU core and memory frequencies on different DNNs. Each figure includes the results of different batch sizes. Due to the limitation of GPU memory size, some large batch sizes are not supported. Notice that GTX2080Ti supports both core and memory frequency scaling. To give a comprehensive result of core frequency scaling, we calculate the geometric mean of the data of each core frequency across all memory frequency sets. For memory frequency scaling, we calculate the geometric mean of the data of each memory frequency across all memory frequency sets.

\subsection{Using implicit GEMM algorithm}
\begin{figure*}[htbp]
	\centering     
	\subfigure[power and performance of training AlexNet]
	{
		\includegraphics[width=0.4\linewidth]{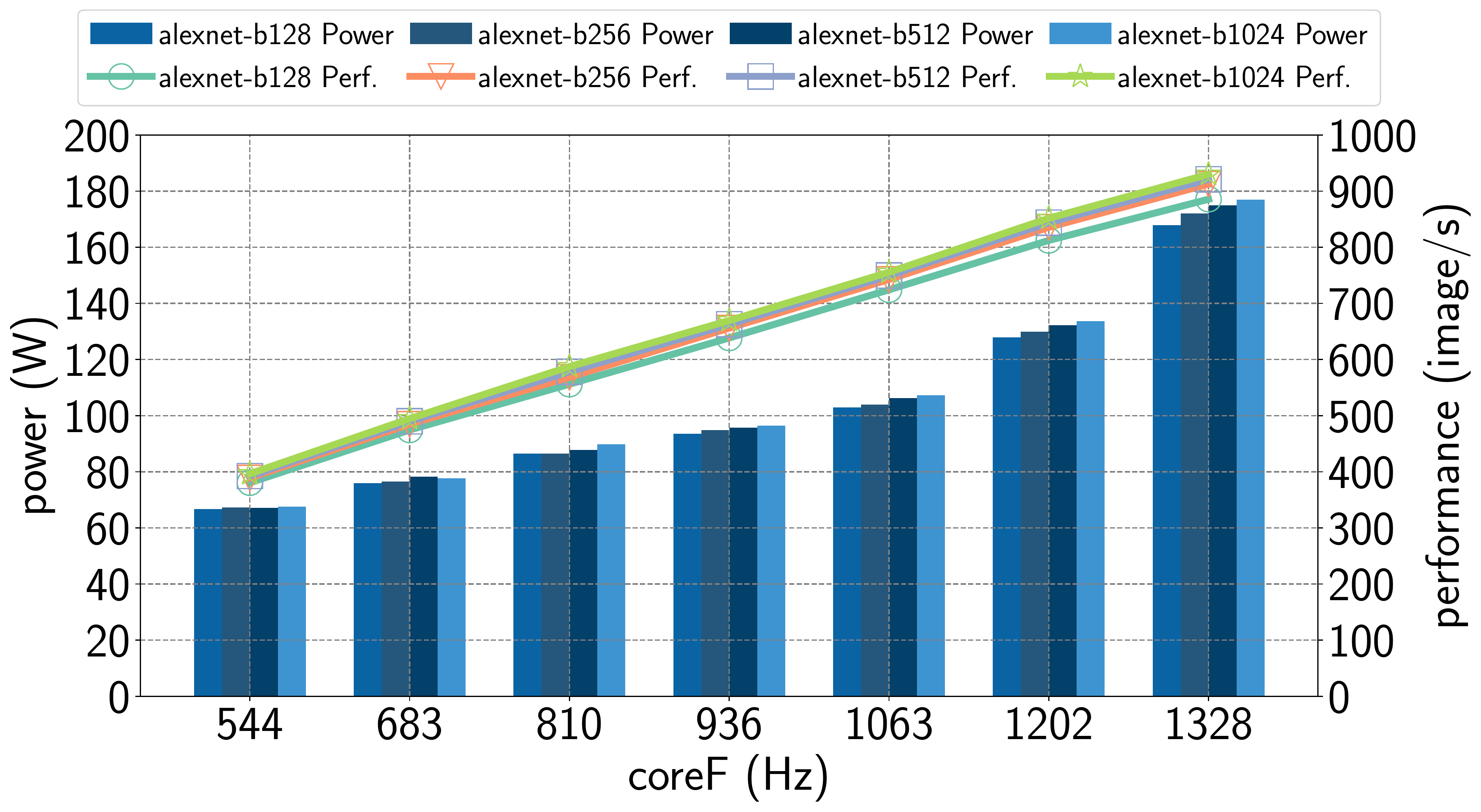}
		\label{fig:power_perf_appendix_p100_ipc_gemm_alexnet}
	}
	\subfigure[power and performance of training GoogleNet]
	{
		\includegraphics[width=0.4\linewidth]{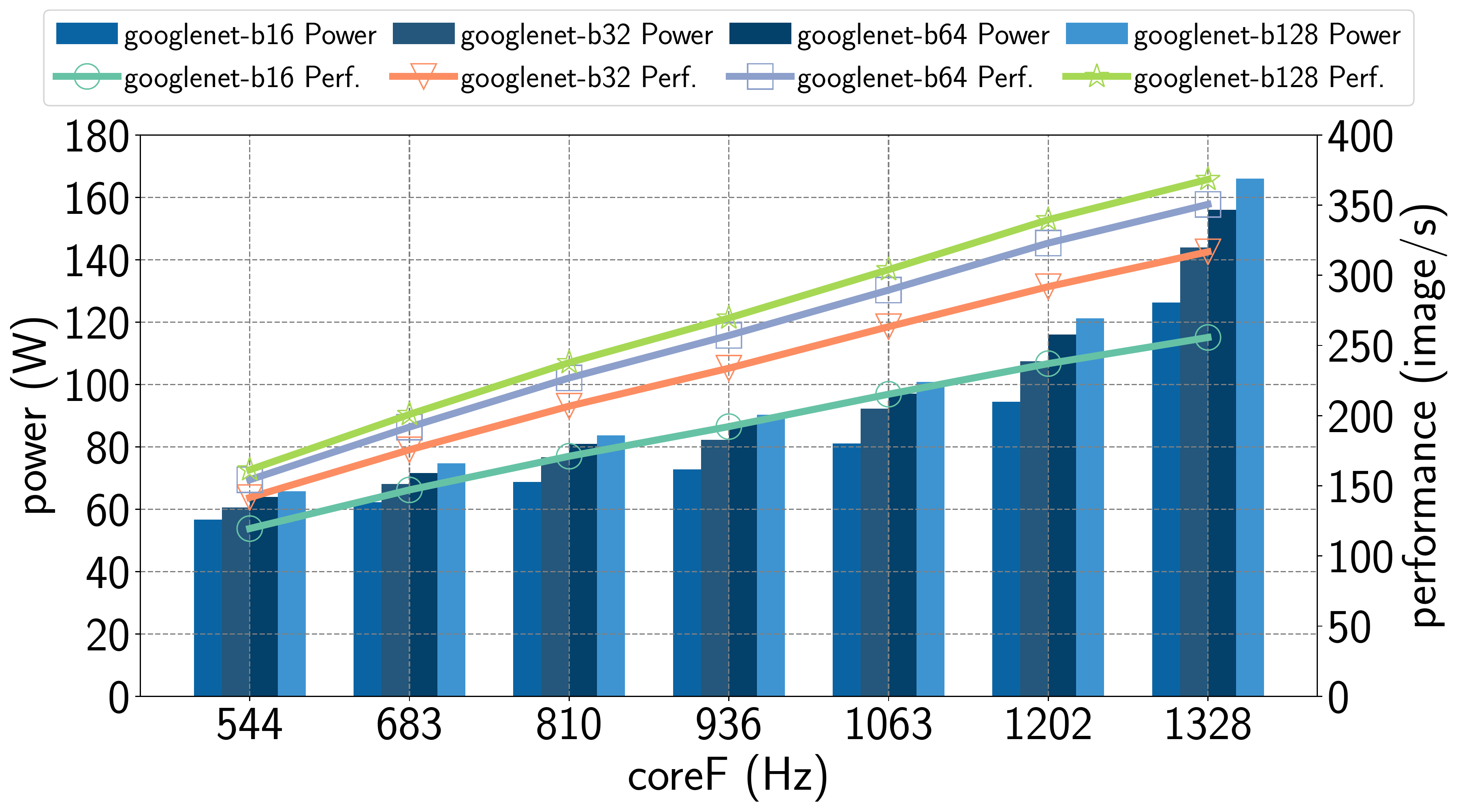}
		\label{fig:power_perf_appendix_p100_ipc_gemm_googlenet}
	}
	\subfigure[power and performance of training VggNet-16]
	{
		\includegraphics[width=0.4\linewidth]{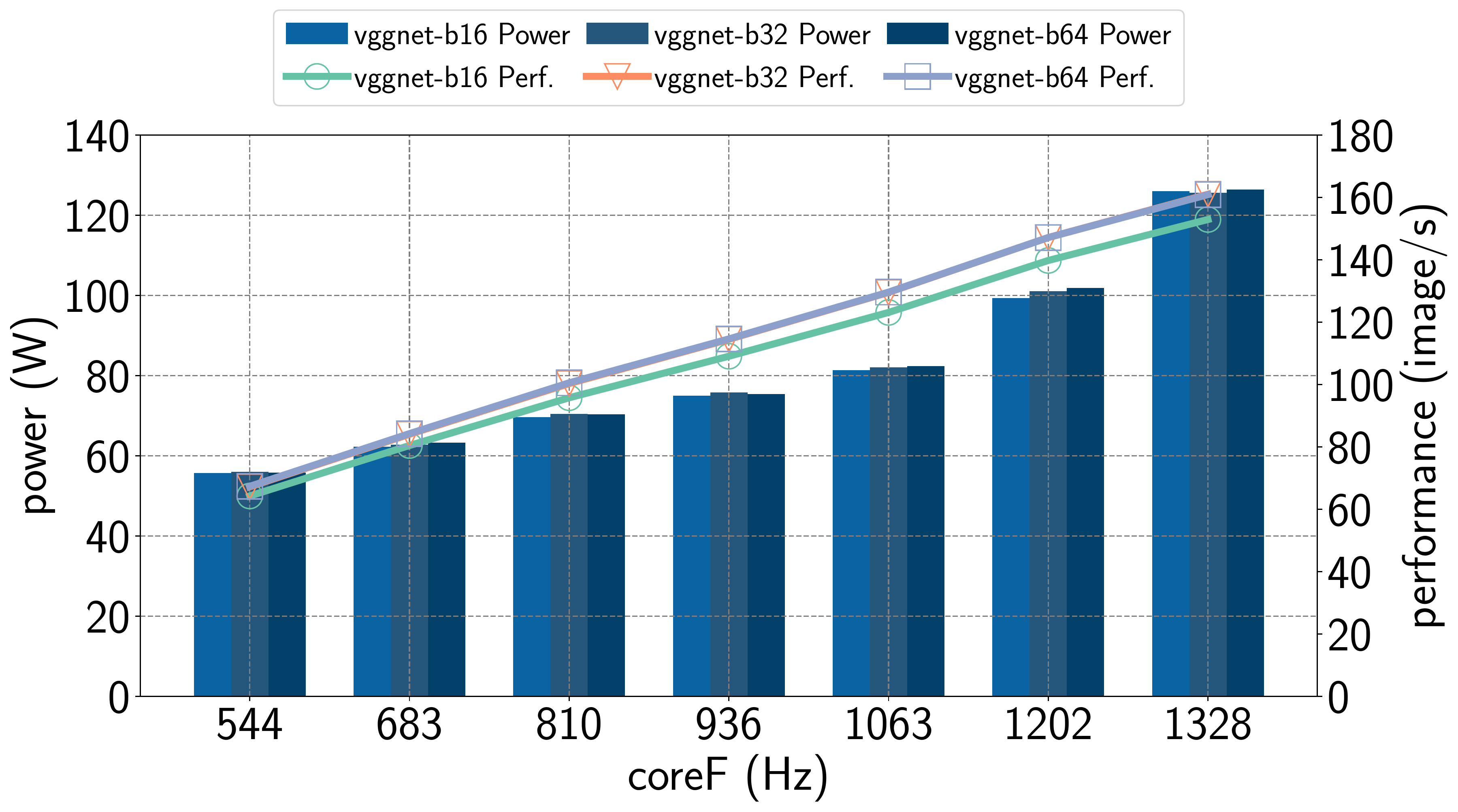}
		\label{fig:power_perf_appendix_p100_ipc_gemm_vggnet}
	}
	\subfigure[power and performance of training ResNet-50]
	{
		\includegraphics[width=0.4\linewidth]{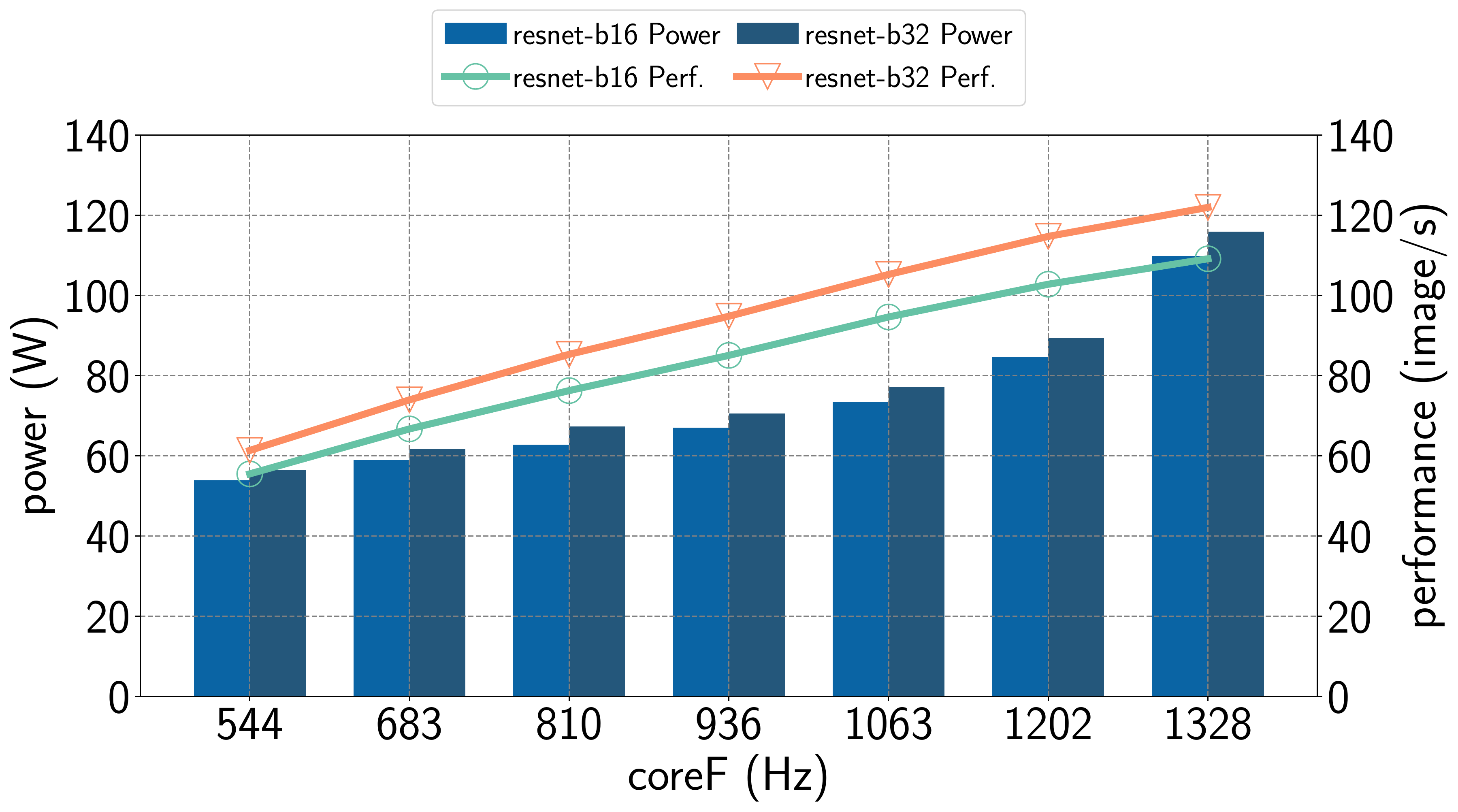}
		\label{fig:power_perf_appendix_p100_ipc_gemm_resnet}
	}
	\caption{training using implicit GEMM on P100 with increase of core frequency}
	\label{fig:ipc_gemm_perf_p100}
\end{figure*}

\begin{figure*}[htbp]
	\centering     
	\subfigure[power and performance of training AlexNet]
	{
		\includegraphics[width=0.4\linewidth]{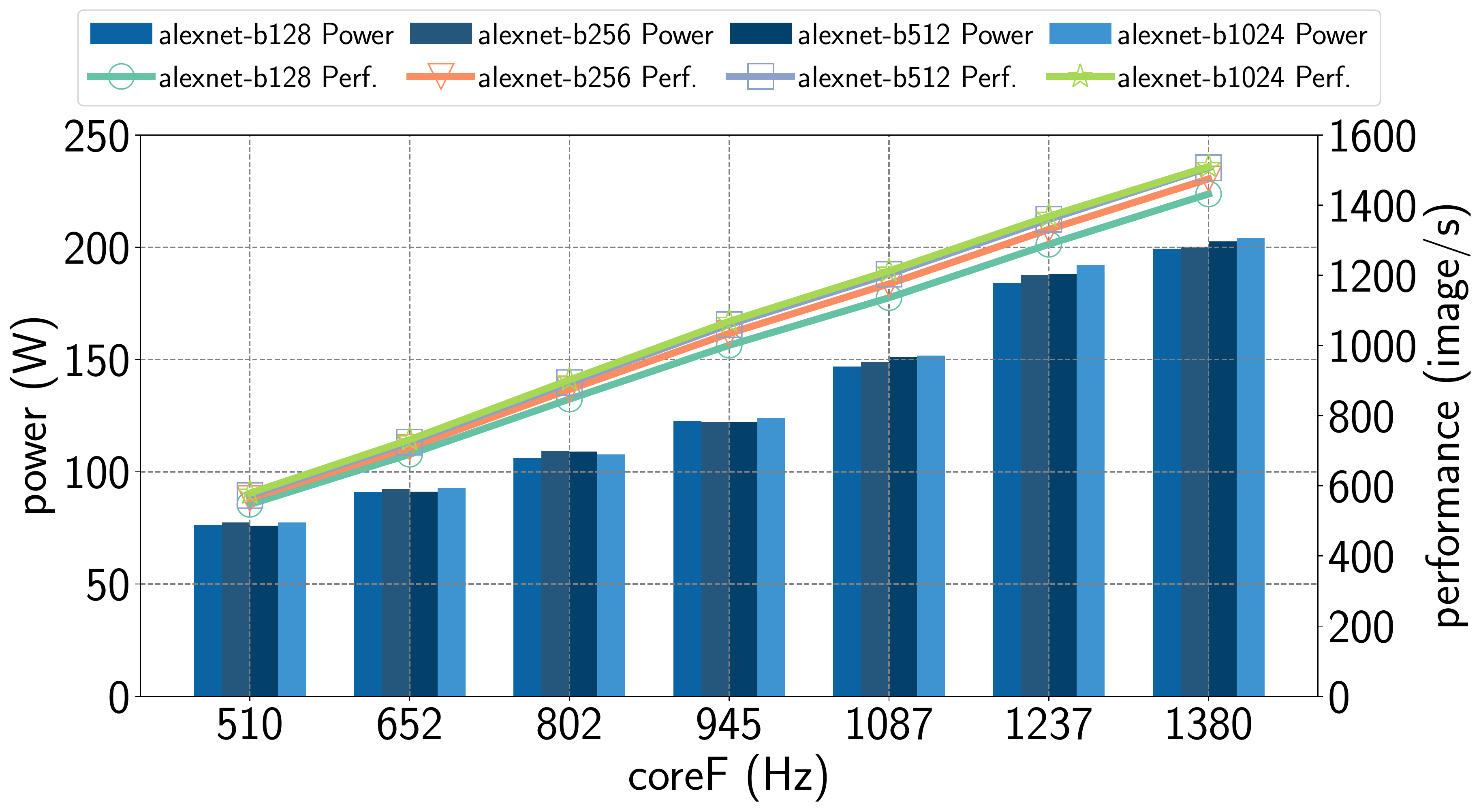}
		\label{fig:power_perf_appendix_v100_ipc_gemm_alexnet}
	}
	\subfigure[power and performance of training GoogleNet]
	{
		\includegraphics[width=0.4\linewidth]{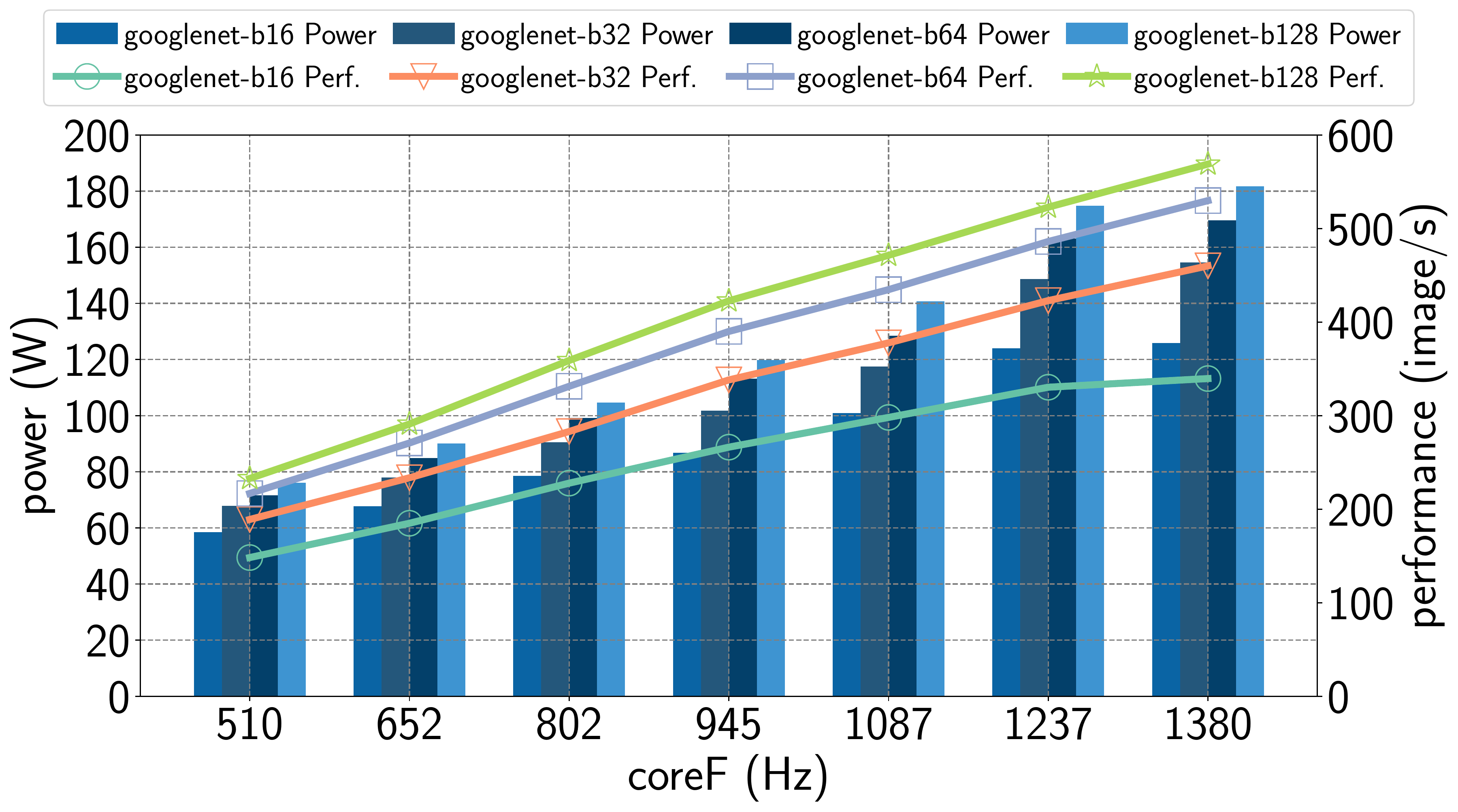}
		\label{fig:power_perf_appendix_v100_ipc_gemm_googlenet}
	}
	\subfigure[power and performance of training VggNet-16]
	{
		\includegraphics[width=0.4\linewidth]{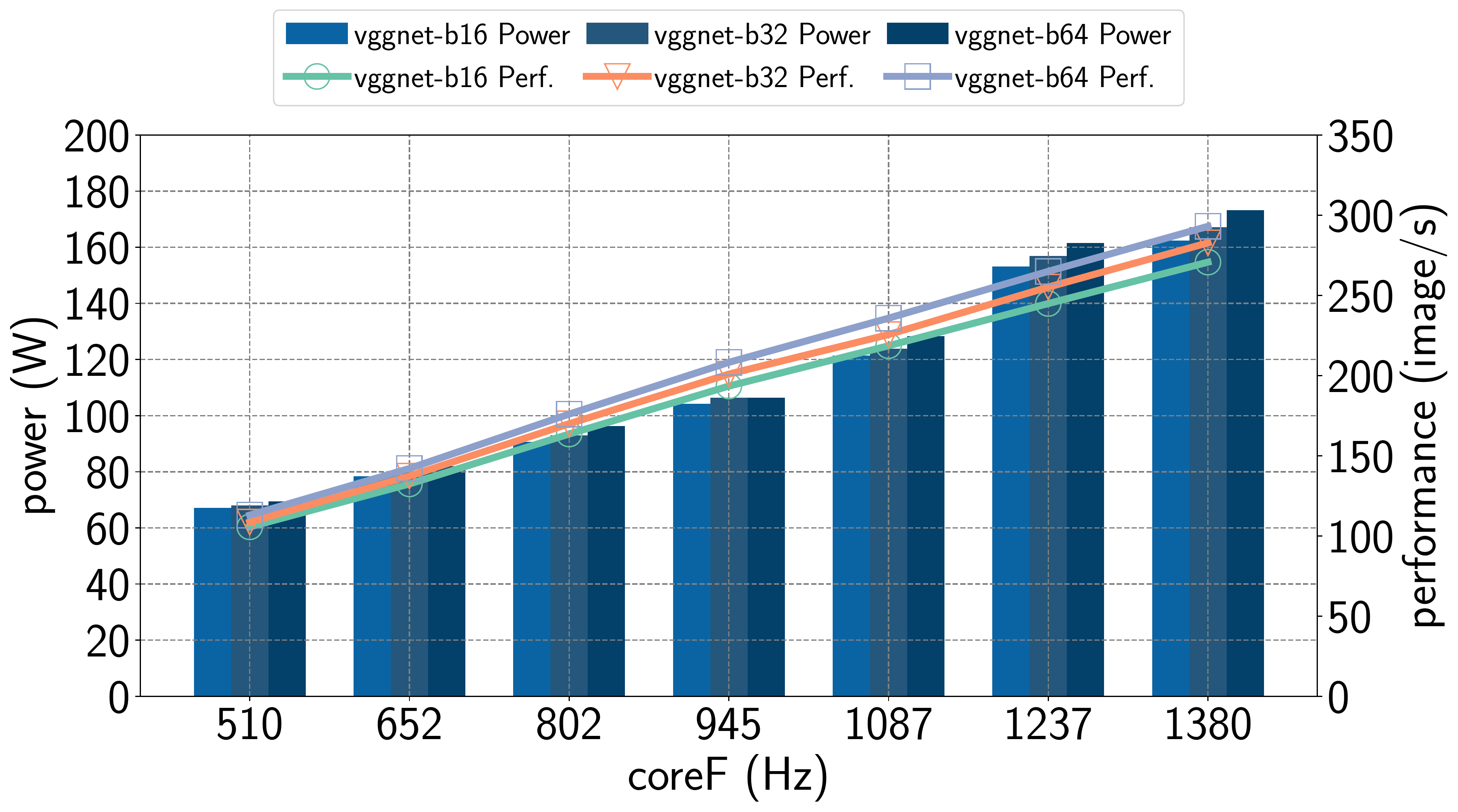}
		\label{fig:power_perf_appendix_v100_ipc_gemm_vggnet}
	}
	\subfigure[power and performance of training ResNet-50]
	{
		\includegraphics[width=0.4\linewidth]{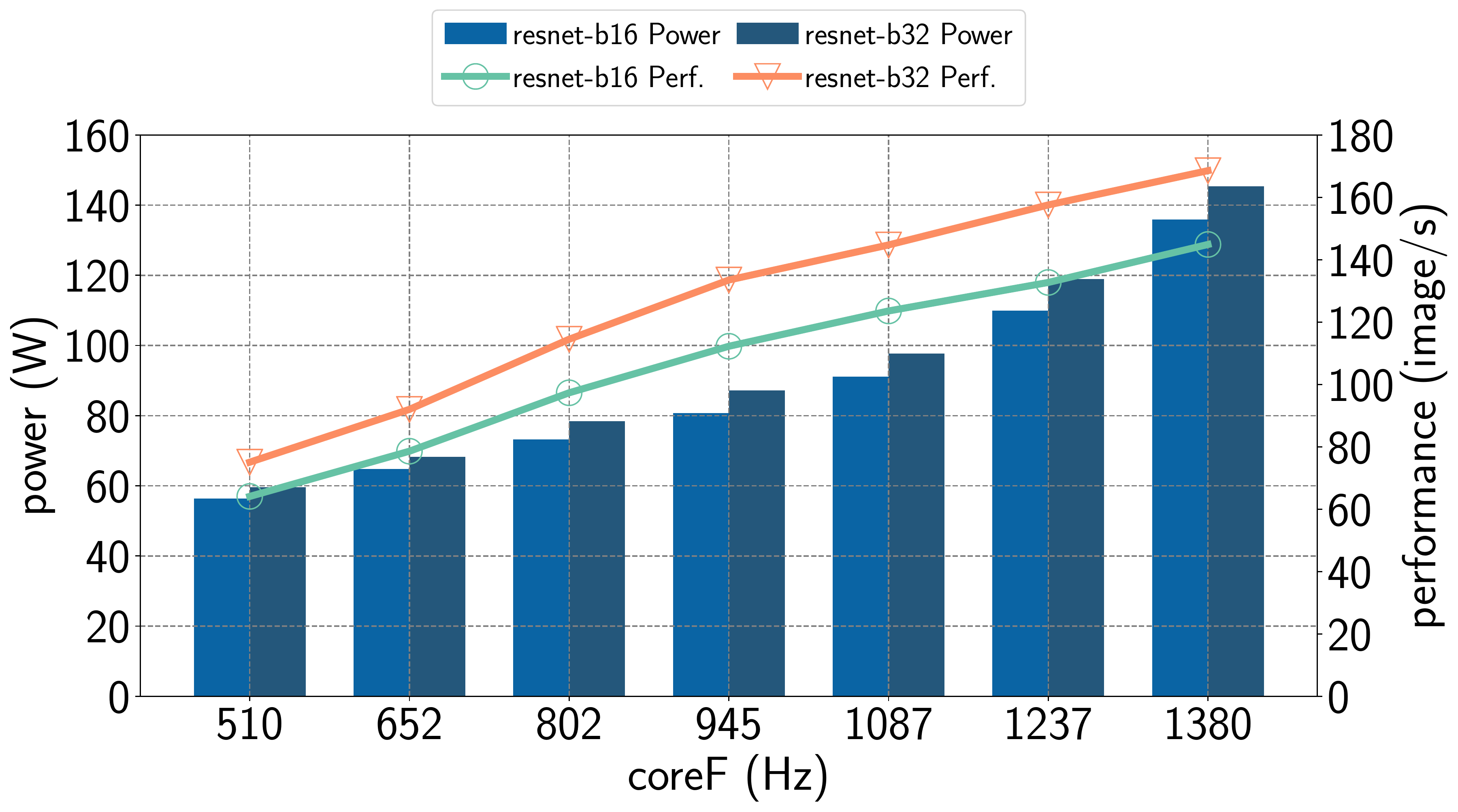}
		\label{fig:power_perf_appendix_v100_ipc_gemm_resnet}
	}
	\caption{training using implicit GEMM on V100 with increase of core frequency}
	\label{fig:ipc_gemm_perf_v100}
\end{figure*}

Figures \ref{fig:ipc_gemm_perf_p100} and \ref{fig:ipc_gemm_perf_v100} demonstrate the results of GEMM algorithm on two Tesla GPUs, P100 and V100. When $f^{core}$ is less than 1,000 MHz, the performance mostly has a faster-growing trend than the power consumption with the increase of the core frequency. However, when $f^{core}$ surpasses 1,000 MHz, the power consumption has a sudden jump, which raises up the total energy consumption again. Thus, the core frequency that achieves the best energy efficiency usually lies in the middle interval. Besides, it is observed that the performance of AlexNet and VggNet-16 rarely changes with different batch sizes, while GoogleNet and ResNet-50 gain higher image processing throughputs with larger batch sizes. Notice that GoogleNet and ResNet-50 have more layers than the other two. Larger batch sizes help GoogleNet and ResNet-50 achieve higher GPU utilization. 

\begin{figure*}[htbp]
	\centering     
	\subfigure[power and performance of training AlexNet]
	{
		\includegraphics[width=0.4\linewidth]{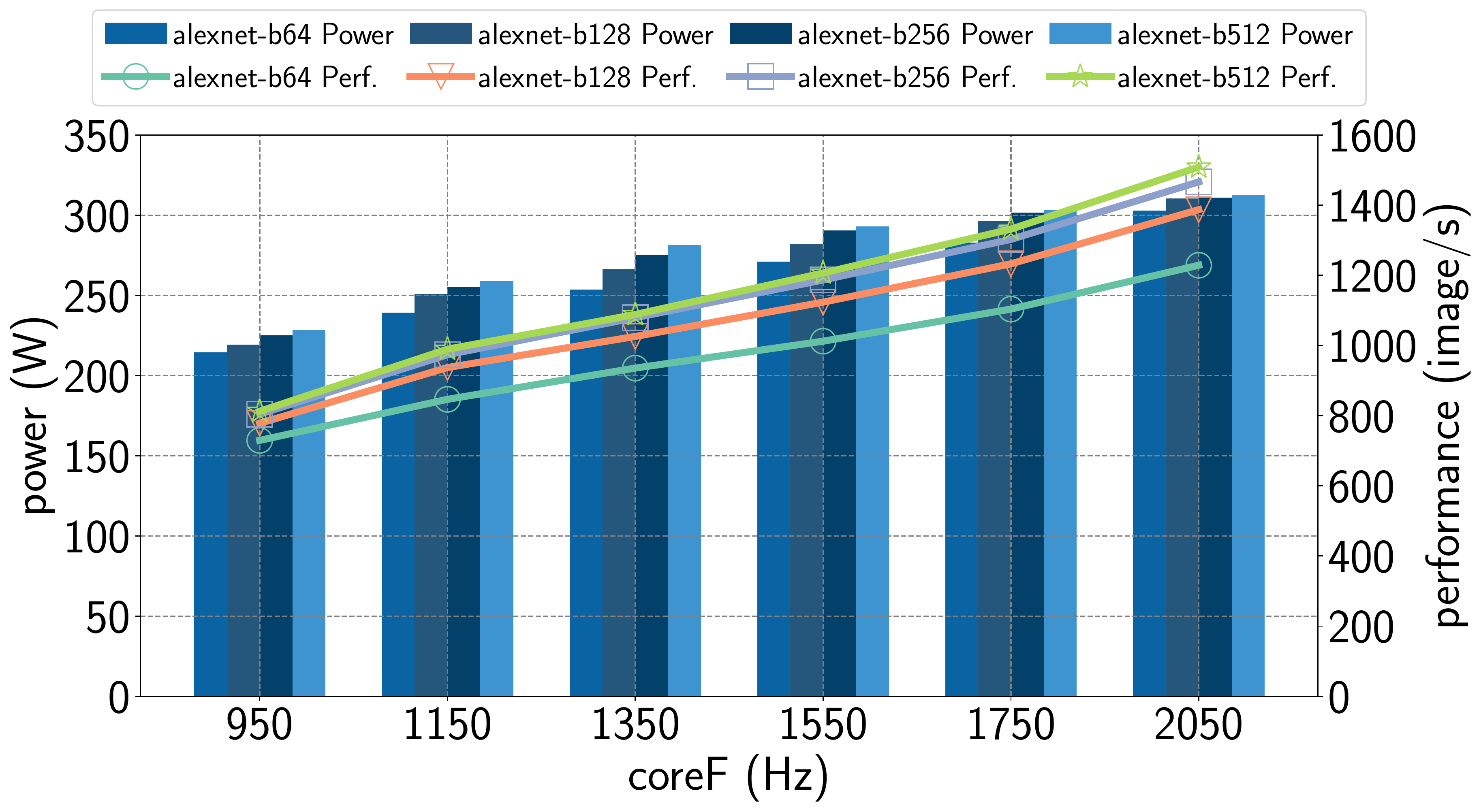}
		\label{fig:power_perf_appendix_gtx2080ti_ipc_gemm_alexnet_coreF}
	}
	\subfigure[power and performance of training GoogleNet]
	{
		\includegraphics[width=0.4\linewidth]{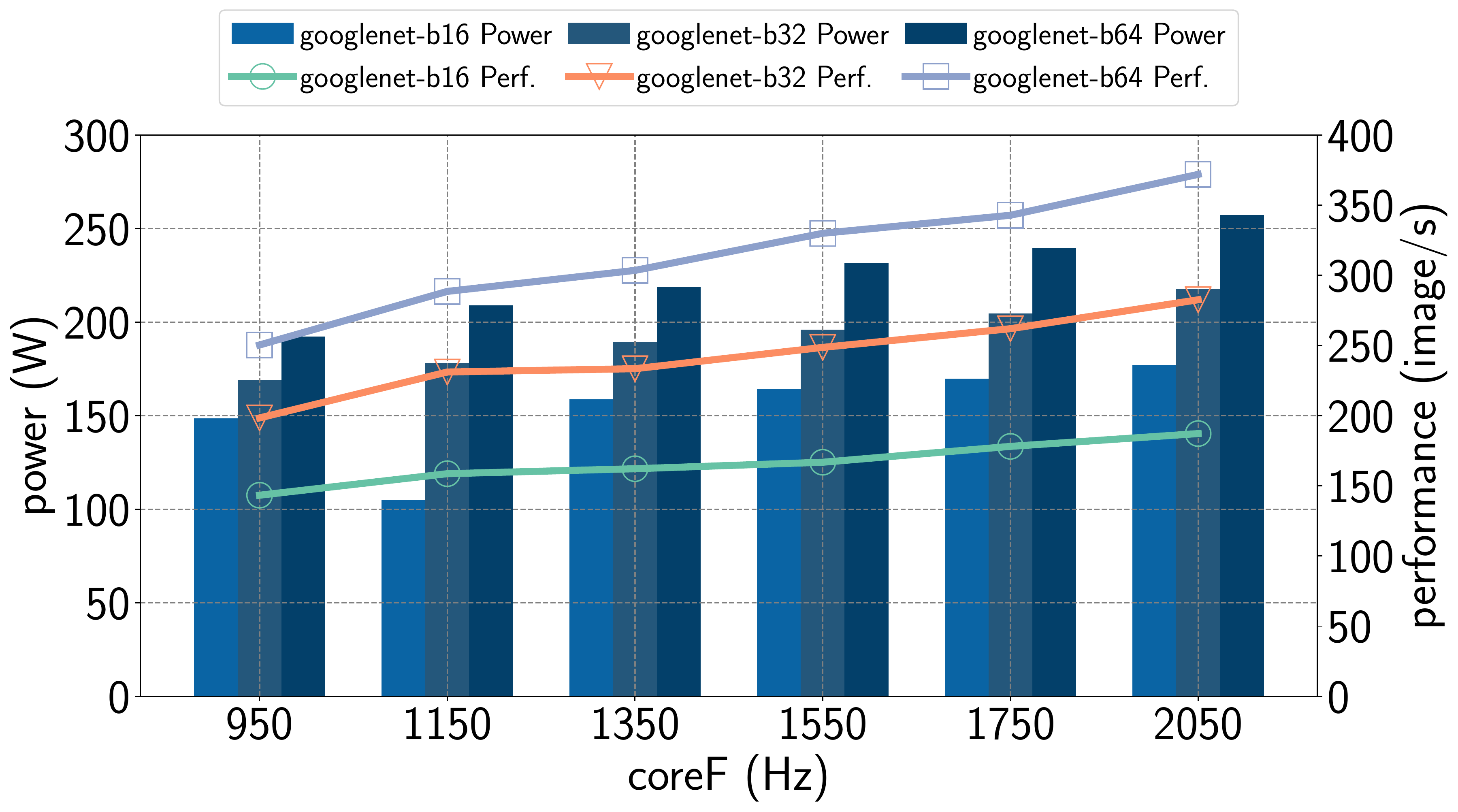}
		\label{fig:power_perf_appendix_gtx2080ti_ipc_gemm_googlenet_coreF}
	}
	\subfigure[power and performance of training VggNet-16]
	{
		\includegraphics[width=0.4\linewidth]{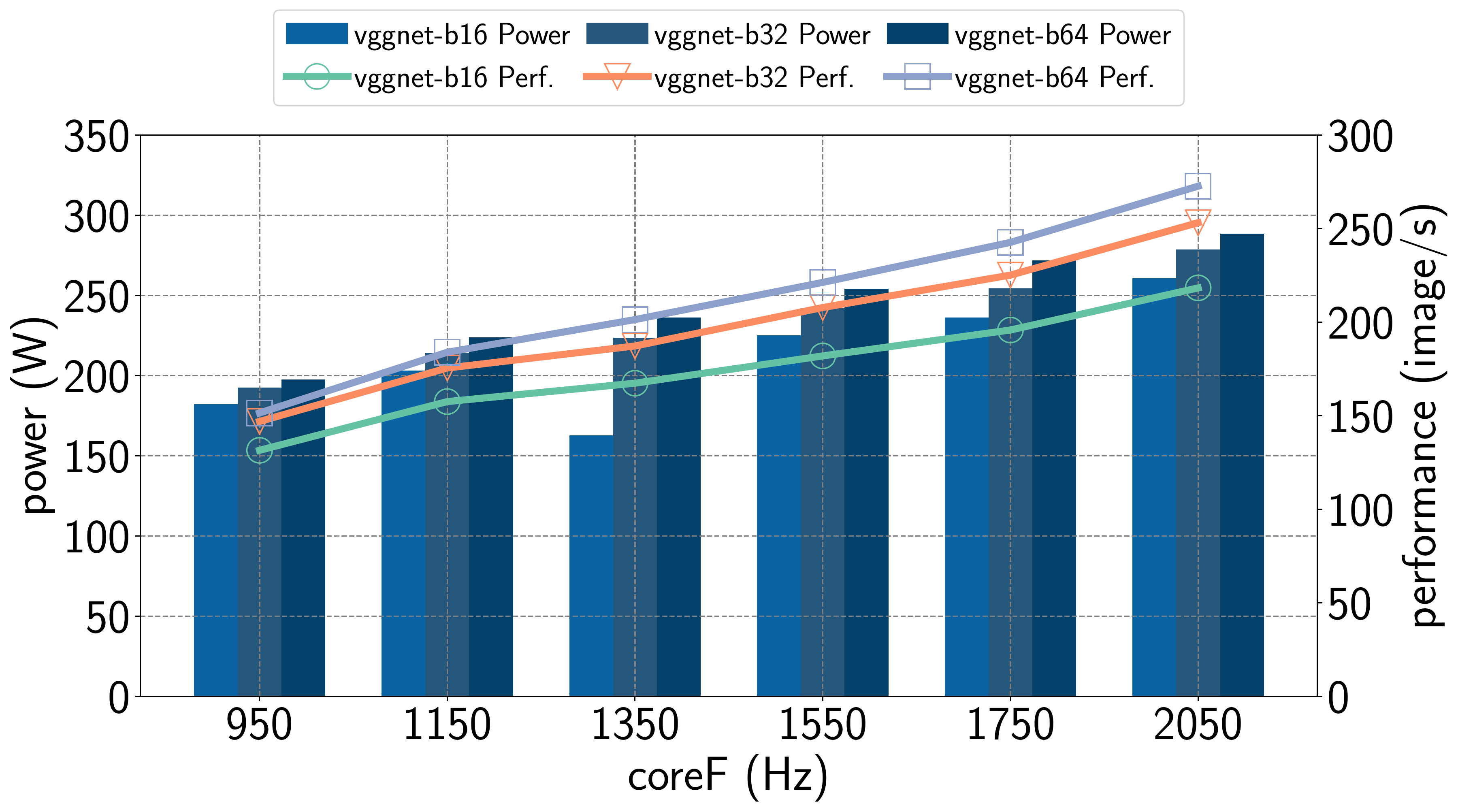}
		\label{fig:power_perf_appendix_gtx2080ti_ipc_gemm_vggnet_coreF}
	}
	\subfigure[power and performance of training ResNet-50]
	{
		\includegraphics[width=0.4\linewidth]{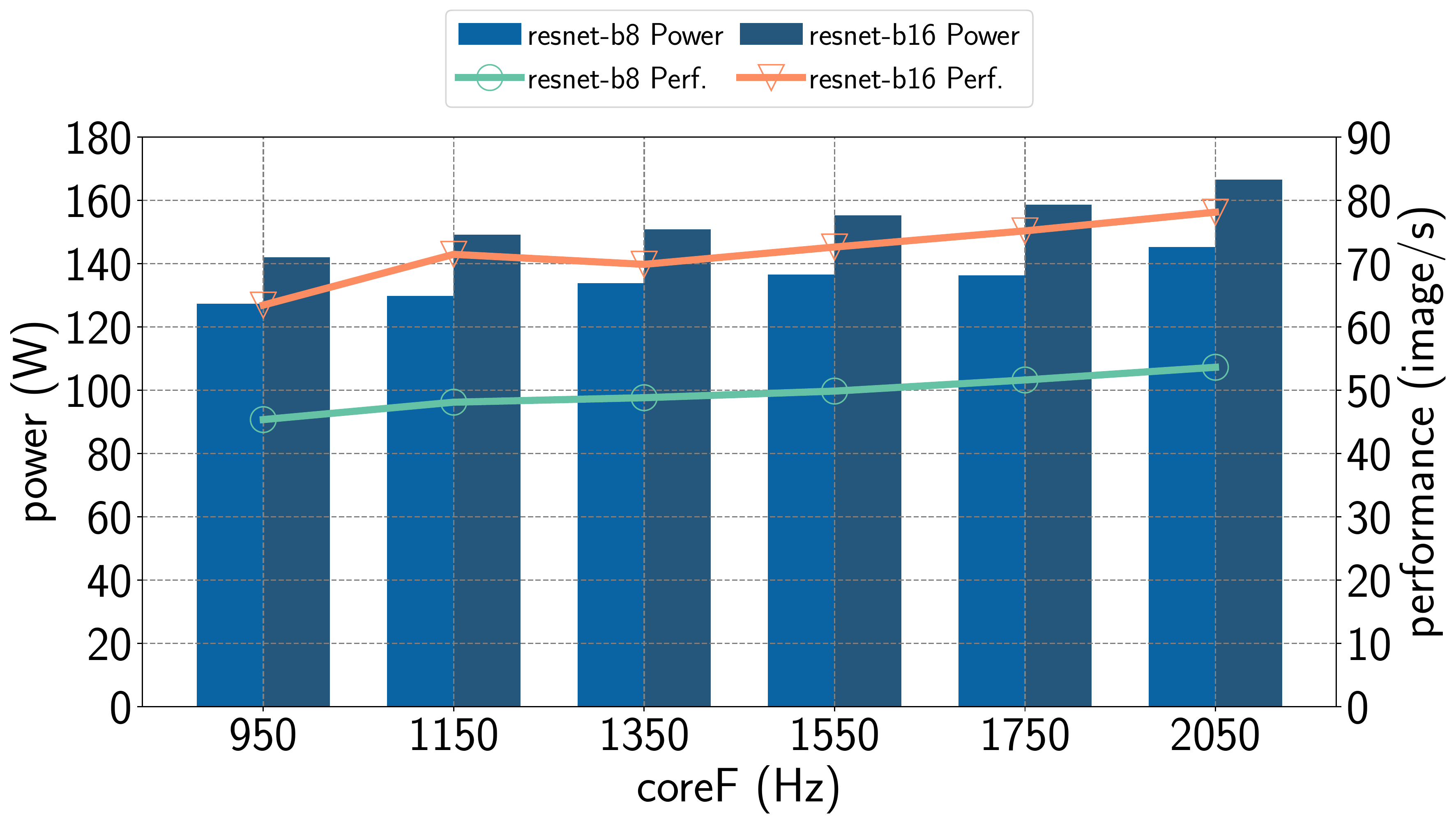}
		\label{fig:power_perf_appendix_gtx2080ti_ipc_gemm_resnet_coreF}
	}
	\caption{training using implicit GEMM on GTX2080Ti with increase of core frequency}
	\label{fig:ipc_gemm_perf_core_gtx2080ti}
\end{figure*}

\begin{figure*}[htbp]
	\centering     
	\subfigure[power and performance of training AlexNet]
	{
		\includegraphics[width=0.4\linewidth]{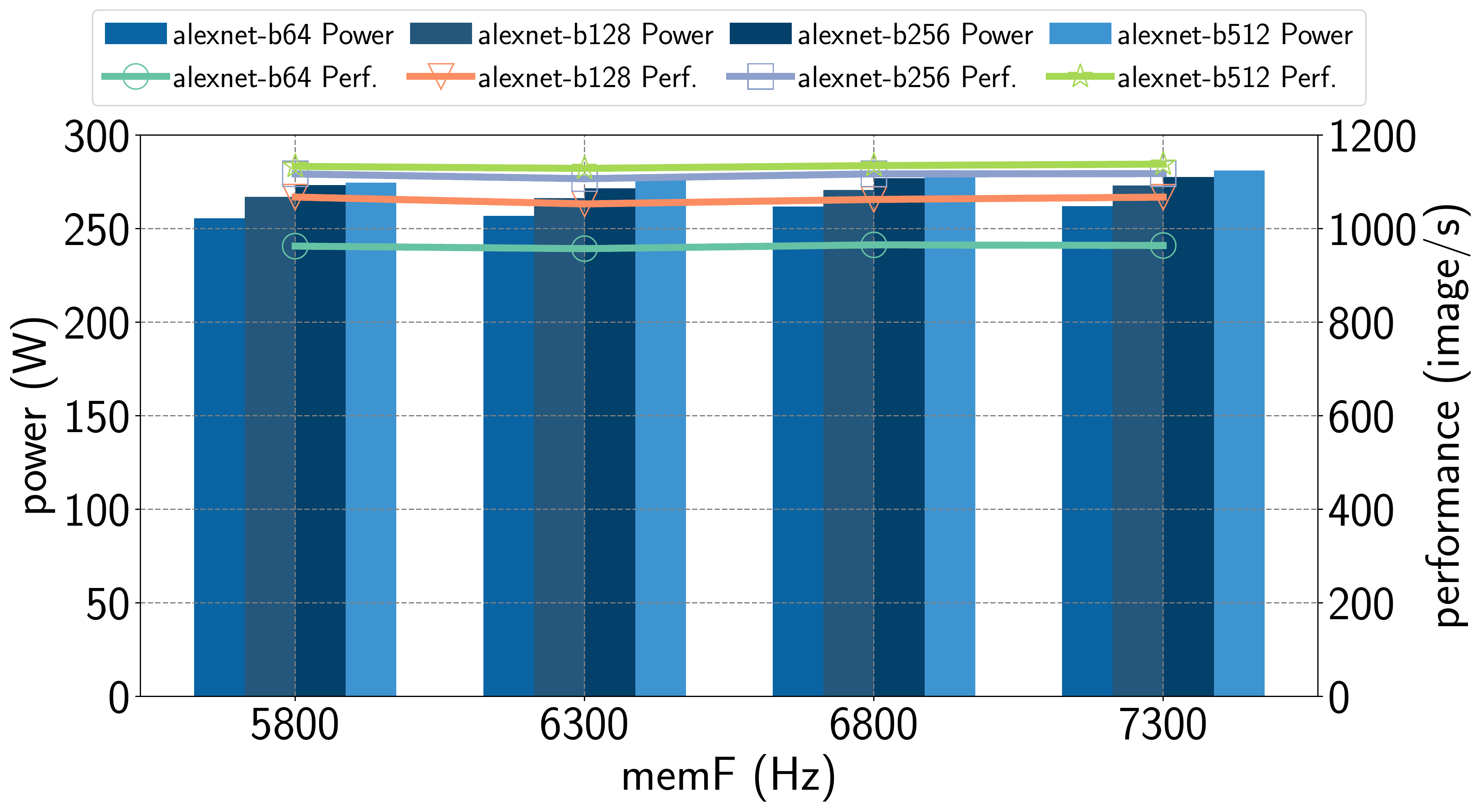}
		\label{fig:power_perf_appendix_gtx2080ti_ipc_gemm_alexnet_memF}
	}
	\subfigure[power and performance of training GoogleNet]
	{
		\includegraphics[width=0.4\linewidth]{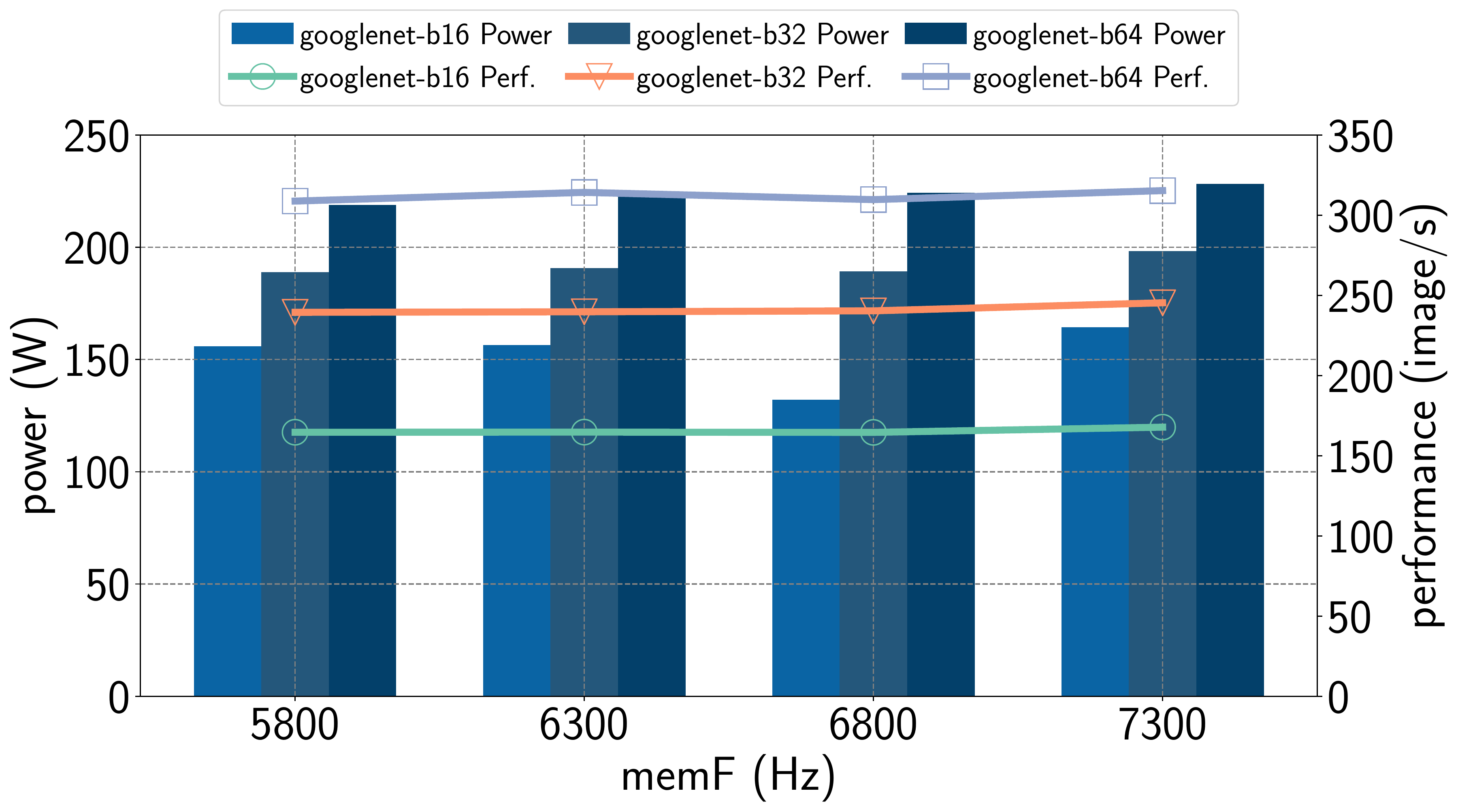}
		\label{fig:power_perf_appendix_gtx2080ti_ipc_gemm_googlenet_memF}
	}
	\subfigure[power and performance of training VggNet-16]
	{
		\includegraphics[width=0.4\linewidth]{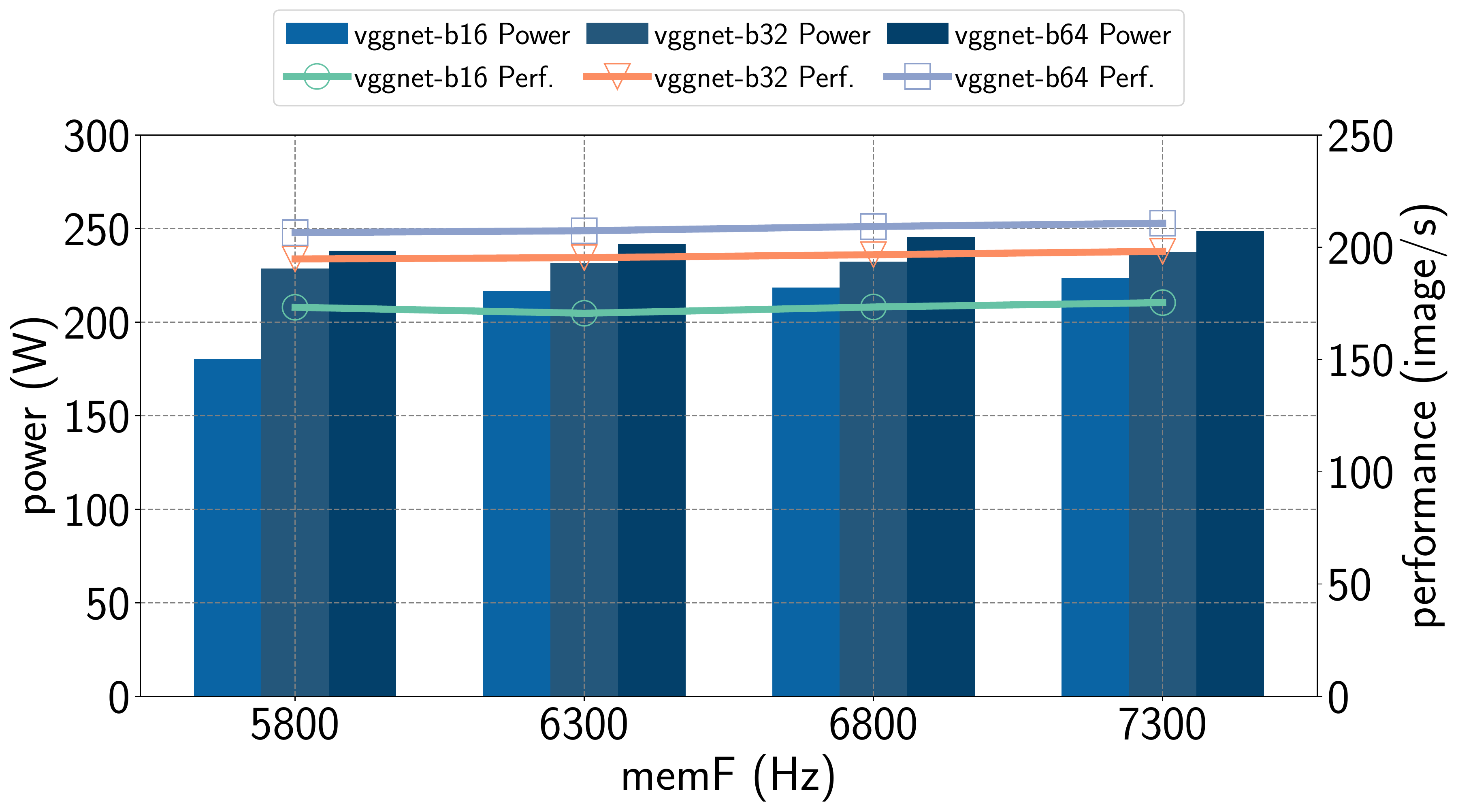}
		\label{fig:power_perf_appendix_gtx2080ti_ipc_gemm_vggnet_memF}
	}
	\subfigure[power and performance of training ResNet-50]
	{
		\includegraphics[width=0.4\linewidth]{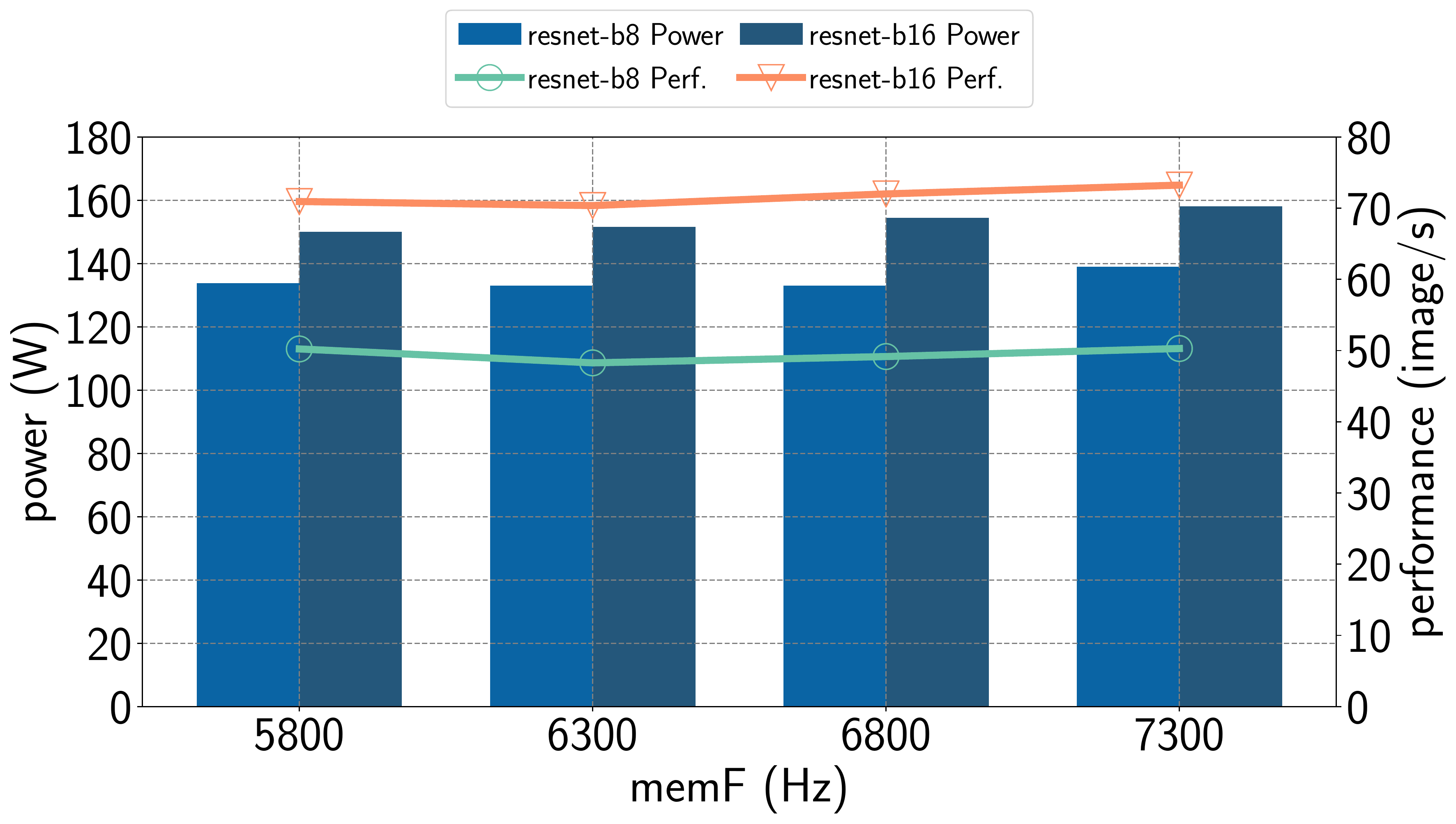}
		\label{fig:power_perf_appendix_gtx2080ti_ipc_gemm_resnet_memF}
	}
	\caption{training using implicit GEMM on GTX2080Ti with increase of memory frequency}
	\label{fig:ipc_gemm_perf_memory_gtx2080ti}
\end{figure*}

Figures \ref{fig:ipc_gemm_perf_core_gtx2080ti} and \ref{fig:ipc_gemm_perf_memory_gtx2080ti} demonstrate the results of GEMM algorithm on GTX 2080Ti. Different from two Tesla GPUs, the performance of GTX 2080Ti always has a faster-growing trend than the power consumption, which results in that the best energy efficiency is mostly achieved at the highest core frequency. On the contrary, increasing the memory frequency hardly affects the performance of DNN training, but leads to higher power consumption. Thus, applying a low memory frequency surprisingly helps conserve energy. 
\subsection{Using Winograd algorithm}
\begin{figure*}[htbp]
	\centering     
	\subfigure[power and performance of training AlexNet]
	{
		\includegraphics[width=0.4\linewidth]{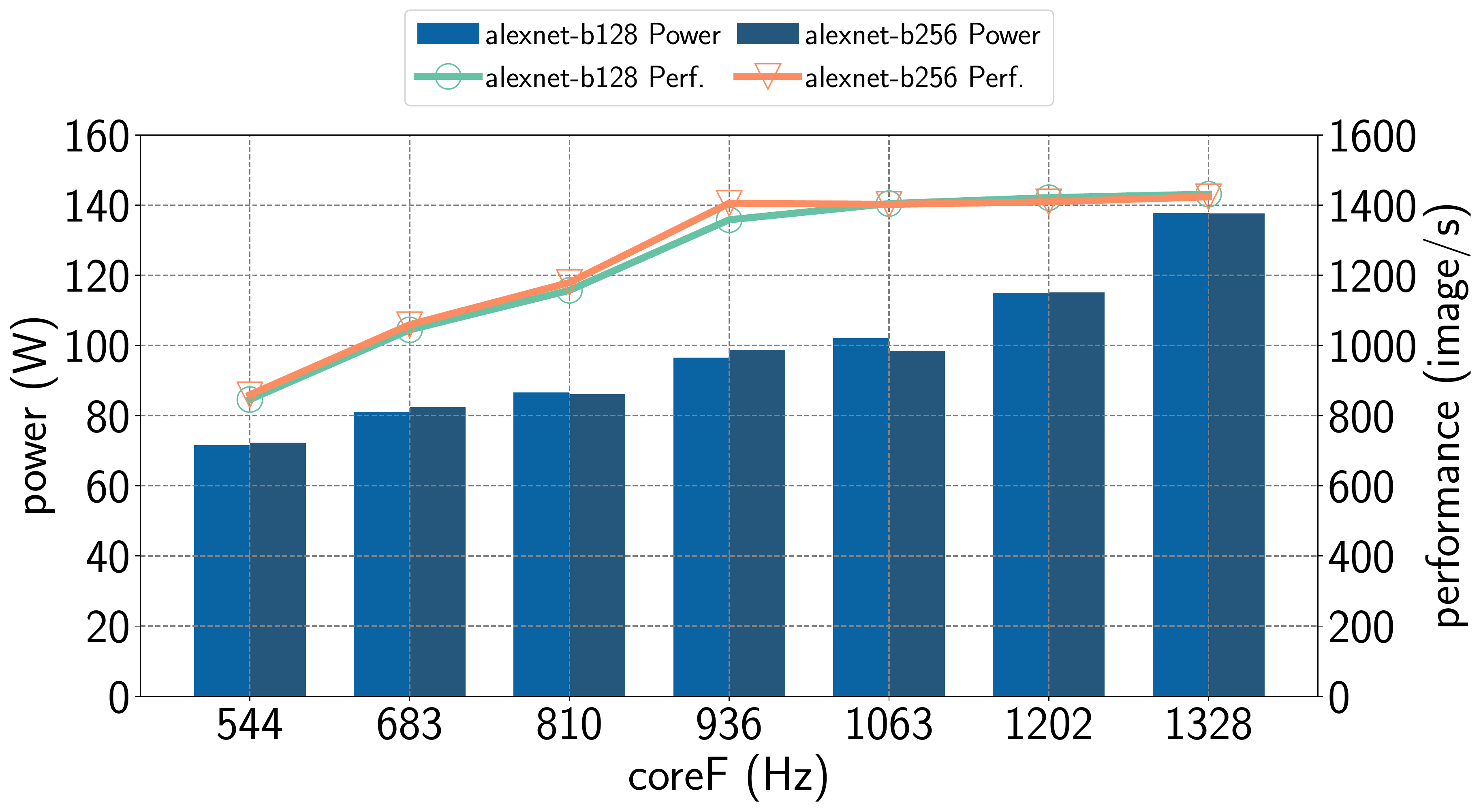}
		\label{fig:power_perf_appendix_p100_winograd_alexnet}
	}
	\subfigure[power and performance of training GoogleNet]
	{
		\includegraphics[width=0.4\linewidth]{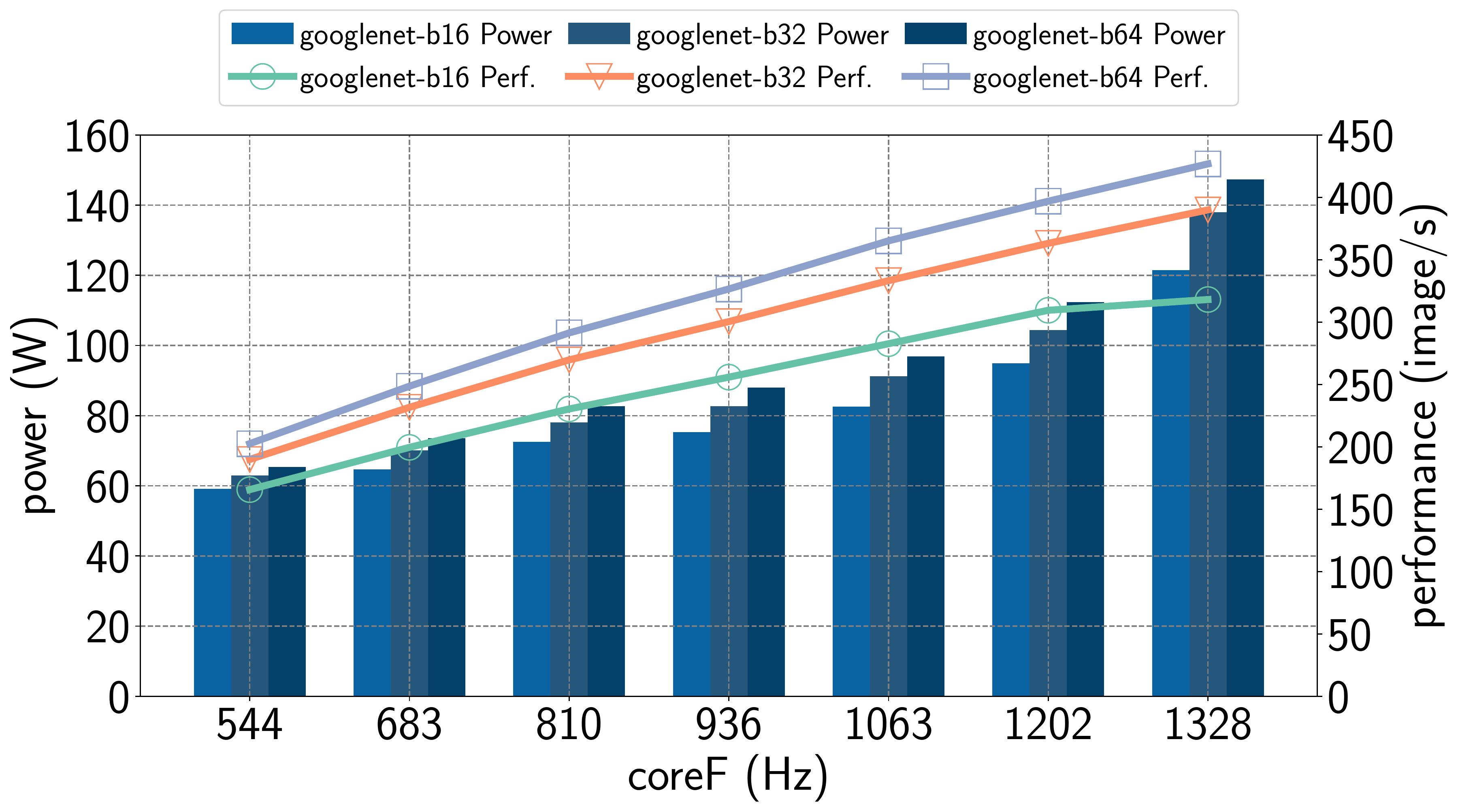}
		\label{fig:power_perf_appendix_p100_winograd_googlenet}
	}
	\subfigure[power and performance of training VggNet-16]
	{
		\includegraphics[width=0.4\linewidth]{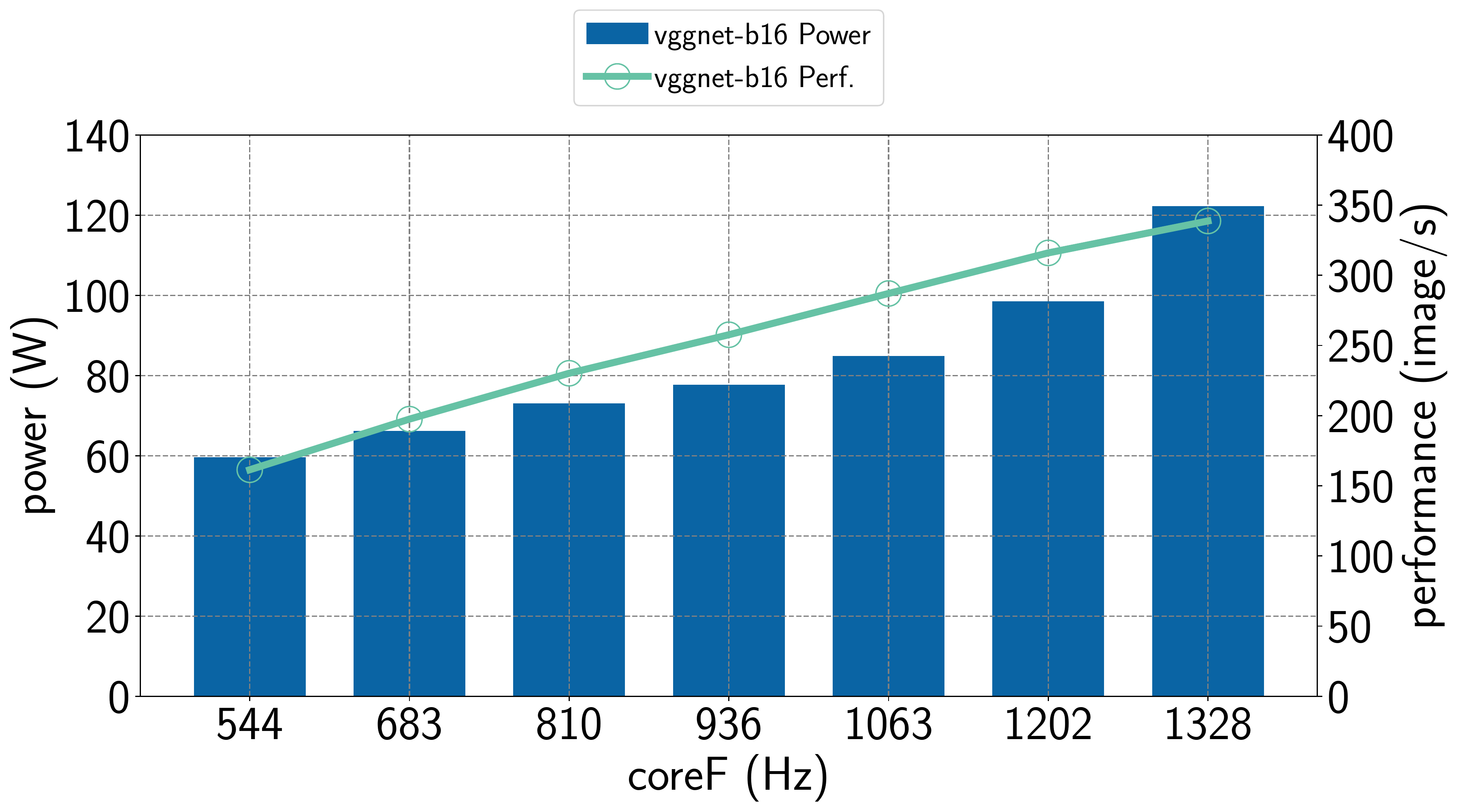}
		\label{fig:power_perf_appendix_p100_winograd_vggnet}
	}
	\subfigure[power and performance of training ResNet-50]
	{
		\includegraphics[width=0.4\linewidth]{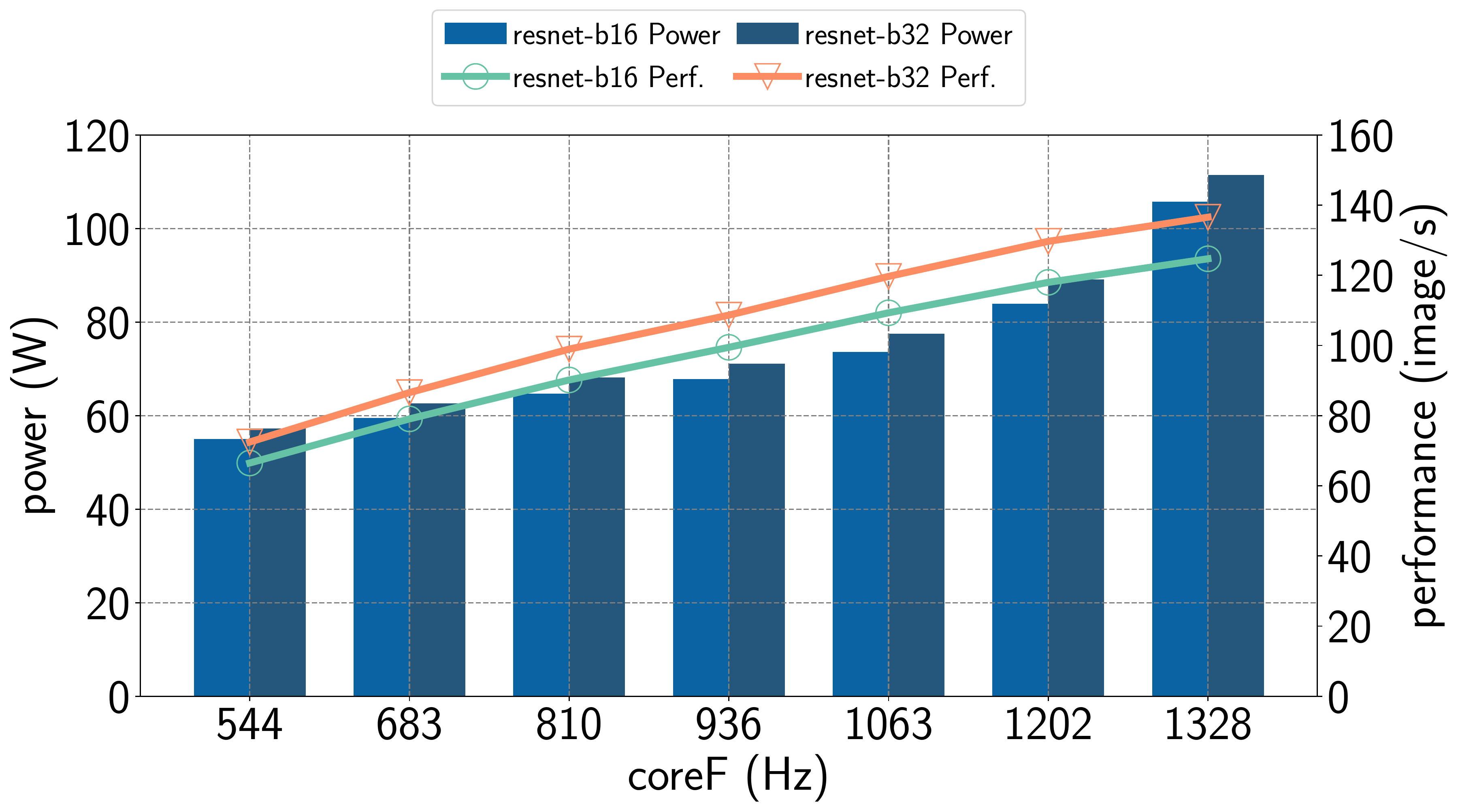}
		\label{fig:power_perf_appendix_p100_winograd_resnet}
	}
	\caption{training using Winograd on P100 with increase of core frequency}
	\label{fig:winograd_perf_p100}
\end{figure*}

\begin{figure*}[htbp]
	\centering     
	\subfigure[power and performance of training AlexNet]
	{
		\includegraphics[width=0.4\linewidth]{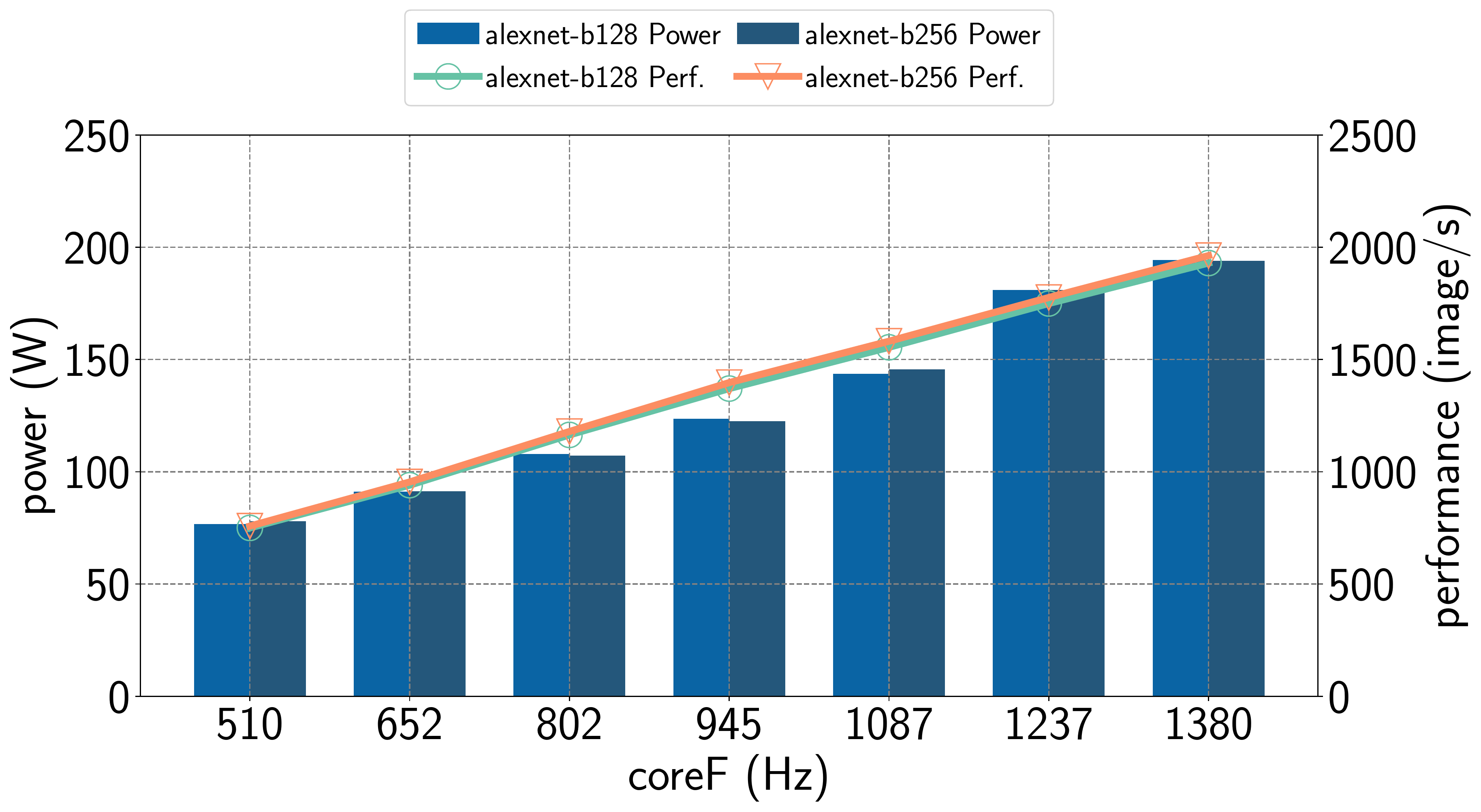}
		\label{fig:power_perf_appendix_v100_winograd_alexnet}
	}
	\subfigure[power and performance of training GoogleNet]
	{
		\includegraphics[width=0.4\linewidth]{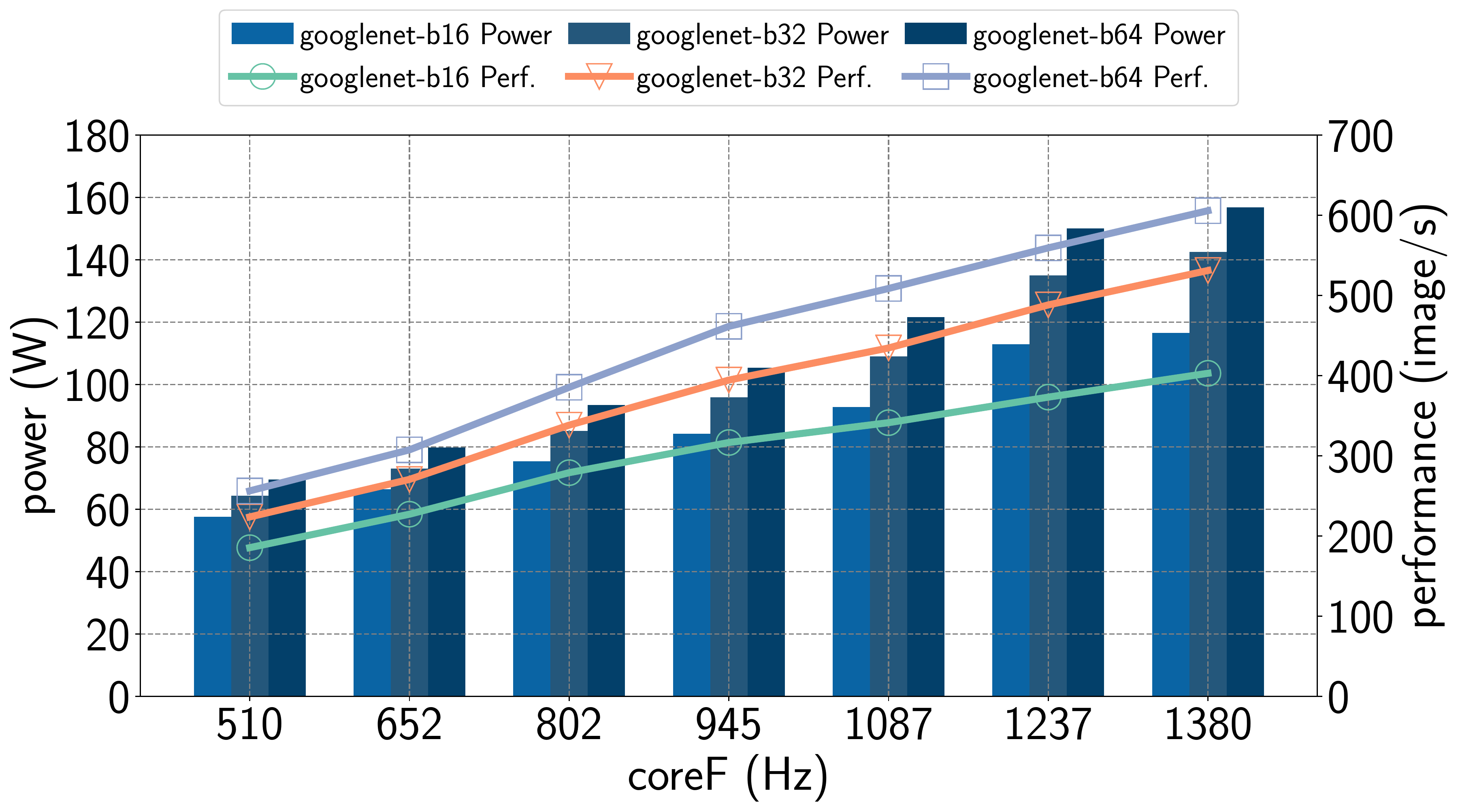}
		\label{fig:power_perf_appendix_v100_winograd_googlenet}
	}
	\subfigure[power and performance of training VggNet-16]
	{
		\includegraphics[width=0.4\linewidth]{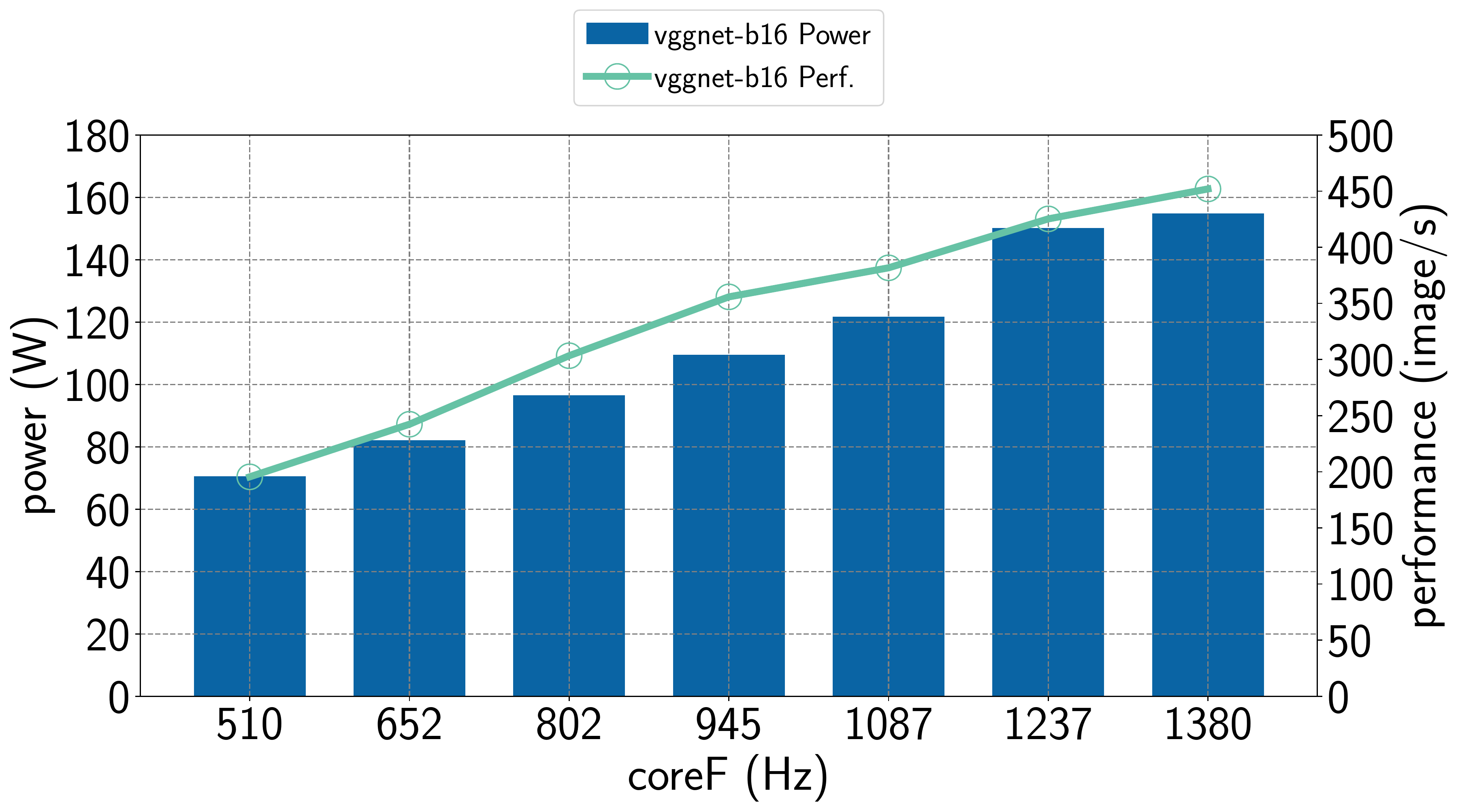}
		\label{fig:power_perf_appendix_v100_winograd_vggnet}
	}
	\subfigure[power and performance of training ResNet-50]
	{
		\includegraphics[width=0.4\linewidth]{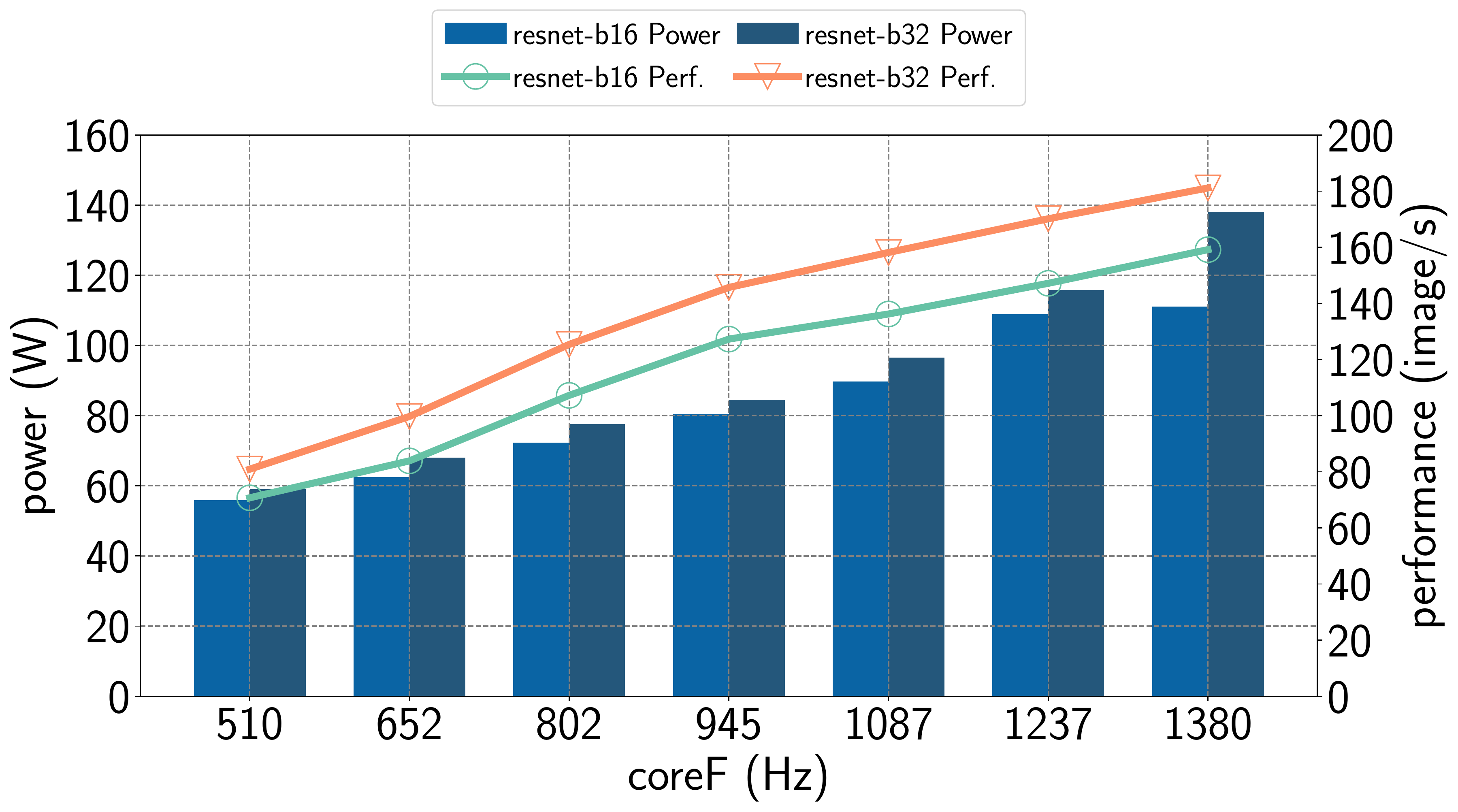}
		\label{fig:power_perf_appendix_v100_winograd_resnet}
	}
	\caption{training using Winograd on V100 with increase of core frequency}
	\label{fig:winograd_perf_v100}
\end{figure*}

Figures \ref{fig:winograd_perf_p100} and \ref{fig:winograd_perf_v100} demonstrate the results of Winograd algorithm on two Tesla GPUs, P100 and V100. Notice that Winograd requires a larger GPU memory to tackle convolution than GEMM does. Only a few batch sizes of four DNNs are supported on GPUs. Similar to GEMM, the core frequency that achieves the best energy efficiency usually lies in the middle interval. Besides, GoogleNet and ResNet-50 achieve higher image processing throughputs with larger batch sizes. Compared to GEMM, Winograd achieves better performance while keeping nearly the same power consumption. Thus, Winograd has better energy efficiency than GEMM, which also meets the results in Figure \ref{fig:conv_algo_compare}. 

\begin{figure*}[htbp]
	\centering     
	\subfigure[power and performance of training AlexNet]
	{
		\includegraphics[width=0.31\linewidth]{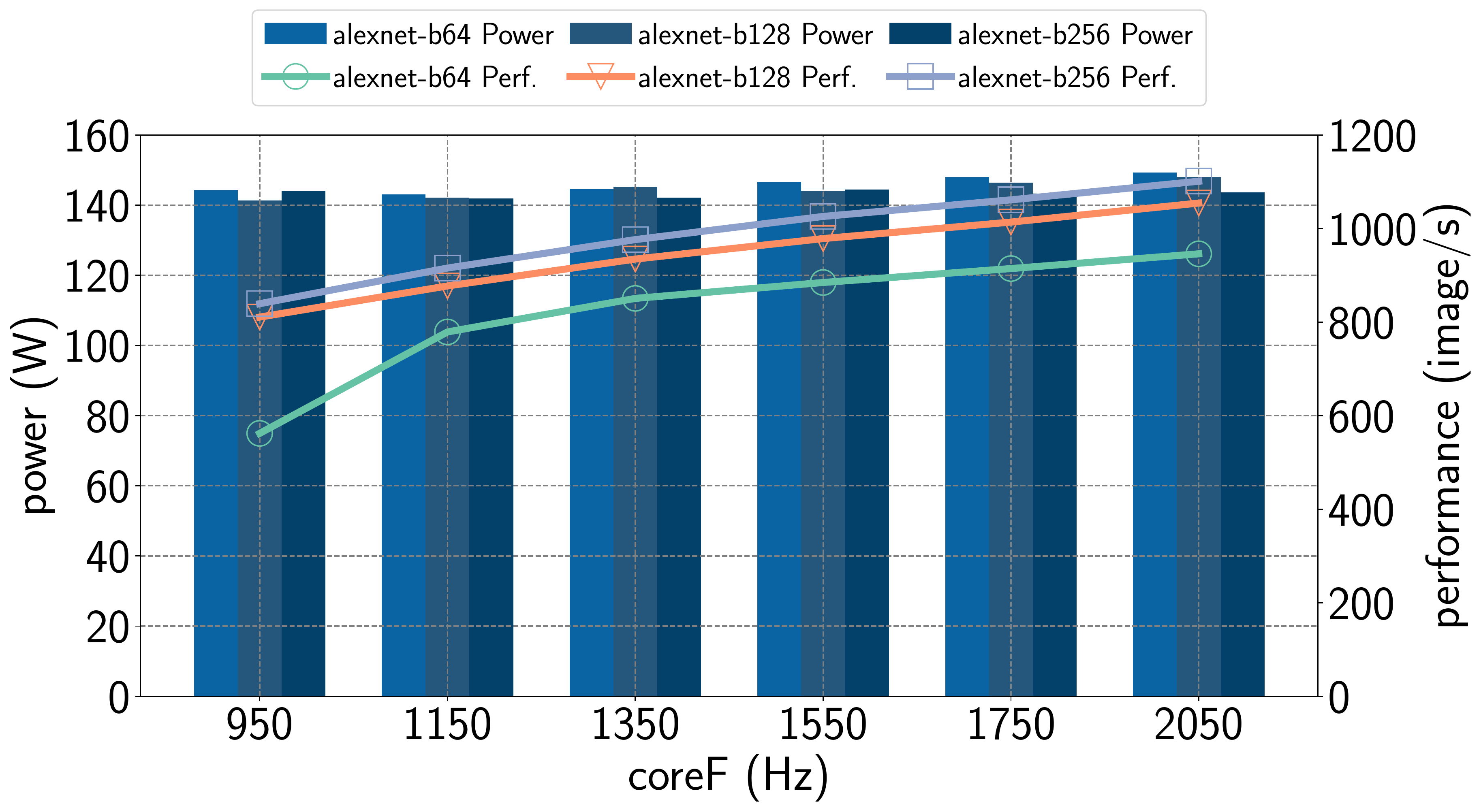}
		\label{fig:power_perf_appendix_gtx2080ti_winograd_alexnet_coreF}
	}
	\subfigure[power and performance of training GoogleNet]
	{
		\includegraphics[width=0.31\linewidth]{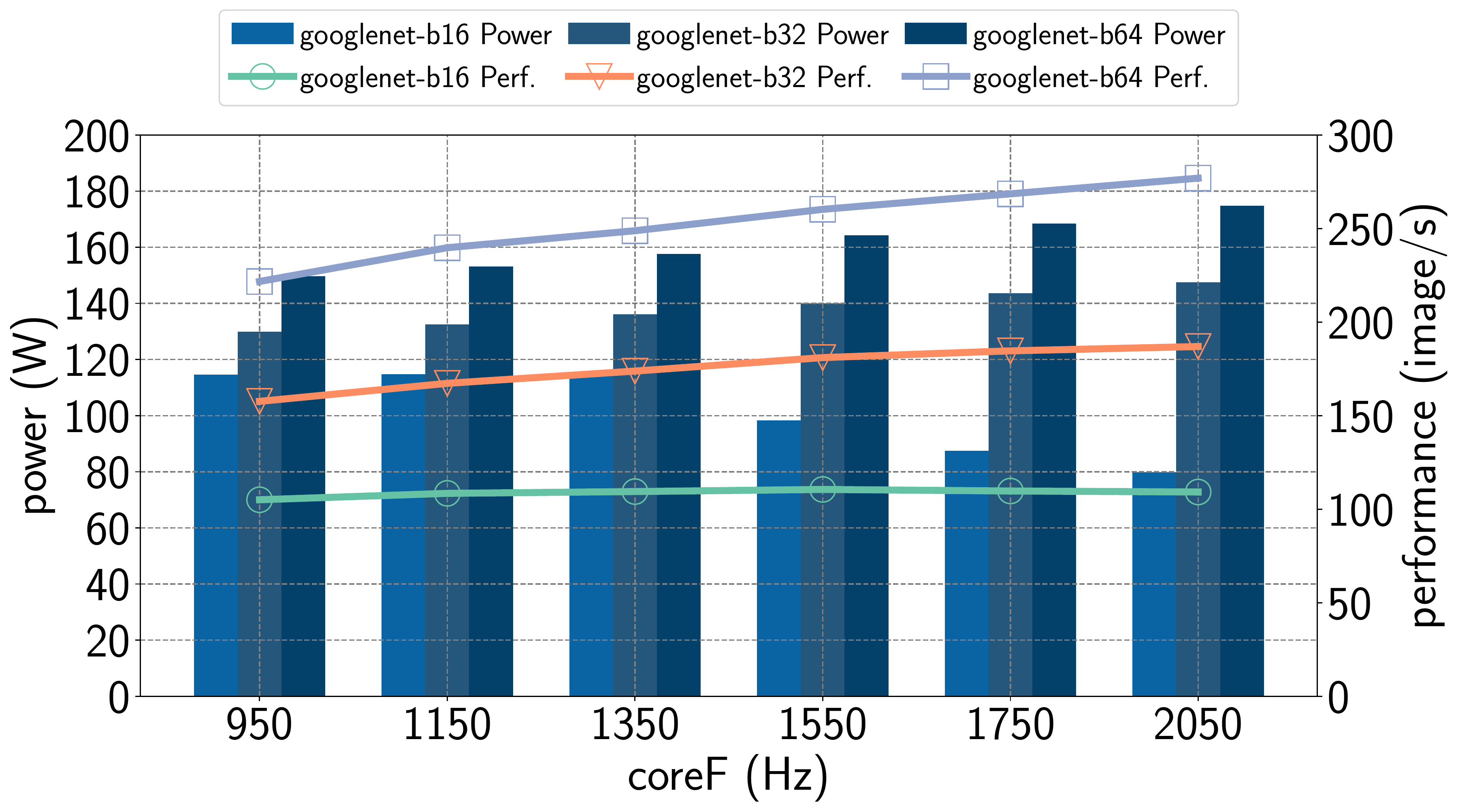}
		\label{fig:power_perf_appendix_gtx2080ti_winograd_googlenet_coreF}
	}
	\subfigure[power and performance of training ResNet-50]
	{
		\includegraphics[width=0.31\linewidth]{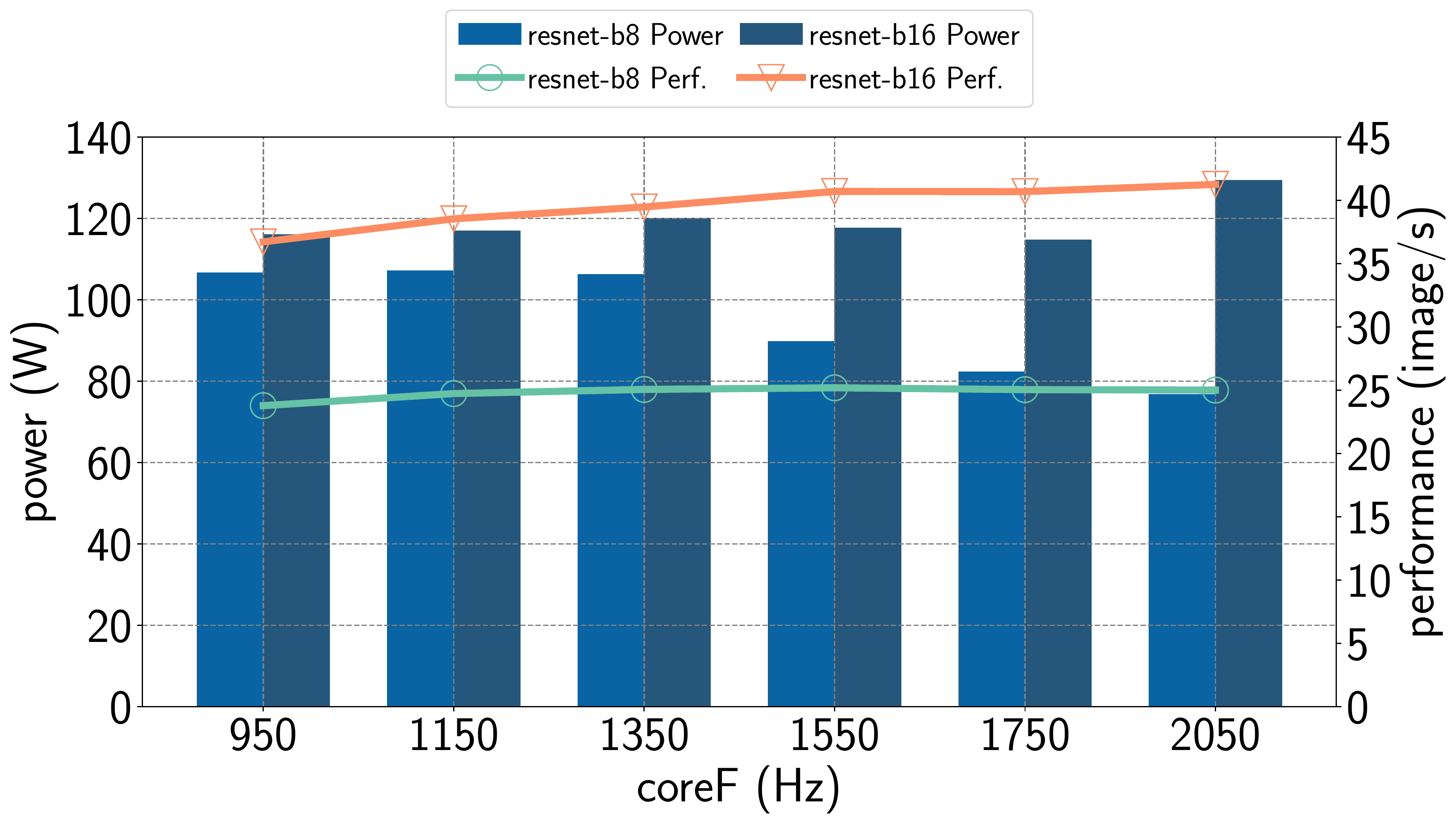}
		\label{fig:power_perf_appendix_gtx2080ti_winograd_resnet_coreF}
	}
	\caption{training using Winograd on GTX2080Ti with increase of core frequency}
	\label{fig:winograd_perf_core_gtx2080ti}
\end{figure*}

\begin{figure*}[htbp]
	\centering     
	\subfigure[power and performance of training AlexNet]
	{
		\includegraphics[width=0.31\linewidth]{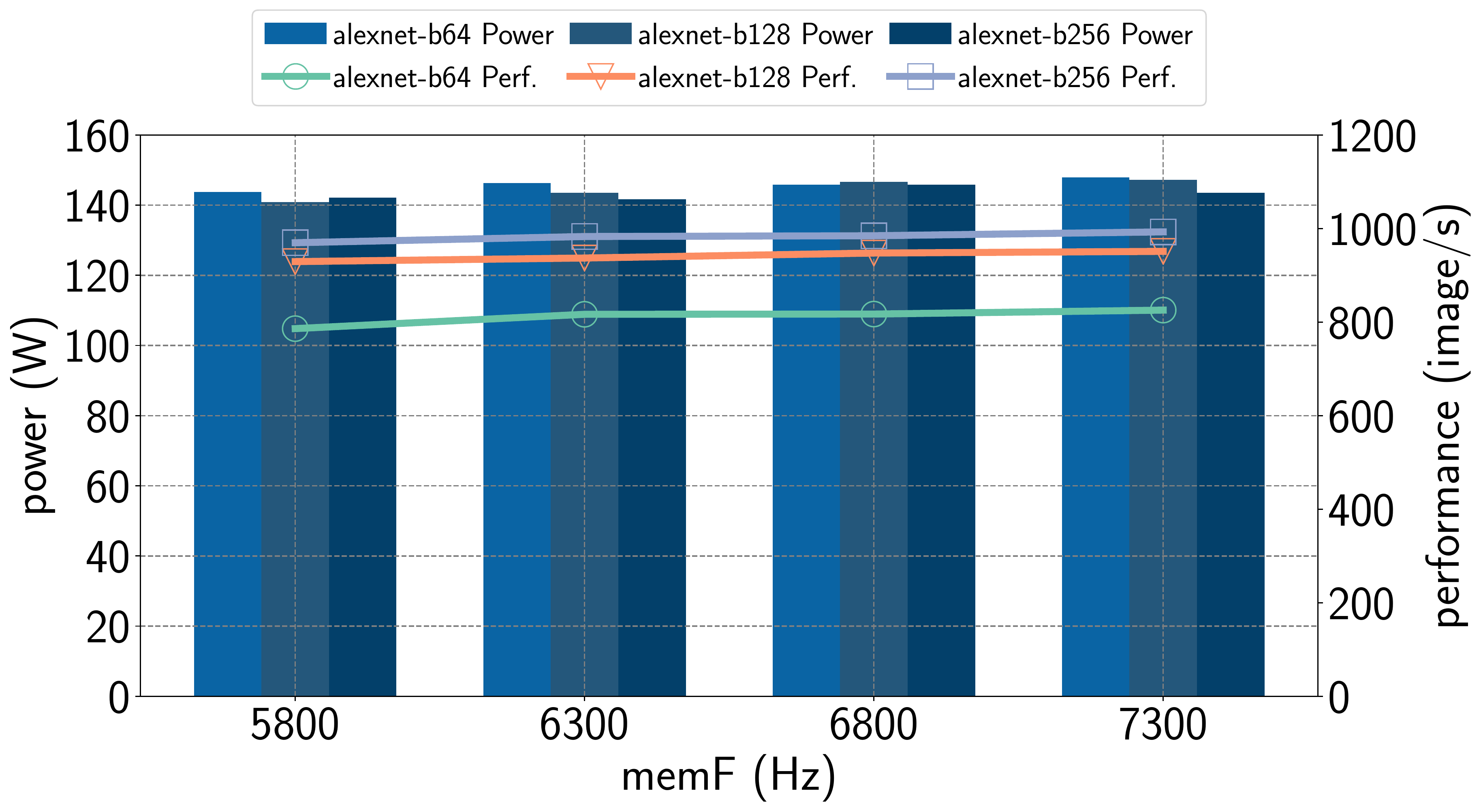}
		\label{fig:power_perf_appendix_gtx2080ti_winograd_alexnet_memF}
	}
	\subfigure[power and performance of training GoogleNet]
	{
		\includegraphics[width=0.31\linewidth]{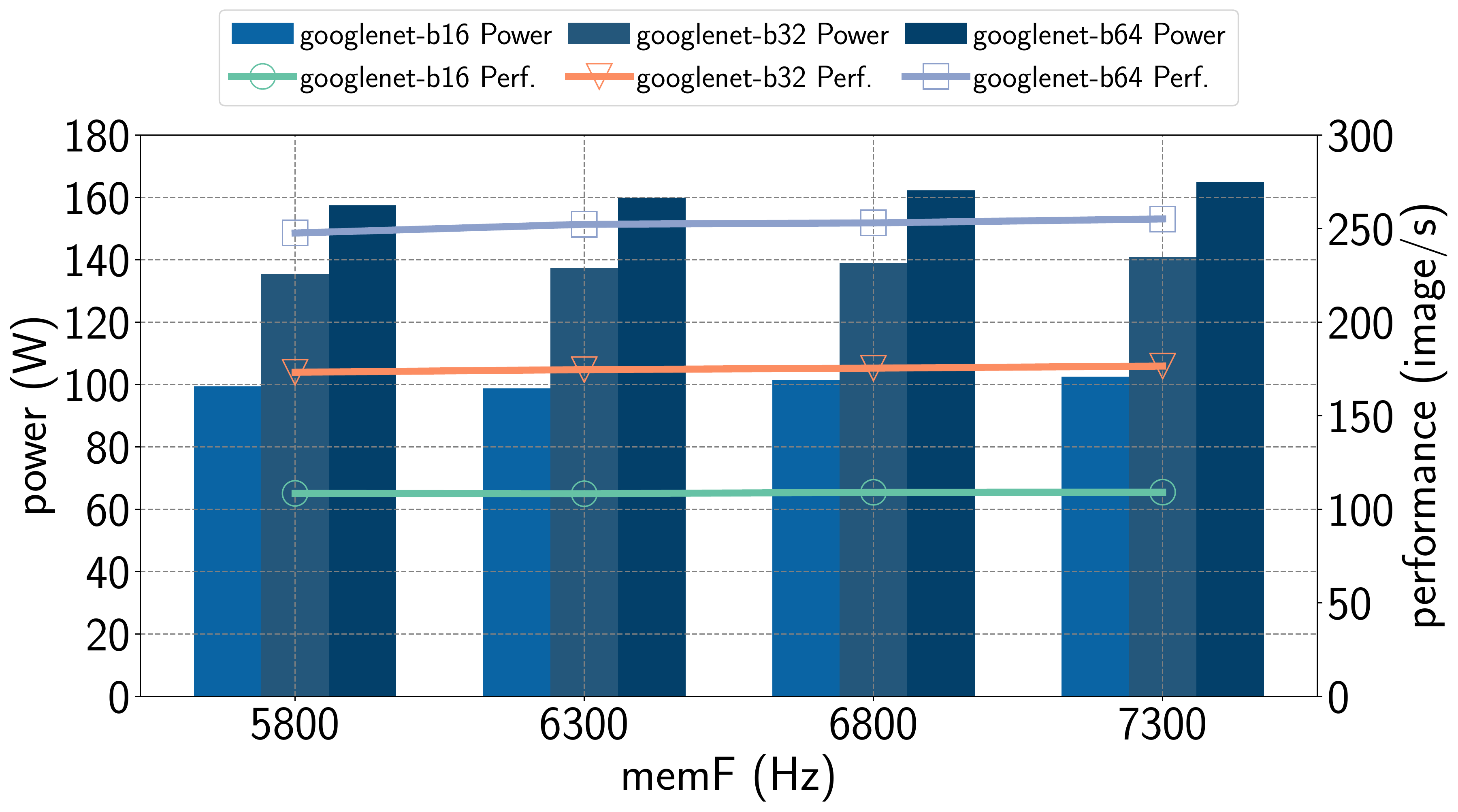}
		\label{fig:power_perf_appendix_gtx2080ti_winograd_googlenet_memF}
	}
	\subfigure[power and performance of training ResNet-50]
	{
		\includegraphics[width=0.31\linewidth]{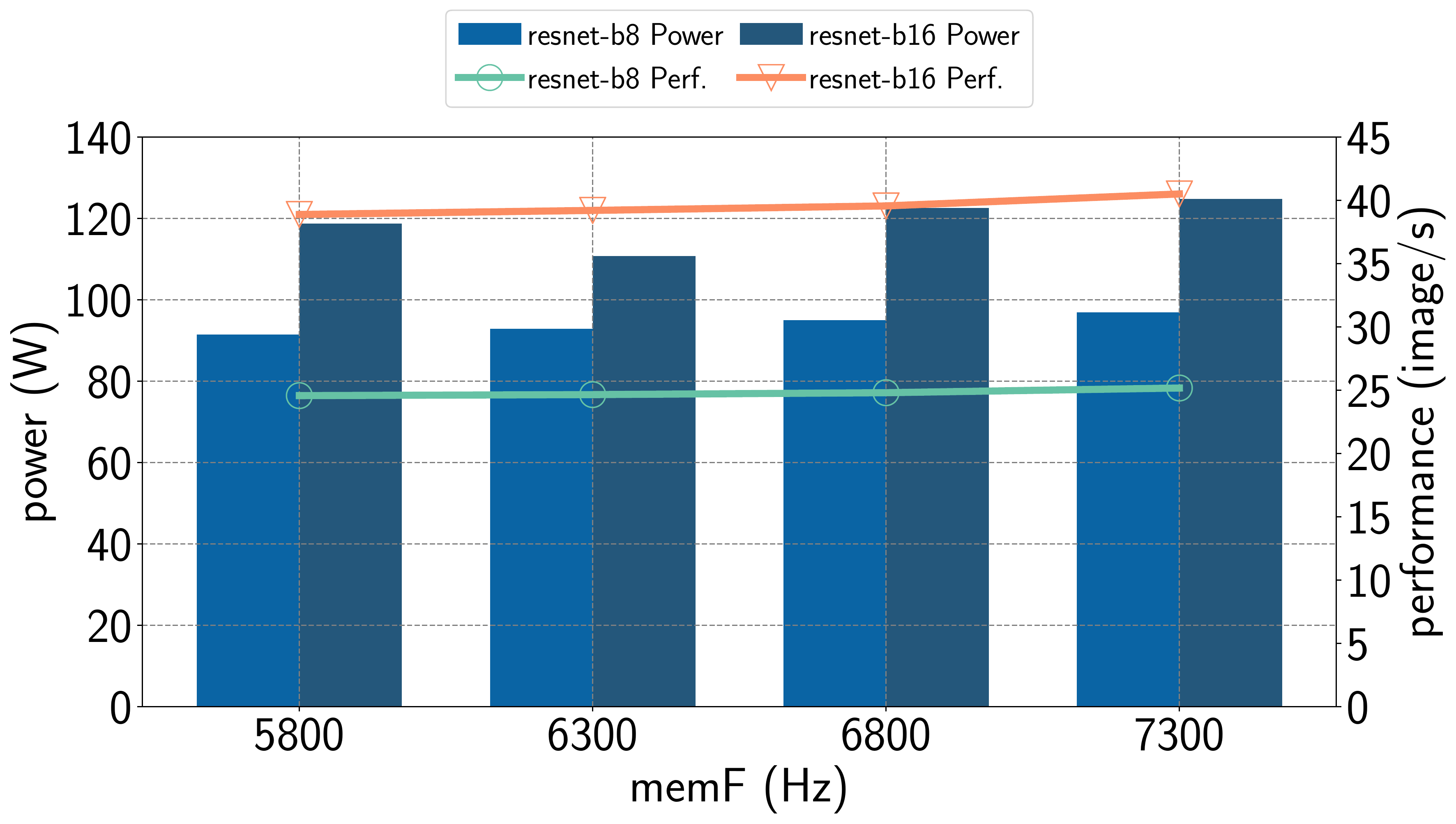}
		\label{fig:power_perf_appendix_gtx2080ti_winograd_resnet_memF}
	}
	\caption{training using Winograd on GTX2080Ti with increase of memory frequency}
	\label{fig:winograd_perf_memory_gtx2080ti}
\end{figure*}

Figures \ref{fig:winograd_perf_core_gtx2080ti} and \ref{fig:winograd_perf_memory_gtx2080ti} demonstrate the results of Winograd algorithm on GTX 2080Ti. Scaling up the core frequency leads to different performance improvement for different DNNs and even different batch sizes. The power consumption of AlexNet remains nearly the same when applying different batch sizes, while that of GoogleNet and ResNet-50 becomes larger with the increase of the batch size. Similar to GEMM, increasing the memory frequency hardly helps conserve energy since the performance cannot benefit from it.  
\subsection{Using FFT algorithm}
\begin{figure*}[htbp]
	\centering     
	\subfigure[power and performance of training AlexNet]
	{
		\includegraphics[width=0.4\linewidth]{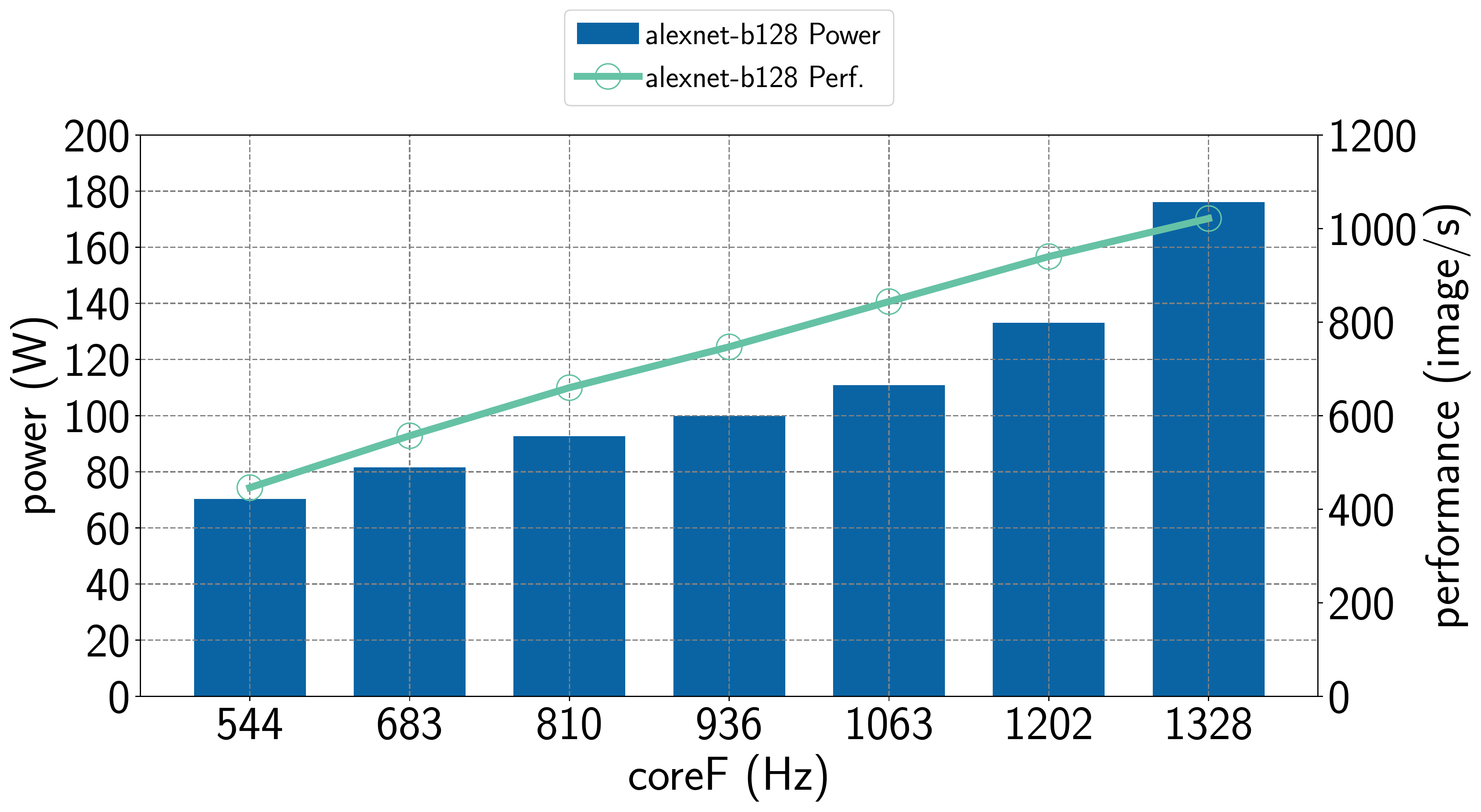}
		\label{fig:power_perf_appendix_p100_fft_tile_alexnet}
	}
	\subfigure[power and performance of training GoogleNet]
	{
		\includegraphics[width=0.4\linewidth]{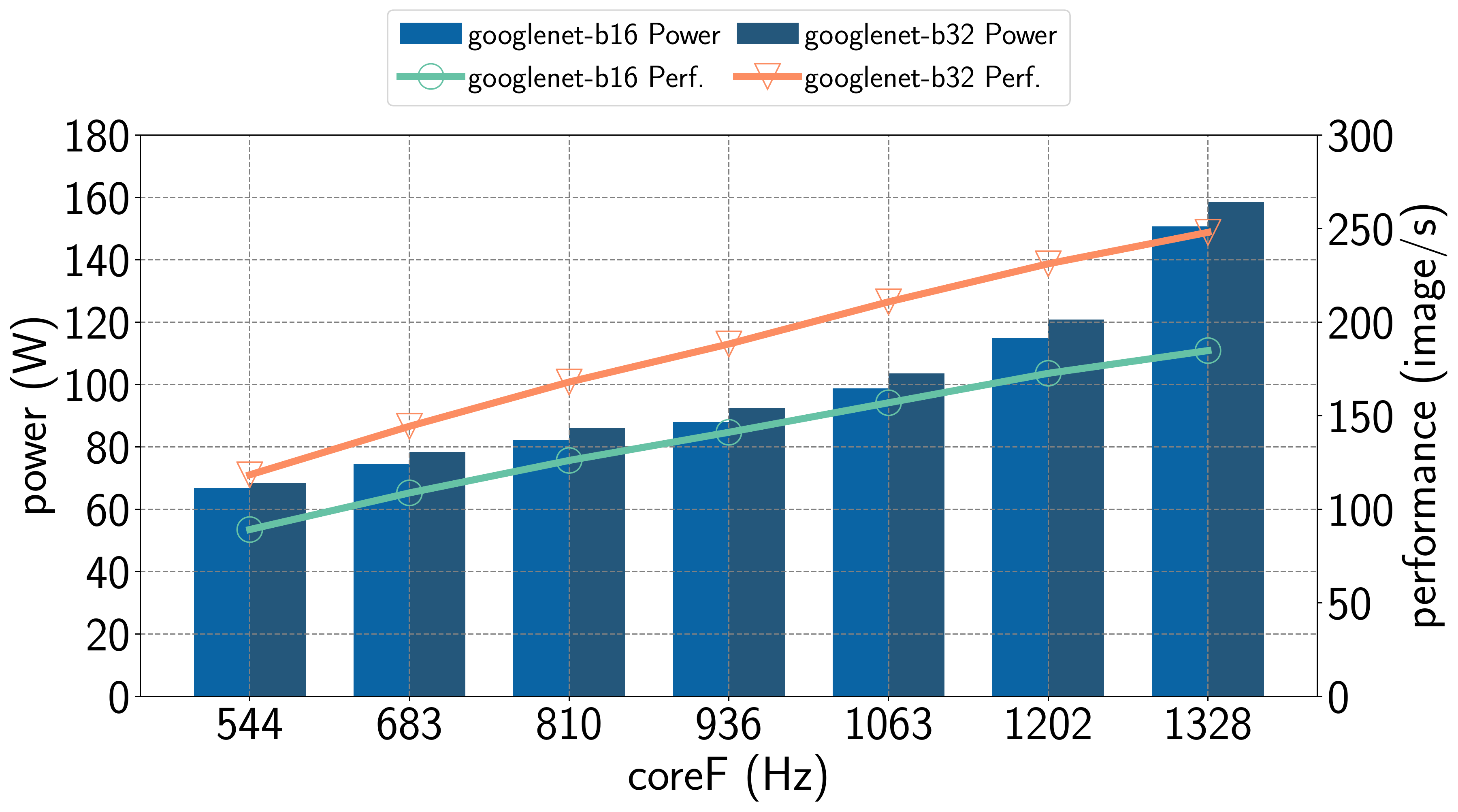}
		\label{fig:power_perf_appendix_p100_fft_tile_googlenet}
	}
	\caption{training using FFT on P100 with increase of core frequency}
	\label{fig:fft_perf_p100}
\end{figure*}

\begin{figure*}[htbp]
	\centering     
	\subfigure[power and performance of training AlexNet]
	{
		\includegraphics[width=0.4\linewidth]{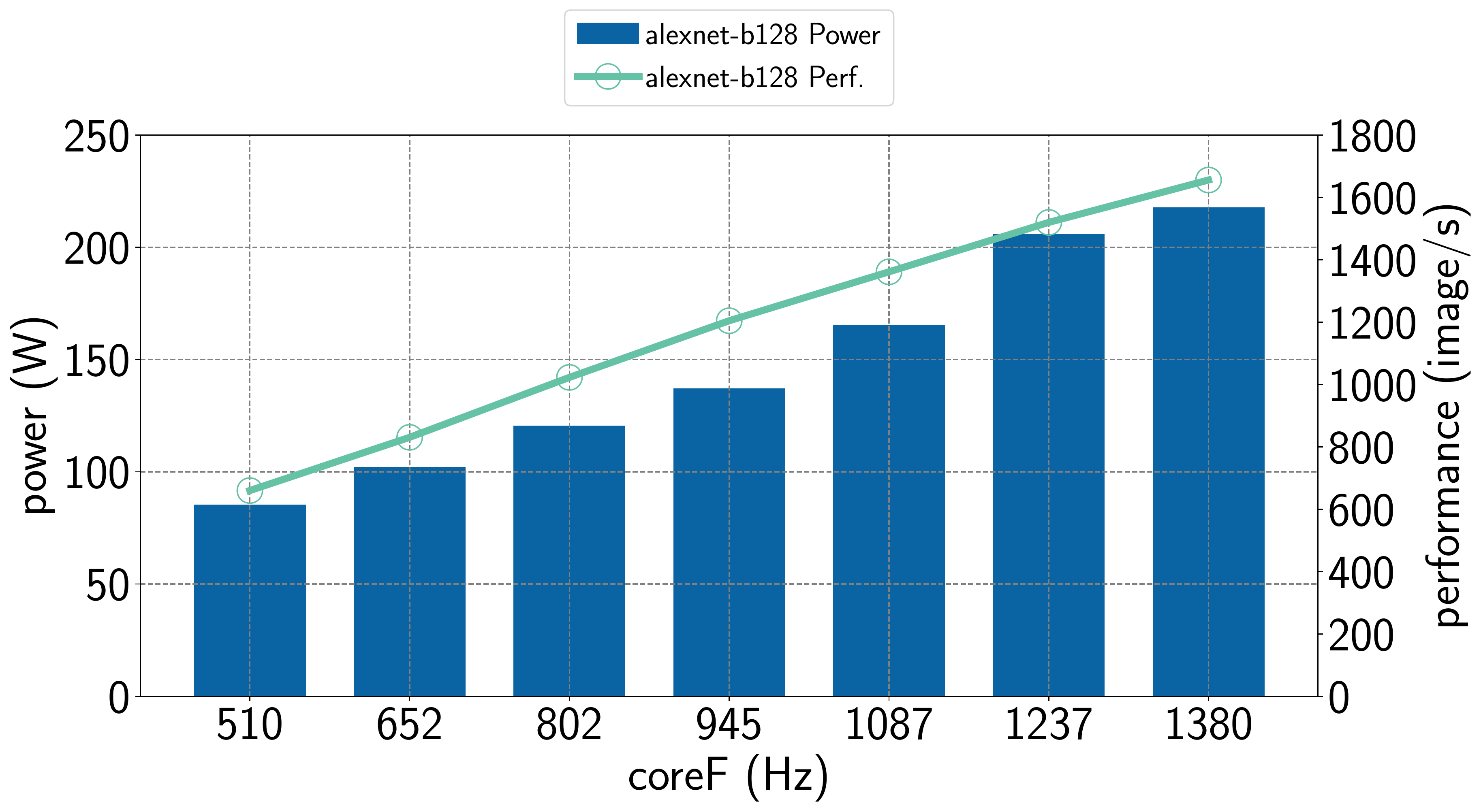}
		\label{fig:power_perf_appendix_v100_fft_tile_alexnet}
	}
	\subfigure[power and performance of training GoogleNet]
	{
		\includegraphics[width=0.4\linewidth]{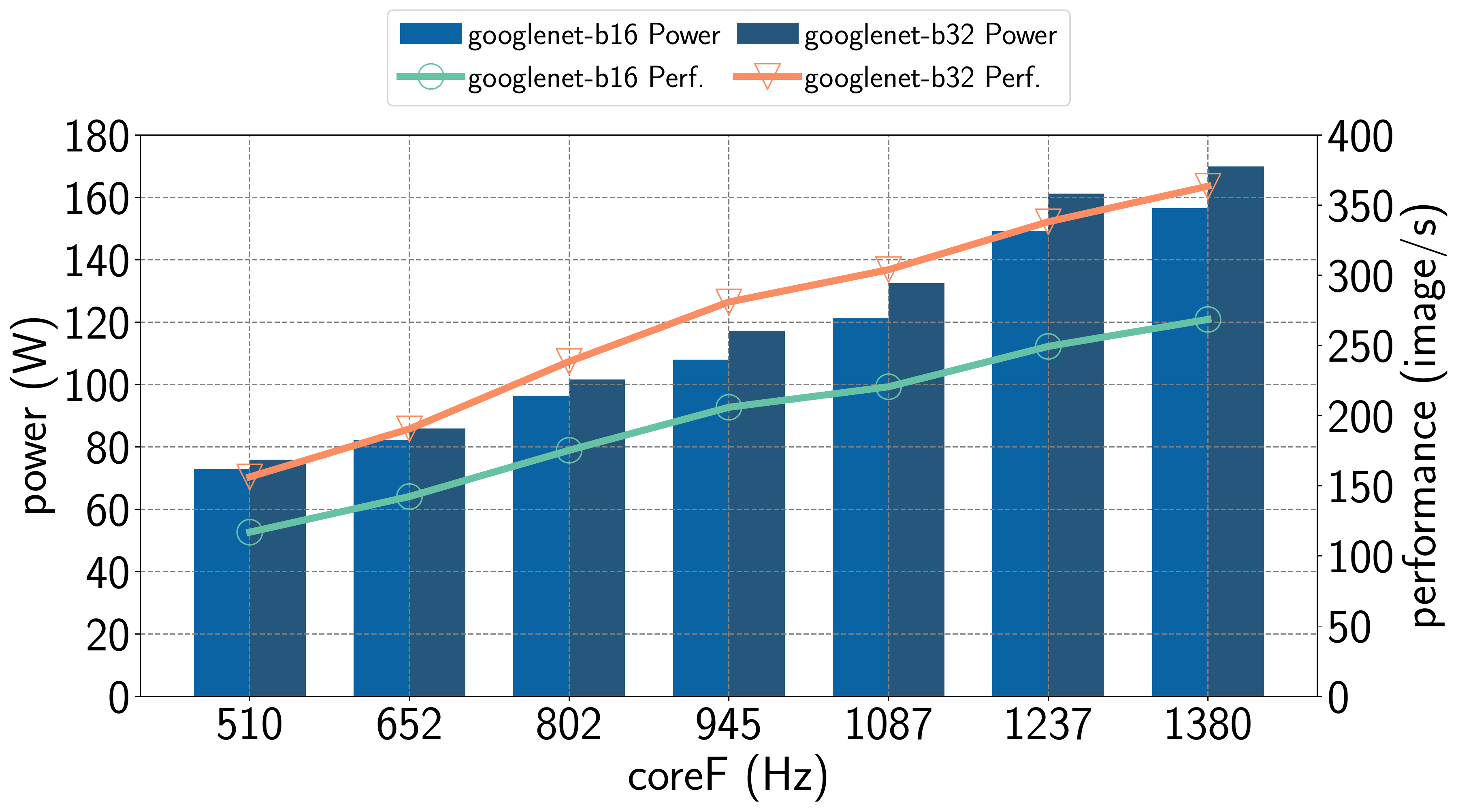}
		\label{fig:power_perf_appendix_v100_fft_tile_googlenet}
	}
	\caption{training using FFT on V100 with increase of core frequency}
	\label{fig:fft_perf_v100}
\end{figure*}

Figures \ref{fig:fft_perf_p100} and \ref{fig:fft_perf_v100} demonstrate the results of FFT algorithm on two Tesla GPUs, P100 and V100. FFT requires the largest GPU memory to tackle convolution among three algorithms. The tested GPUs can only process small batch sizes when applying FFT for DNN training. Similar to the previous two algorithms, the performance grows faster than the power for most cases with the increase of the core frequency. GoogleNet also achieves a higher image processing throughput with a larger batch size. 

\begin{figure*}[htbp]
	\centering     
	\subfigure[power and performance of training AlexNet]
	{
		\includegraphics[width=0.4\linewidth]{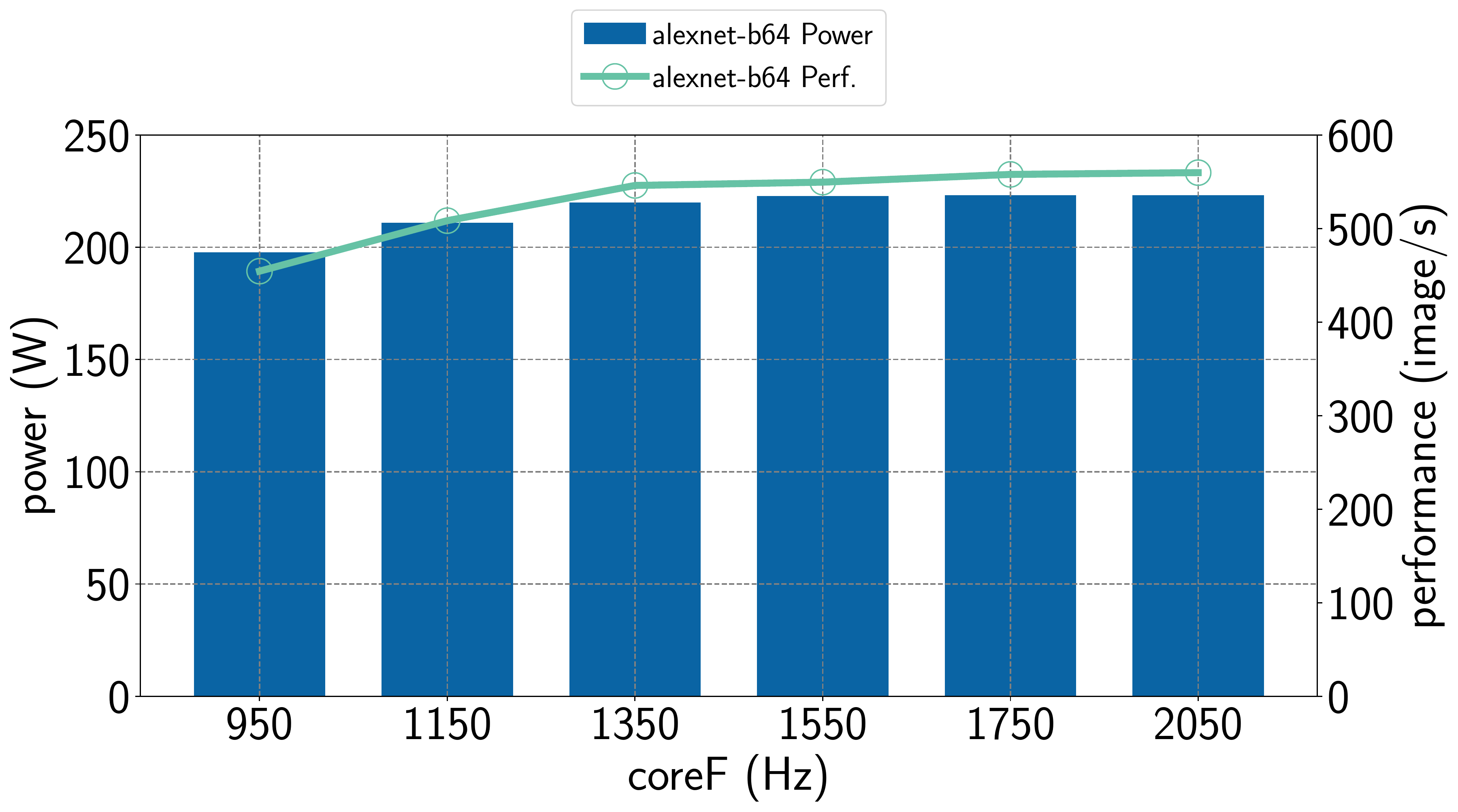}
		\label{fig:power_perf_appendix_gtx2080ti_fft_tile_alexnet_coreF}
	}
	\subfigure[power and performance of training GoogleNet]
	{
		\includegraphics[width=0.4\linewidth]{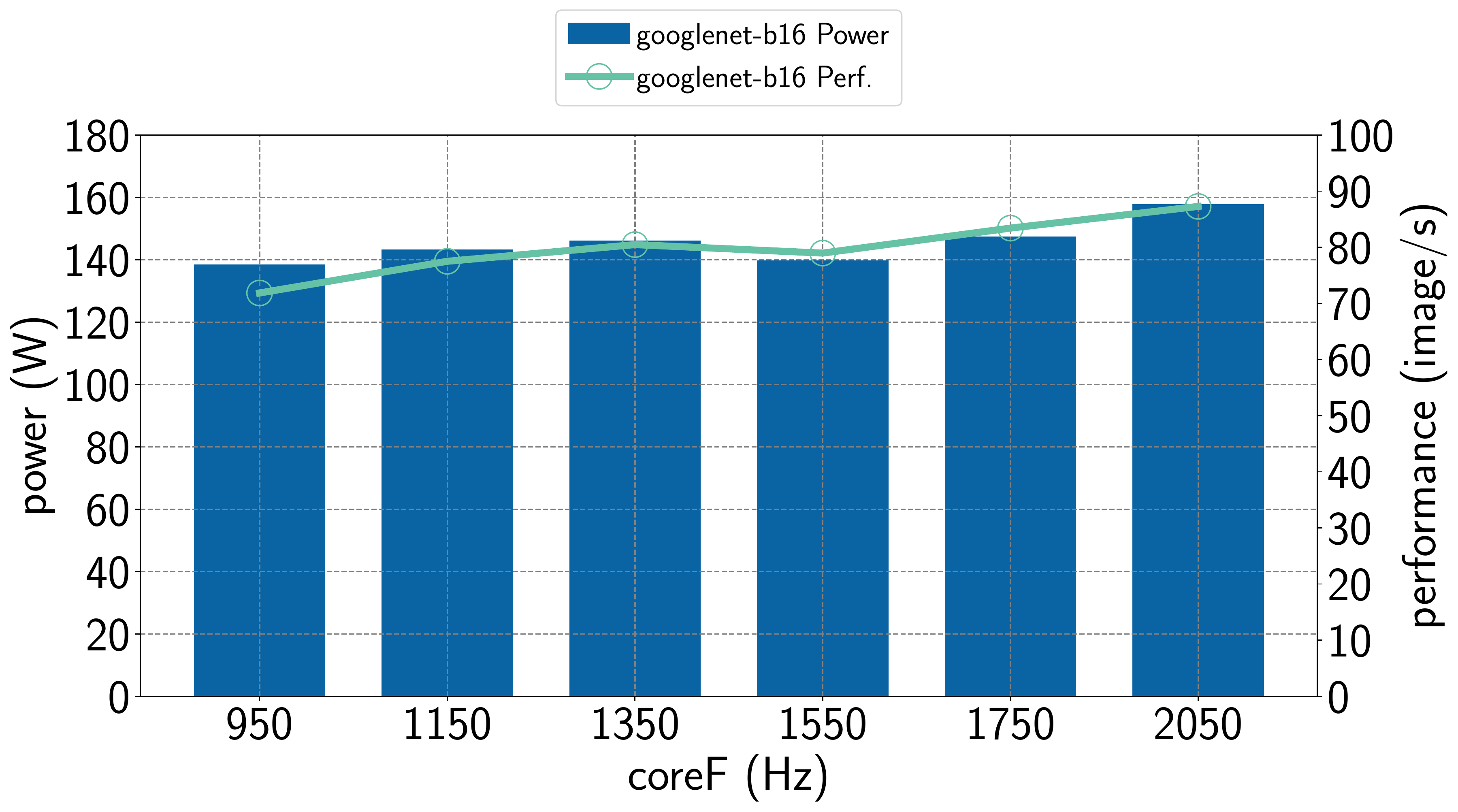}
		\label{fig:power_perf_appendix_gtx2080ti_fft_tile_googlenet_coreF}
	}
	\caption{training using FFT on GTX2080Ti with increase of core frequency}
	\label{fig:fft_perf_core_gtx2080ti}
\end{figure*}

\begin{figure*}[htbp]
	\centering     
	\subfigure[power and performance of training AlexNet]
	{
		\includegraphics[width=0.4\linewidth]{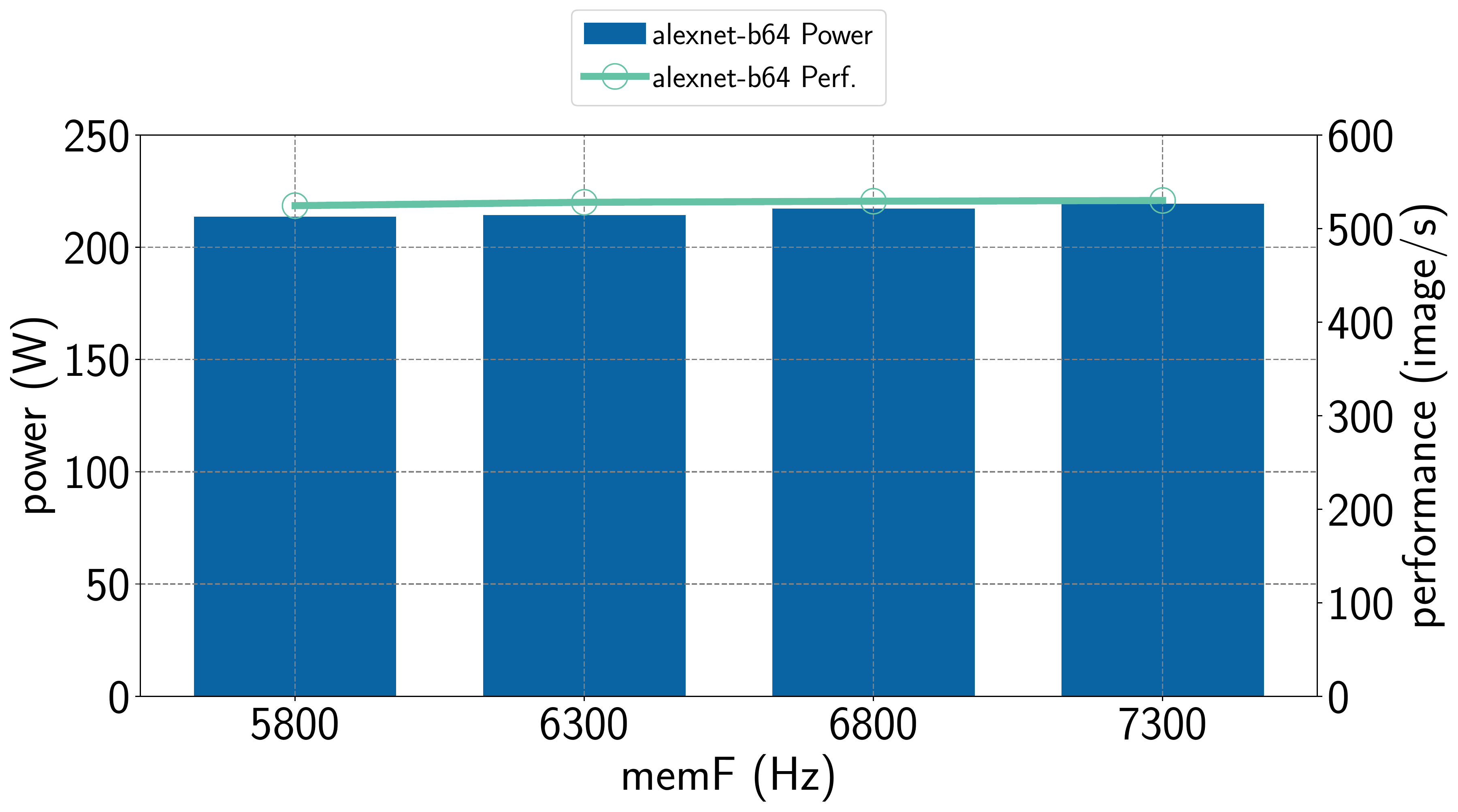}
		\label{fig:power_perf_appendix_gtx2080ti_fft_tile_alexnet_memF}
	}
	\subfigure[power and performance of training GoogleNet]
	{
		\includegraphics[width=0.4\linewidth]{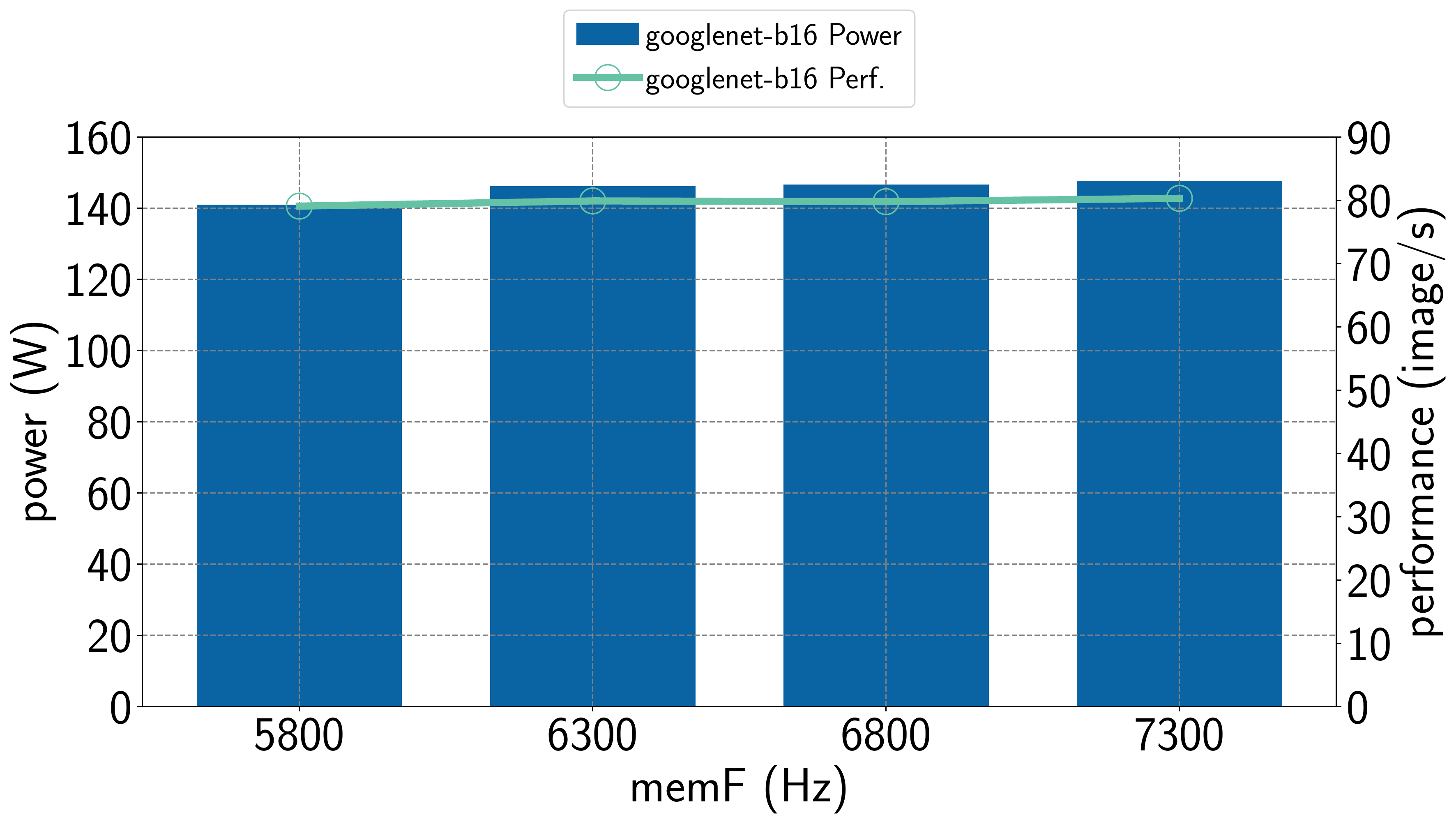}
		\label{fig:power_perf_appendix_gtx2080ti_fft_tile_googlenet_memF}
	}
	\caption{training using FFT on GTX2080Ti with increase of memory frequency}
	\label{fig:fft_perf_memory_gtx2080ti}
\end{figure*}

Figures \ref{fig:fft_perf_core_gtx2080ti} and \ref{fig:fft_perf_memory_gtx2080ti} demonstrate the results of FFT algorithm on GTX 2080Ti. Unfortunately, no matter for the performance and the power consumption, scaling up both the core and memory frequencies brings few benefits. It seems that the current implementation of FFT-based convolution on Turing GPUs still cannot fully utilize the computational resources. Besides, FFT generally works slower than the other two algorithms when the convolutional kernel size is small. Since the convolution layers of four tested DNNs features small kernel sizes, it is difficult for FFT to beat GEMM and Winograd in DNN applications. 

\end{document}